\documentclass[pra,aps,twocolumn,showpacs]{revtex4-2}
\usepackage{amsmath,amssymb,graphicx}
\usepackage{float}
\usepackage{comment}
\usepackage{bm}
\usepackage{bbold}

\usepackage{graphicx}% Include figure files
\usepackage{dcolumn}% Align table columns on decimal point
\usepackage{bm}% bold math
\usepackage{hyperref}% add hypertext capabilities
\usepackage{xcolor}
\usepackage{braket}
\usepackage{textcomp}
\usepackage{comment}
\usepackage{inputenc}
\begin{document}

\title{Photonic quantum metrology with variational quantum optical non-linearities
}

\author{A.\ Muñoz de las Heras}
\email{alberto.munoz@iff.csic.es}
\affiliation{%
Institute of Fundamental Physics IFF-CSIC, Calle Serrano 113b, 28006 Madrid, Spain}%
\author{C.\ Tabares}
\affiliation{%
Institute of Fundamental Physics IFF-CSIC, Calle Serrano 113b, 28006 Madrid, Spain}%
\author{J.\ T.\ Schneider}
\affiliation{%
Institute of Fundamental Physics IFF-CSIC, Calle Serrano 113b, 28006 Madrid, Spain}%
\author{L.\ Tagliacozzo}
\affiliation{%
Institute of Fundamental Physics IFF-CSIC, Calle Serrano 113b, 28006 Madrid, Spain}%
\author{D.\ Porras}
\affiliation{%
Institute of Fundamental Physics IFF-CSIC, Calle Serrano 113b, 28006 Madrid, Spain}%
\author{A.\ González-Tudela}
\affiliation{%
Institute of Fundamental Physics IFF-CSIC, Calle Serrano 113b, 28006 Madrid, Spain}%

\date{\today}

\begin{abstract}

Photonic quantum metrology harnesses quantum states of light, such as NOON or Twin-Fock states, to measure unknown parameters beyond classical precision limits. Current protocols suffer from two severe limitations that preclude their scalability: the exponential decrease in fidelities (or probabilities) when generating states with large photon numbers due to gate errors, and the increased sensitivity of such states to noise. Here, we develop a deterministic protocol combining quantum optical non-linearities and variational quantum algorithms that provides a substantial improvement on both fronts. First, we show how the variational protocol can generate metrologically-relevant states with a small number of operations which does not significantly depend on photon-number, resulting in exponential improvements in fidelities when gate errors are considered. Second, we show that such states offer a better robustness to noise compared to other states in the literature. Since our protocol harnesses interactions already appearing in state-of-the-art setups, such as cavity QED, we expect that it will lead to more scalable photonic quantum metrology in the near future.

\end{abstract}

\maketitle

%%%%%%%%%%%%%%%%%%%%%%%%%%%%%%%%%%%%%%%%%%%%
\section{Introduction} 
\label{sec:Introduction}

Quantum metrology capitalizes on quantum resources 
to improve measurement precision beyond classical limits~\cite{Bollinger1996,Leibfried2004,Giovannetti2004,Giovannetti2006,giovannetti11a,Toth_2014,Degen2017}.
Classically, the estimation error of an unknown parameter $\varphi$ using $N$ probes is bound by the standard quantum limit (SQL) $\Delta\varphi \geq 1/\sqrt{N}$.
However, entangled probes can offer a quadratic improvement over the SQL, reaching the so-called Heisenberg limit (HL) $\Delta\varphi=1/N$. 
In the photonic scenario of phase estimation~\cite{Dowling2015,DEMKOWICZDOBRZANSKI2015,Pirandola2018,Polino2020}, quantum states of light 
such as NOON~\cite{Holland1993} or Twin-Fock states (TFS), i.e., the same Fock state at each arm of a Mach--Zehnder interferometer (MZI)~\cite{campos03a}, overcome the SQL (even reaching the HL in the case of NOON states). Proof-of-principle experiments have already shown the potential of this approach, but so far restricted to up to five photons~\cite{Afek2010, Israel2012,Rozema2014}. The underlying reason behind such low numbers is that photonic quantum metrology suffers from two limitations: i) State-of-the-art methods to generate metrologically-relevant states involve a number of operations~\cite{Vogel1993,Law1996,Gonzalez-Tudela2015} or interaction time~\cite{Uria2020,Uria2023} increasing with the number of photons. This ultimately yields an exponential fidelity decrease with photon number when gate errors are considered. A way of improving fidelities consists in using post-selection~\cite{Guerlin2007,Deleglise2008,Sayrin2011,deng2023heisenberglimited,Kok2002,Walther2004,Afek2010, Israel2012,Wang2016b,Wang2019a}, but at the price of vanishingly small probabilities with a growing photon number. ii) The resource entangled states, such as NOON ones, suffer from an increased sensitivity to noise, e.g., decoherence and photon loss in the channel, spoiling their quantum advantage even if generated accurately. Thus, innovative ideas are required to scale photonic quantum metrology protocols beyond proof-of-principle realizations.

Recently, variational quantum algorithms (VQAs)~\cite{Cerezo2021,Bharti2022} have emerged as a tool to make the best out of current quantum hardware, which is noisy and thus can perform a limited number of coherent operations.
The key idea of these hybrid algorithms is to use a classical optimizer to find the set of parameters of a parametrized quantum circuit (PQC) implemented on the hardware such that it minimizes a given cost function. Recent works on spin systems have shown how these VQAs can also be useful in the context of quantum metrology~\cite{Kaubruegger2019,sun2023variational,Meyer2021, Ma2021, le2023variational,Koczor2020,Beckey2022,Yang2022,Kaubruegger2023}, e.g., by using the quantum Fisher information (QFI)~\cite{Koczor2020,Beckey2022,Yang2022} as cost function. However, in the photonic context this potential of variational approaches for quantum metrology has been scarcely explored, and limited to either linear systems~\cite{KRISNANDA2021,cimini2023variational} or PQCs with fixed non-linearities~\cite{Steinbrecher2019}.

\begin{figure*}[tb]
    \centering
    \includegraphics[width=\linewidth]{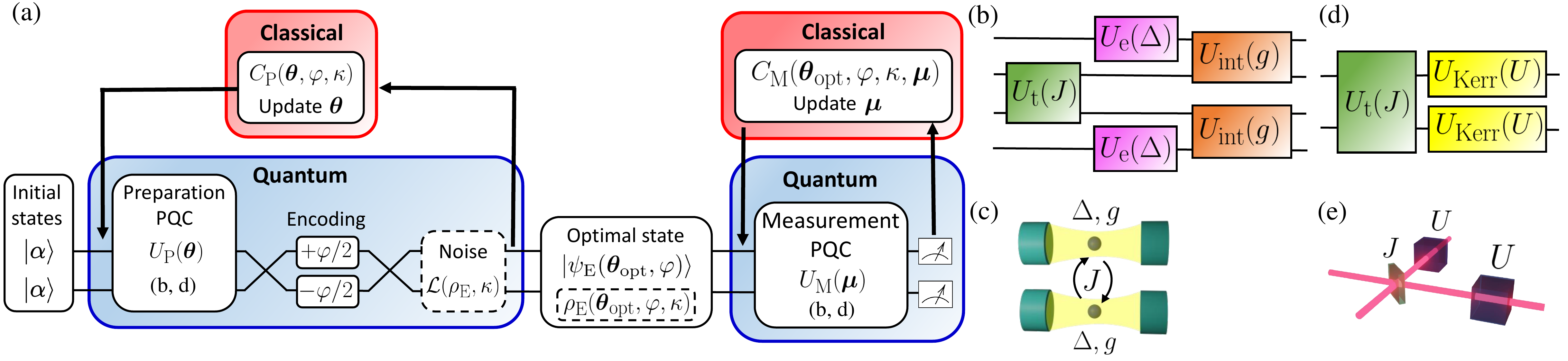}
    \caption{(a) Overview of the variational optimization protocol. Two identical coherent states $\ket{\alpha}$ are the input of a variational quantum algorithm (VQA) aimed at finding the optimal state for the estimation of a phase $\varphi$. This consists of a quantum part including a parametrized quantum circuit (PQC) described by a unitary $U_{\rm P}$, a Mach--Zehnder interferometer (MZI) encoding the phase difference between its two arms, and (optionally) a non-unitary evolution accounting for noise with decay rate $\kappa$ described by a Lindbladian $\mathcal{L}$ acting on the density matrix of the system $\rho_{\rm E}$. The classical part of the VQA is an optimizer that changes the parameters $\boldsymbol{\theta}$ of the PQC in search of the minimum of the cost function $C_{\rm P}$. The resulting optimal state $\ket{\psi_{\rm E}}$ ($\rho_{\rm E}$ in the noisy case) evaluated at the optimal parameters $\boldsymbol{\theta}_{\rm opt}$ is the input of the second part of the VQA, which employs the PQC to prepare optimal measurements as well as a classical optimizer that aims at minimizing the cost function $C_{\rm M}$ by varying the parameters $\boldsymbol{\mu}$ of the second PQC. (b) A single layer of the emitters ansatz, consisting of a tunneling unitary $U_{\rm t}$ depending on the tunneling amplitude $J$, a detuning unitary $U_{\rm e}$ for each mode depending on the cavity-emitters detuning $\Delta$, and an interaction unitary $U_{\rm int}$ for each mode depending on the light-matter coupling strength $g$. 
    The upper and lower modes of the scheme correspond to the emitters, while the two central ones are the photonic modes. 
    (c) A possible implementation of the emitters ansatz in a cavity-QED setup. (d) A single layer of the Kerr ansatz, consisting of a tunneling unitary $U_{\rm t}$ depending on the tunneling amplitude $J$ and a Kerr unitary $U_{\rm Kerr}$ for each mode depending on the non-linearity strength $U$. (e) A possible implementation of the Kerr ansatz in a photonic setup.}
    \label{fig:scheme}
\end{figure*}

In this work, we combine VQAs with state-of-the-art quantum optical non-linearities to design an algorithm that overcomes the limitations of current protocols. In particular, we consider two types of PQCs (the \textit{ansätze}) each formed by two coupled cavity systems but featuring different types of non-linearities: the coupling to a two-level system that appears in cavity QED and a Kerr-type one. Our method employs the QFI as cost function to find the optimal parameters that transform unentangled coherent states into states that approximately saturate the HL. Importantly, we find that the number of operations required is independent of the photon number with both types of non-linearities, which guarantees that the fidelity of the generated states will not decrease with the photon number, unlike in existing protocols. For the two ansätze, we consider the impact of noise and show that the generated states feature a larger robustness than NOON and TFS. In a second step of the VQA, we consider photon number measurements and maximize the classical Fisher information (CFI) to find the optimal measurement within that scheme.  Our variational approach can be applied following two different strategies: \textit{in situ}~\cite{Kimble1998,cimini2023variational,on2023programmable}, i.e., optimizing the PQC directly on the quantum hardware, or \textit{in silicon}, i.e., simulating the PQC on a classical computer and then running the quantum hardware with the optimal parameters\cite{Steinbrecher2019,Sun2021}. Codes to reproduce the results of this manuscript are available in~\cite{GitHub}.

%%%%%%%%%%%%%%%%%%%%%%%%%%%%%%%%%%%%%%%%%%%%
\section{The algorithm} 
\label{sec:The algorithm}

Let us initially restrict ourselves to the noiseless case. Our VQA can be divided into two steps: preparation and measurement, as sketched in Fig.~\ref{fig:scheme}(a) (more details can be found in Appendix~\ref{sec:algorithm}).
In the preparation stage, a PQC described by a unitary operator $U_{\rm P}(\boldsymbol{\theta})$ is applied to two cavity modes.
The initial state of each cavity is a coherent state with mean-photon number $|\alpha|^2=N/2$~\cite{gerry2005introductory}, such that the mean number of photons summing both arms is $N$~\footnote{This choice of initial states is motivated by the ease with which coherent states can be experimentally prepared. In Appendix~\ref{sec:squeezed} we also consider squeezed coherent states as inputs, but there was no significant improvement over the results}.
The resulting state $\ket{\psi_{\rm P}(\boldsymbol{\theta})}$ is then sent through a MZI consisting of a symmetric beamsplitter followed by the encoding of a phase difference $\varphi$ between the two modes and by another symmetric beamsplitter, resulting in a state $\ket{\psi_{\rm E}(\boldsymbol{\theta},\varphi)}$.
To optimize the preparation of probe states, one needs to maximize the QFI $\mathcal{F}_{\rm Q}$, thus setting the cost function to $C_{\rm P}(\boldsymbol{\theta}, \varphi) = -\mathcal{F}_{\rm Q}$.
Then, this quantity is fed to the classical optimizer, which in turn updates the parameters $\boldsymbol{\theta}$. The lower bound on the estimation error on $\varphi$ is given by the quantum Cramér--Rao bound $(\Delta\varphi)^2\geq\mathcal{F}^{-1}_{\rm Q}$~\cite{Polino2020}. The closer $\mathcal{F}^{-1}_{\rm Q}$ is to the HL, the larger the metrological potential obtained with the PQC.

To calculate the QFI, we use the approximation introduced in Ref.~\cite{Beckey2022}, see Appendix~\ref{sec:algorithm} for more details. This requires evaluating the fidelity between the states $\ket{\psi_{\rm E}(\boldsymbol{\theta},\varphi)}$ and $\ket{\psi_{\rm E}(\boldsymbol{\theta},\varphi+\delta)}$, where $\delta\rightarrow 0$ is a small phase difference. In principle, this demands a number of measurements as well as an amount of computation growing exponentially with the system size~\cite{Cramer2010}. In Appendix~\ref{sec:measuring qfi} we review some of the most promising techniques to measure the QFI---or, equivalently, the fidelity between two quantum states---aimed at alleviating this problem. Besides, the small size of the platform considered in our manuscript (hosting only two photonic modes, which implies a Hilbert space dimension scaling quadratically with $N$) will further reduce the complexity of fidelity measurements.

Since the QFI is maximized irrespective of the measurement scheme, in the second step of the protocol we assess the best way to extract the information within a specific measurement type.
In particular, we first apply a unitary $U_{\rm M}(\boldsymbol{\mu})$ to the output state $\ket{\psi_{\rm E}(\boldsymbol{\theta}_{\rm opt}, \varphi)}$ resulting from the previous optimization and then we consider a measurement in the photon-number basis. The role of the unitary, which we label as measurement PQC, is to enable the algorithm to find the best possible combination of modes before measuring. The optimal parameters $\boldsymbol{\mu}_{\rm opt}$ are found by maximizing the CFI $\mathcal{F}_{\rm C}$, which we use as the new cost function of our algorithm $C_{\rm M} (\boldsymbol{\theta}_{\rm opt}, \varphi, \boldsymbol{\mu}) = -\mathcal{F}_{\rm C}$. To compute the CFI, we construct the density matrix $\rho_{\rm M}(\boldsymbol{\theta}_{\rm opt}, \varphi, \boldsymbol{\mu})=\ket{\psi_{\rm M}(\boldsymbol{\theta}_{\rm opt}, \varphi, \boldsymbol{\mu})}\bra{\psi_{\rm M}(\boldsymbol{\theta}_{\rm opt}, \varphi, \boldsymbol{\mu})}$, where $\ket{\psi_{\rm M}(\boldsymbol{\theta}_{\rm opt}, \varphi, \boldsymbol{\mu})}=U_{\rm M}(\boldsymbol{\mu})\ket{\psi_{\rm E}(\boldsymbol{\theta}_{\rm opt}, \varphi)}$. With this expression and its derivative  $\partial\rho_{\rm M}(\boldsymbol{\theta}_{\rm opt}, \varphi, \boldsymbol{\mu})/\partial\varphi$, we calculate the CFI (see Appendix~\ref{sec:algorithm}) and optimize it. An optimal measurement should give $\mathcal{F}_{\rm C}=\mathcal{F}_{\rm Q}$; such hierarchy is again summarized by the quantum Cramér--Rao bound $(\Delta\varphi)^2 \geq \mathcal{F}^{-1}_{\rm C} \geq \mathcal{F}^{-1}_{\rm Q}$~\cite{Polino2020}.

%%%%%%%%%%%%%%%%%%%%%%%%%%%%%%%%%%%%%%%%%%%%
\section{Physical platforms \& ansätze} 
\label{sec:Physical platforms and ansatze}

A crucial element of VQAs is the chosen PQC, since it determines the solution space that the algorithm can explore. In our case, the PQC will be defined by the two architectures represented in Fig.~\ref{fig:scheme}(b-e). Their common ingredient are two coupled single-mode cavities described by bosonic annihilation (creation) operators $a^{(\dagger)}_{1,2}$, whose coupling Hamiltonian reads $H_{\rm t} = J \left(a^\dagger_{2}a_{1} + a^\dagger_{1}a_{2}\right)$, where $J$ is the tunneling rate. The difference stems in the source of non-linearity. 

On the one hand (see Fig.~\ref{fig:scheme}(b, c)), we consider a non-linearity coming from the coupling to a two-level emitter, like in cavity QED setups~\cite{Kimble1998,Walther_2006,Reiserer2015,Blais2021}. This platform enables encoding three different variational parameters per layer, namely, $J$, the emitter-cavity detuning $\Delta$, and the coupling strength $g$, whose corresponding Hamiltonians read $H^{(i)}_{\rm e} = \Delta \sigma^\dagger_i \sigma_i$ and $H^{(i)}_{\rm int} = g \left(\sigma^\dagger_i a_i + \sigma_i a^\dagger_i\right)$, respectively.
Here, $\sigma_i = \ket{g}\bra{e}$ takes the emitter $i$ from its excited state $\ket{e}$ to its ground state $\ket{g}$.
In this case, the unitary describing the PQC can be written as
\begin{align}
    U^{\rm(emit)}_{\rm P, M} = \prod^d_{j=1} U^{(2,j)}_{\rm int}U^{(1,j)}_{\rm int}U^{(2,j)}_{\rm e}U^{(1,j)}_{\rm e}U^{(j)}_{\rm t}.
\end{align}
Above, $d$ is the number of layers of the PQC, $U^{(j)}_{\rm t}=e^{-iT^{(J)}_j H_{\rm t}}$, $U^{(i,j)}_{\rm e}=e^{-iT^{(\Delta)}_j H^{(i)}_{\rm e}}$, and $U^{(i,j)}_{\rm int}=e^{-iT^{(g)}_j H^{(i)}_{\rm int}}$, where $T^{(J, \Delta, g)}_j$ is the physical time in which each term is applied. $\boldsymbol{\theta}, \boldsymbol{\mu} = \{\tilde{J}_1, \tilde{\Delta}_1, \tilde{g}_1, ... ,\tilde{J}_d, \tilde{\Delta}_d, \tilde{g}_d\}$ are the  variational parameters, each given by $\tilde{J}_j = J_j T^{(J)}_{j}$, $\tilde{\Delta}_j = \Delta_j T^{(\Delta)}_{j}$, and $\tilde{g}_j = g_j T^{(g)}_{j}$. The number of gates is $N_{\rm gates}=3d$.

On the other hand (see Fig.~\ref{fig:scheme}(d, e)), we consider the Kerr-type Hamiltonian $H^{(i)}_{\rm Kerr} = U/2\times a^\dagger_i a_i\left(a^\dagger_i a_i-1\right)$ arising, for example, from $\chi^{(3)}$ non-linearities in non-linear crystals~\cite{butcher_cotter_1990}.
With these interactions, we can write the ansatz of the Kerr non-linear circuit as follows:
\begin{align}
    U^{\rm(Kerr)}_{\rm P, M} = \prod^d_{j=1} U^{(2,j)}_{\rm Kerr}U^{(1,j)}_{\rm Kerr}U^{(j)}_{\rm t}.
\end{align}
where $\boldsymbol{\theta}, \boldsymbol{\mu} = \{\tilde{J}_1, \tilde{U}_1, ... ,\tilde{J}_d, \tilde{U}_d\}$ are the variational parameters (two per layer), each given by $\tilde{J}_j = J_j T^{(J)}_{j}$ and $\tilde{U}_j = U_j T^{(U)}_{j}$. $T^{(J, U)}_j$ is the physical time in which each term is applied. The number of gates is $N_{\rm gates}=2d$. Compared to earlier works~\cite{Steinbrecher2019}, we use the Kerr non-linearity as a variational parameter to check if it can provide an advantage over fixed $U$-ansätze. A discussion on the physical realization of tunable optical non-linearities is included in Appendix~\ref{sec:implementation}, both for the Kerr non-linearity and the photon-emitter interaction. 

Ideally, we would like the VQA to use an ansatz with as few gates as possible. The reason is that, e.g., if we assume a constant error per gate $\varepsilon$, the overall fidelity of state generation after performing $N_{\rm gates}$ will be $(1-\varepsilon)^{N_{\rm gates}}$.

\begin{figure}[tb]
    \centering
    \includegraphics[width=\linewidth]{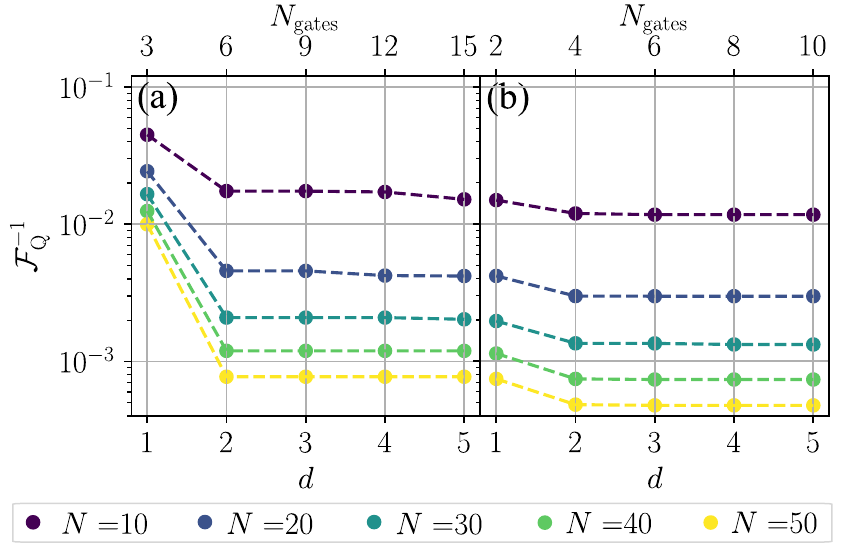}
    \caption{Scaling of the inverse of the QFI $\mathcal{F}^{-1}_{\rm Q}$ as a function of the number of layers $d$ (and gates, $N_{\rm gates}$) of the PQC, for different values of the mean number of photons $N$. (a) Emitters ansatz. (b) Kerr ansatz.}
    \label{fig:scaling d}
\end{figure}

%%%%%%%%%%%%%%%%%%%%%%%%%%%%%%%%%%%%%%%%%%%%
\section{Noiseless results} 
\label{sec:Noiseless results}

In Fig.~\ref{fig:scaling d} we show the convergence of our algorithm for the QFI with respect to the number of layers $d$ (and gates, $N_{\rm gates}$) of the preparation PQC  for the emitters (panel a) and the Kerr non-linearity (panel b) ansätze. Both panels show  $\mathcal{F}^{-1}_{\rm Q}$ as $d$ is varied for different mean number of photons $N$ ranging from $N=10$ to $N=50$. 
For both ansätze convergence is rapidly obtained with only two layers, making our protocol extraordinarily efficient in terms of circuit depth. What is more important, the value of $d$ at which convergence is attained does not depend on $N$, at least for the range of $N$ studied. This is in stark contrast with state-of-art protocols~\cite{Law1996,Uria2020} in which $N_\mathrm{gates}\sim N$, and thus the fidelity decays exponentially the number of  photons.

\begin{figure}[tb]
    \centering
    \includegraphics[width=\linewidth]{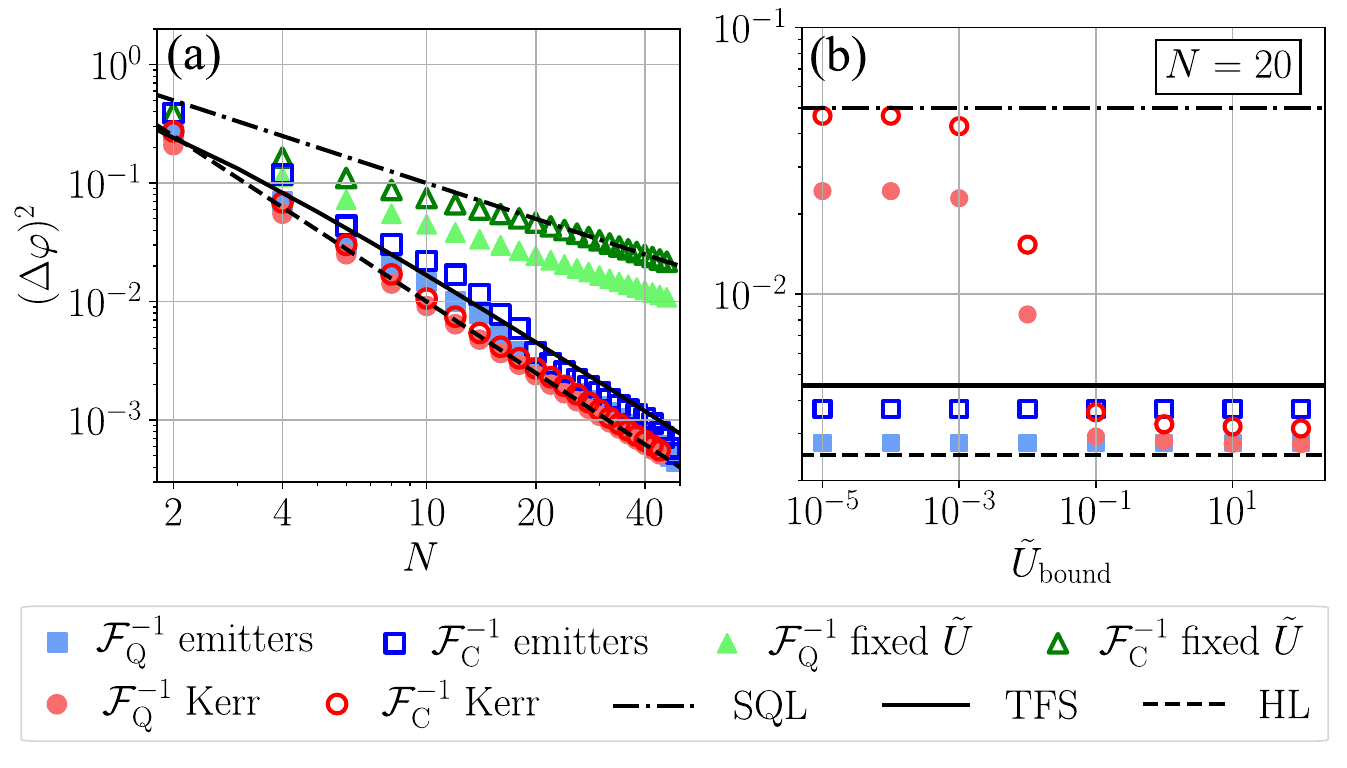}
    \caption{Estimation error $(\Delta\varphi)^2$ in the noiseless scenario. Panel (a) shows our results as a function of the mean number of photons $N$. In panel (b), we address the effect of a bound in the Kerr non-linearity strength $\tilde{U}_{\rm bound}$ for a mean number of photons $N=20$. In both panels, blue squares, red circles, and green triangles correspond to the results of the emitters ansatz, the Kerr ansatz with unrestricted $\tilde{U}$, and the Kerr ansatz with fixed $\tilde{U}=2\pi$, respectively: filled (void) markers are the inverse of the QFI (CFI) $\mathcal{F}^{-1}_{\rm Q(C)}$ in the two cases. Dashed-dotted/solid/dashed lines signal the SQL/TFS/HL scaling. In both panels, we employed a circuit depth $d=5$.}
    \label{fig:noiseless}
\end{figure}

Once we guarantee the convergence of the preparation step, we study whether optimal probe states and optimal measurements can be obtained with our protocol.
Our results are shown in Fig.~\ref{fig:noiseless}. In panel (a) we plot the estimation error $(\Delta\varphi)^2$ as a function of the mean number of photons $N$ in the VQA protocol for a circuit depth $d=5$~\footnote{In Figs.~\ref{fig:noiseless}-\ref{fig:noise} we employ a circuit depth $d=5$ larger than the number of layers $d=2$ necessary to attain convergence because we found a slight improvement with respect to the latter case. Results for $d=2$ can be found in Appendix~\ref{sec:d=2}.
}. 
The values of $\mathcal{F}^{-1}_{\rm Q}$ obtained with both ansätze are smaller than $(\Delta\varphi)^2$ for TFS with $N/2$ Fock states in each arm~\cite{Holland1993}. While the optimal states produced by the Kerr ansatz saturate the HL for small $N$ and remain very close to it as $N$ grows, those generated by the emitters ansatz start approaching the HL at $N\gtrsim 20$. A linear fit reveals that our results follow a scaling $\mathcal{F}^{-1}_{\rm Q}\sim 1/N^{\beta}$ very similar to that of the HL: $\beta = 2.0$ for the emitters ansatz and $\beta=1.95$ for the Kerr one.
More details on the nature of the states prepared by our VQA, as well as a comparision with NOON states and TFS, are provided in Appendix~\ref{sec:states}.
Regarding $\mathcal{F}^{-1}_{\rm C}$, the results of the Kerr ansatz are very close to the respective values of $\mathcal{F}^{-1}_{\rm Q}$. In the case of the emitters ansatz, the CFI closely follows the QFI, although complete saturation is not attained. In any case, both ansätze are able to prepare almost optimal measurements. We benchmarked these results with those of the Kerr ansatz featuring a fixed value of the non-linearity variational parameter $\tilde{U}=2\pi$, as in Ref.~\cite{Steinbrecher2019}.
In this case, $\mathcal{F}^{-1}_{\rm Q,C}$ tend to the classical $1/N$ scaling, signaling that the tunability of the non-linearity strength is a crucial factor to obtain metrological advantage. 

To implement our protocol in an experiment, one needs to verify that the optimal values of the variational parameters are within the reach of state-of-the-art optical platforms. In particular, very strong optical non-linearities may not be physically realizable. In Appendix~\ref{sec:parameters}, we show that for our tunable non-linearity ansätze the maximum values of $\tilde{g}$ and $\tilde{U}$ required are of the order of $1$. Since $T^{(J,\Delta,g,U)}_i$ are limited by the coherence time $\kappa^{-1}$, $g$ and $U$ must be smaller than the typical decoherence rates in the systems.
Such coherence times are within the reach of certain cavity-QED platforms in both the microwave and optical regimes~\cite{GU2017,Meschede1985,Thompson1992,Lodahl2015,SanchezMunoz2018}.
However, in Kerr optical cavities the current record is held by polaritonic systems with only $U/\kappa\sim 10^{-2}$~\cite{Delteil2019,Munoz-Matutano2019}, while microwave resonators reach $U/\kappa\sim 10^{2}$~\cite{Koch2007,Kirchmair2013,Puri2017}.
This limitation motivated us to study the effect of restricted Kerr non-linearities in the optimization. For that, we introduce a bound $\tilde{U}\in[-\tilde{U}_{\rm bound},\tilde{U}_{\rm bound}]$ in the range of parameters that the optimizer can explore and study the dependence of $\mathcal{F}_{\rm Q}$ and $\mathcal{F}_{\rm C}$ on $\tilde{U}_{\rm bound}$. The results are displayed in Fig.~\ref{fig:noiseless}(b) for fixed $N=20$, showing two distinct behaviors for $\tilde{U}_{\rm bound}\lesssim 10^{-3}$ and $\tilde{U}_{\rm bound}\gtrsim 10^{-1}$ with a continuous crossover in between. In the former regime, which is the one realistically achievable with state-of-the-art optical platforms, the bound prevents the optimizer from exploring the region of the Hilbert space where the minimum of the cost function lies, and the resulting values of QFI and CFI are the ones that would be obtained using coherent states at the input of the Mach--Zehnder interferometer (see Appendix~\ref{sec:kerr below threshold}). However, above a critical value of $\tilde{U}_{\rm bound}$, the optimizer finds a very similar solution to that obtained using the emitters ansatz. In Appendix~\ref{sec:critical bound N}, we study the dependence of this critical $\tilde{U}_{\rm bound}$ on different mean-photon numbers, showing that it does not depend strongly on $N$ for the photon numbers we can explore.

%%%%%%%%%%%%%%%%%%%%%%%%%%%%%%%%%%%%%%%%%%%%
\section{Effect of noise} 
\label{sec:Effect of noise}

As a last step, we extend the previous study to a more realistic situation including noise in the quantum channels.
Formally, we do it by constructing the density matrix $\rho_{\rm E}(\boldsymbol{\theta}_{\rm opt}, \varphi)=\ket{\psi_{\rm E}(\boldsymbol{\theta}_{\rm opt}, \varphi)}\bra{\psi_{\rm E}(\boldsymbol{\theta}_{\rm opt}, \varphi)}$ after the MZI and letting it experience a non-unitary evolution according to the Lindblad master equation $ \dot{\rho}_{\rm E} = \kappa\sum^2_{i=1} \left[ L_i \rho_{\rm E}(t) L^\dagger_{i} -\frac{1}{2}\{L^\dagger_i  L_i, \rho_{\rm E}(t) \} \right]$, where $\{L_i\}$ is the set of jump operators describing the noise channel and $\kappa$ is the loss rate, which for simplicity we assume to be equal for all channels.
As a further simplification, we include noise of two types only on the photonic degrees of freedom: amplitude (i.e., photon loss) and phase damping (i.e., decoherence)~\cite{gardiner2004quantum}. For the former, the set of jump operators is $\{a_1,a_2\}$, resulting in a decay of the photonic population in both modes. In the latter, the set of jump operators is $\{a^\dagger_1 a_1, a^\dagger_2 a_2\}$. This in turn preserves the diagonal elements of the density matrix (i.e., the occupation probabilities) while producing a decay in its off-diagonal elements (i.e., erasing the coherences).

\begin{figure}[tb]
    \centering
    \includegraphics[width=\linewidth]{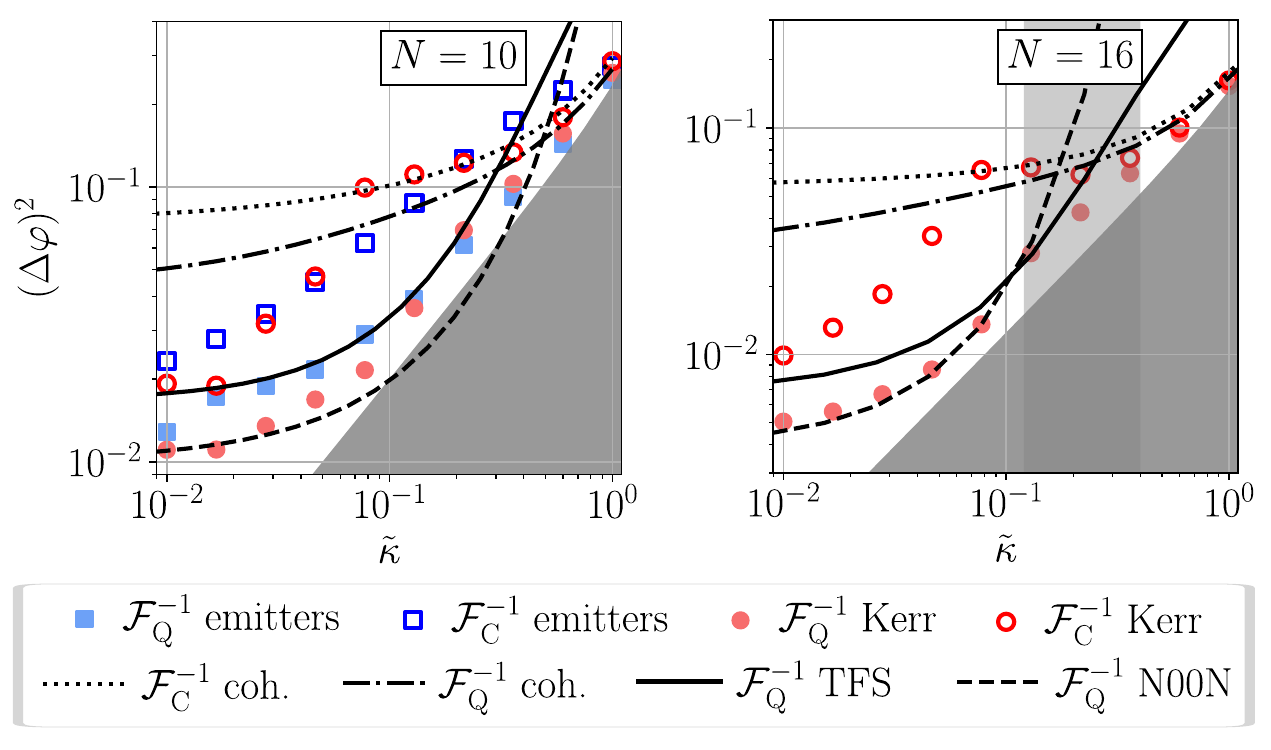}
    \caption{Estimation error $(\Delta\varphi)^2$ as a function of the dimensionless noise parameter $\tilde{\kappa}$ for mean photon numbers $N=10$ (panel a) and $N=16$ (panel b). Blue squares (red circles) correspond to the results of the emitters (Kerr) ansatz: filled (void) markers are the inverse of the quantum (classical) Fisher information $\mathcal{F}^{-1}_{\rm Q(C)}$ in the two cases. Dashed-dotted/solid/dashed lines signal the values of $\mathcal{F}^{-1}_{\rm Q}$ obtained for coherent states/TFS/NOON states without preparation PQC. The dotted line corresponds to $\mathcal{F}^{-1}_{\rm C}$ for coherent states without measurement PQC. 
    In the light grey shaded area of panel (b) the optimal states attain values of $\Delta\varphi$ which are sizeably smaller than those of TFS.
    The dark grey shaded area is beyond the asymptotic bound for dephasing noise~\cite{DD2012}.
    In both panels, we employed the circuit depth $d=5$.}
    \label{fig:noise}
\end{figure}

Fig.~\ref{fig:noise} shows the results of our algorithm in the presence of both noise channels for a circuit depth $d=5$ and fixed photon numbers $N=10$ (panel a) and $N=16$ (panel b) as a function of the dimensionless noise factor $\tilde{\kappa}=\kappa T_{\kappa}$, where $T_{\kappa}$ is the typical noise timescale. For $N=16$ only the results for the Kerr ansatz are shown, as the calculation for the emitters one does not reach such value of $N$ due to the numerical overhead introduced by the emitters degrees of freedom.
We benchmark the optimal values obtained with our VQA with those given by coherent states with $\alpha=\sqrt{N/2}$, TFS $\ket{N/2}\otimes\ket{N/2}$, and NOON states $(\ket{N,0}+\ket{0,N})/\sqrt{2}$ alone without the preparation and measurement PQCs.
In general, as $\tilde{\kappa}$ increases, the values of $\mathcal{F}^{-1}_{\rm Q}$ obtained with both the emitters and the Kerr ansätze are lifted upwards, attaining values similar to those of NOON and TFS up to the value of $\tilde{\kappa}$ where these states surpass the coherent ones.
On the contrary, for larger values of $\tilde{\kappa}$ the states generated by our VQA maintain a metrological advantage over coherent states. This improvement becomes larger with growing $N$, as can be seen in panel (b), which features a range of $\tilde{\kappa}$ (shaded in light grey) in which the variationally computed value of $\mathcal{F}^{-1}_{\rm Q}$ is 
sizeably
smaller than that of TFS, which are considered noise-robust states~\cite{Polino2020,Dorner2009}. This places the states generated by our protocol amongst the most noise-resilient ones. As $\tilde{\kappa}$ grows such improvement diminishes, as it is expected for all generation protocols. The dark grey shaded area represents the region beyond the asymptotic bound (for large $N$) in the presence of dephasing noise~\cite{DD2012}. For $\tilde{\kappa}\gtrsim 10^{-1}$ the results of our optimal states are close to such a bound.
As for the CFI, it turns out to be more susceptible to noise, and even for small values of $\tilde{\kappa}$ it deviates significantly from the corresponding QFI, reaching the CFI of coherent states for smaller $\tilde{\kappa}$. This implies that photon number measurements are not a good choice in a noisy situation.

%%%%%%%%%%%%%%%%%%%%%%%%%%%%%%%%%%%%%%%%%%%%
\section{Conclusions \& Outlook} 
\label{sec:Conclusions and Outlook}

Summing up, using a variational approach, we propose a method to generate metrologically-relevant photonic states 
which offers an exponential advantage over standard deterministic protocols when gate errors are considered.
By comparing the performance of both Kerr and emitter non-linearities, we predict that the emitter ansatz will perform better in platforms with limited Kerr non-linearities. We also showed that the tunable character of the non-linearity is essential to reach a Heisenberg scaling in the estimation error. Interestingly, our method is able to find states which provide a metrological advantage in the presence of moderate values of noise beyond 
other noise-resilient states considered in the literature, such as Twin-Fock states. In future works, we plan to extend our algorithm beyond the two-mode scenario in order to study multi-parameter estimation~\cite{Meyer2021}, as well as to apply it to other relevant problems in metrology such as electric field estimation~\cite{Brownnutt2015}.

\begin{acknowledgements}
  The authors acknowledge support from the Proyecto Sin\'ergico CAM 2020 Y2020/TCS-6545 (NanoQuCo-CM), the CSIC Research Platform on Quantum Technologies PTI-001 and from Spanish projects PID2021-127968NB-I00 and TED2021-130552B-C22 funded by  MCIN/AEI/10.13039/501100011033/FEDER, UE and MCIN/AEI/10.13039/501100011033, respectively. AMH acknowledges support from Fundación General CSIC's ComFuturo program, which has received funding from the European Union's Horizon 2020 research and innovation program under the Marie Skłodowska-Curie grant agreement No. 101034263. AGT also acknowledges support from a 2022 Leonardo Grant for Researchers and Cultural Creators, and BBVA Foundation. The authors also acknowledge Centro de Supercomputación de Galicia (CESGA) who provided access to the supercomputer FinisTerrae for performing numerical simulations. The authors thank Martí Perarnau-Llobet, Juan José García-Ripoll, and Geza Giedke for insightful discussions.
\end{acknowledgements}

%%%%%%%%%%%%%%%%%%%%%%%%%%%%%%%%%%% APPENDICES
%%%%%%%%%%%%%%%%%%%%%%%%%%%%%%%%%%

\appendix

\section{Description of the Variational Quantum Algorithm}
\label{sec:algorithm}

In this Section we describe in more detail the VQA proposed in our work. In general, we envision a quantum system featuring two photonic modes. Each of them initially hosts a coherent state
\begin{align}
    \ket{\alpha} = e^{-|\alpha|^2/2}\sum^{\infty}_{n=0} \frac{\alpha^n}{\sqrt{n!}}\ket{n}
\end{align}
with a mean-photon number $|\alpha|^2 = N/2$. In practice, the infinite sum above is truncated at $N$, ensuring that the maximum number of photons in the system is $2N$. The initial state in the Kerr ansatz (which does not include emitters) therefore is $\ket{\psi_0}=\ket{\alpha}_1\otimes\ket{\alpha}_2$, where the subscripts $1,2$ refer to the photonic mode. In the emitters ansatz, we assume that the two-level emitters are initially in their ground state $\ket{g}$, thus giving an initial state $\ket{\psi_0} = \ket{\alpha}_1\otimes\ket{\alpha}_2\otimes\ket{g}_1\otimes\ket{g}_2$. The subscripts in the state of the emitters refer to the photonic mode to which they are coupled.

In the preparation stage, a parametrized quantum circuit (PQC) described by a unitary $U_{\rm P}(\boldsymbol{\theta})$ is applied to the initial state. As explained in the Main Text, the applied unitary depends on the ansatz. The resulting state is $\ket{\psi_{\rm P}(\boldsymbol{\theta})} = U_{\rm P}(\boldsymbol{\theta})\ket{\psi_0}$.

Such state is then sent through a Mach--Zehnder interferometer (MZI) consisting of a symmetric beamsplitter (described by a unitary $U_{\rm BS} = \exp{[-i(a^\dagger_{2}a_{1} + a^\dagger_{1}a_{2})\pi/4]}$) followed by the encoding of a phase difference $\varphi$ between the two modes (given by a unitary $U_{\rm E}(\varphi) = \exp{[i\varphi (a^\dagger_{1}a_{1} - a^\dagger_{2}a_{2})/2]}$) and another symmetric beamsplitter, resulting in a state $\ket{\psi_{\rm E}(\boldsymbol{\theta},\varphi)}  = U_{\rm BS} U_{\rm E}(\varphi) U_{\rm BS} \ket{\psi_{\rm P}(\boldsymbol{\theta})}$. A second copy of $\ket{\psi_{\rm P}(\boldsymbol{\theta})}$ is also sent through the MZI, but in this case the encoded phase is $\varphi + \delta$, where $\delta\ll 1$ is a small parameter.
This allows us to calculate the QFI later on. Without loss of generality, in all our calculations we took $\varphi = \pi/3$ and $\delta=10^{-2}$.

In order to account for noise in the preparation and encoding stages, we need to construct the density matrices $\rho_{\rm E}(\boldsymbol{\theta},\varphi) = \ket{\psi_{\rm E}(\boldsymbol{\theta},\varphi)}\bra{\psi_{\rm E}(\boldsymbol{\theta},\varphi)}$ and $\rho_{\rm E}(\boldsymbol{\theta},\varphi+\delta) = \ket{\psi_{\rm E}(\boldsymbol{\theta},\varphi+\delta)}\bra{\psi_{\rm E}(\boldsymbol{\theta},\varphi+\delta)}$, which experience a non-unitary evolution according to a Lindblad master equation. We consider two noise channels: amplitude damping and phase damping, as explained in the Main Text, each featuring a loss rate $\kappa$. However, working with density matrices is computationally expensive since it requires squaring the dimension of the Hilbert space. Therefore, in practice, when we have $\kappa>0$ we work with vectorized density matrices following the Choi--Jamiolkowski isomorphism \cite{CHOI1975}, but in the noiseless case ($\kappa=0$) we avoid it and deal directly with state vectors.

In the noisy case the resulting state is given by $\rho_{\rm E}(\boldsymbol{\theta},\varphi,\kappa) = e^{\mathcal{L}_{pd}T_{\kappa}}e^{\mathcal{L}_{ad}T_{\kappa}}\rho_{\rm E}(\boldsymbol{\theta},\varphi)e^{\mathcal{L}^\dagger_{ad}T_{\kappa}}e^{\mathcal{L}^\dagger_{pd}T_{\kappa}}$, where $T_{\kappa}$ is a typical noise timescale which for simplicity we assume equal for both noise channels, and $\mathcal{L}_{\rm ad,pd}$ are the Lindbladian super-operators for amplitude damping and phase damping given by 
\begin{align}
    \mathcal{L}_{\rm ad,pd} &= \kappa\sum^2_{i=1} \Big[ L^{\rm (ad,pd)}_i \rho(t) L^{\rm (ad,pd)^\dagger}_{i} 
    \nonumber\\
    &-\frac{1}{2}\{L^{\rm (ad,pd)^\dagger}_i  L^{\rm (ad,pd)}_i, \rho(t) \} \Big].
\label{eq:master eq}
\end{align}
The sum above is over the set of jump operators belonging to the two noise channels, given by $\{L^{\rm (ad)}\}=\{a_1,a_2\}$ and $\{L^{\rm (pd)}\}=\{a^\dagger_1 a_1, a^\dagger_2 a_2\}$, where $a_i$ is the annihilation operator acting on the photonic mode $i$.

The QFI $\mathcal{F}_{\rm Q}$ is then calculated following the approximate formula~\cite{Beckey2022}
\begin{align}
    \mathcal{F}_{\rm Q} = 8 \frac{1-F(\varphi,\varphi+\delta)}{\delta^2},
\label{eq:QFI}
\end{align}
where $F(\varphi,\varphi+\delta)$ is the fidelity between the states in which the phases $\varphi$ and $\varphi+\delta$ were encoded. This is calculated as
\begin{align}
    F(\varphi,\varphi+\delta) &= |\braket{\psi_{\rm E}(\boldsymbol{\theta},\varphi)|\psi_{\rm E}(\boldsymbol{\theta},\varphi+\delta)}|,\\
    F(\varphi,\varphi+\delta) &= \sqrt{\sqrt{\rho_{\rm E}(\boldsymbol{\theta},\varphi,\kappa)} \rho_{\rm E}(\boldsymbol{\theta},\varphi+\delta,\kappa)
    \sqrt{\rho_{\rm E}(\boldsymbol{\theta},\varphi,\kappa)}}
\end{align}
in the noiseless and in the noisy case~\cite{Jozsa1994}, respectively. Remember that Eq.\eqref{eq:QFI} is valid in the limit $\delta\rightarrow 0$.

In a realistic implementation, this procedure can be implemented by considering two different state evolutions in the MZI: i) applying opposite phase shifts $\pm \delta /2$ in the two channels of the MZI; and ii) leaving the two channels unperturbed, without applying any phase shift. The resulting states are then used to calculate the QFI according to Eq.~\eqref{eq:QFI}. Therefore, $\delta$ plays the role of the parameter to be estimated. While in our calculations we considered a phase shift $\varphi = \pi /3$, that choice is arbitrary and considering $\varphi = 0$ does not change the results, because the value of the QFI does not depend on the choice of $\varphi$~\cite{Koczor2020}. We give more information on how to experimentally estimate the QFI and the fidelity in Sec.~\ref{sec:measuring qfi}.

\begin{figure}[tb]
    \centering
    \includegraphics[width=0.8\linewidth]{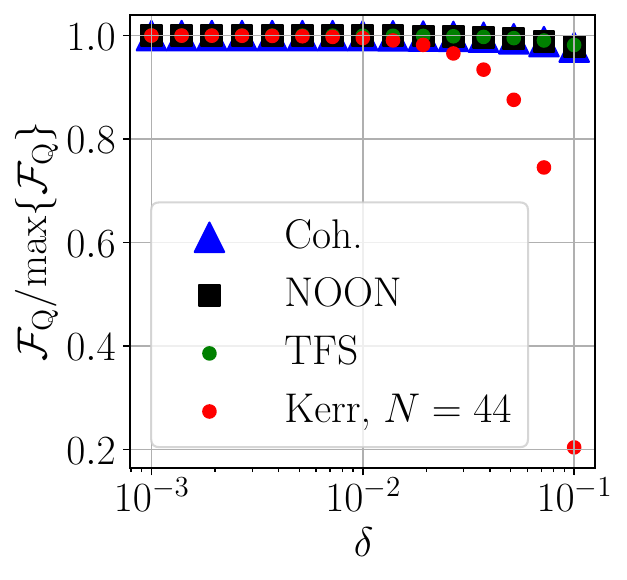}
    \caption{QFI $\mathcal{F}_{\rm Q}$ normalized in units of its converged value $\rm{max}\{\mathcal{F}_{\rm Q}\}$ for coherent states, NOON states, TFS, and an optimal Kerr state with mean photon number N=44 obtained with our VQA as a function of $\delta$.}
    \label{fig:convergence_delta}
\end{figure}

Also, to make sure that our results are converged, we plotted the QFI attained by several states as a function of $\delta$ (see Fig.~\ref{fig:convergence_delta}). From such results it is evident that $\delta=10^{
-2}$ is a safe choice.

To maximize the QFI we choose a cost function $C_{\rm P}(\boldsymbol{\theta}, \varphi, \kappa) = -\mathcal{F}_{\rm Q}$. This quantity is fed to the classical optimizer, which in turn updates the parameters $\boldsymbol{\theta}$. We initialize the preparation PQC close to the identity matrix (i.e., $\boldsymbol{\theta}$ finite but close to $0$), and employ COBYLA~\cite{Powell1994} as the classical optimizer, since it was the one giving the best results in a reasonable convergence time. When the optimization converges, we get the output state evaluated at the optimal parameters $\rho_{\rm E}(\boldsymbol{\theta}_{\rm opt}, \varphi, \kappa)$ in the noisy case and $\ket{\psi_{\rm E}(\boldsymbol{\theta}_{\rm opt}, \varphi)}$ in the noiseless one.

Such state is sent through the measurement PQC, which is characterized by a unitary $U_{\rm M}(\boldsymbol{\mu})$, resulting in a state described by $\rho_{\rm M}(\boldsymbol{\theta}_{\rm opt}, \varphi, \kappa, \boldsymbol{\mu}) = U_{\rm M}(\boldsymbol{\mu}) \rho_{\rm E}(\boldsymbol{\theta}_{\rm opt}, \varphi, \kappa) U_{\rm M}(\boldsymbol{\mu})^\dagger$ in the noisy case and $\ket{\psi_{\rm M}(\boldsymbol{\theta}_{\rm opt}, \varphi, \boldsymbol{\mu})}=U_{\rm M}(\boldsymbol{\mu})\ket{\psi_{\rm E}(\boldsymbol{\theta}_{\rm opt}, \varphi)}$ in the noiseless one.

At this point, we construct the density matrix for the noiseless case as $\rho_{\rm M}(\boldsymbol{\theta}_{\rm opt}, \varphi, \kappa=0, \boldsymbol{\mu}) = \ket{\psi_{\rm M}(\boldsymbol{\theta}_{\rm opt}, \varphi, \boldsymbol{\mu})} \bra{\psi_{\rm M}(\boldsymbol{\theta}_{\rm opt}, \varphi, \boldsymbol{\mu})}$.
Then, the derivative $\partial\rho_{\rm M}(\boldsymbol{\theta}_{\rm opt}, \varphi, \kappa, \boldsymbol{\mu})/\partial\varphi$ is computed by means of the chain rule. The density matrix and its derivative are used to calculate the CFI,
\begin{align}
    \mathcal{F}_{\rm C} = \sum_{n}\frac{1}{\rho_{\rm M, nn}(\boldsymbol{\theta}_{\rm opt}, \varphi, \kappa, \boldsymbol{\mu})}\left(\frac{\partial\rho_{\rm M, nn}(\boldsymbol{\theta}_{\rm opt}, \varphi, \kappa, \boldsymbol{\mu})}{\partial\varphi}\right)^2,
\end{align}
where the sum is carried over the diagonal elements $n,n$. To maximize the classical Fisher information (CFI), we set a cost function $C_{\rm M} (\boldsymbol{\theta}_{\rm opt}, \varphi, \kappa, \boldsymbol{\mu}) = -\mathcal{F}_{\rm C}$. This value is fed to the classical optimizer, which in turn returns the optimal measurement parameters $\boldsymbol{\mu}_{\rm opt}$. As for the preparation PQC, the measurement PQC is initialized close to the identity matrix (i.e., $\boldsymbol{\mu}$ finite but close to $0$), and we employed COBYLA as the classical optimizer.

As one can see, the two optimizations for the QFI and the CFI are carried separately in our VQA.

\section{Starting from Squeezed Coherent States}
\label{sec:squeezed}

\begin{figure*}[tb]
    \centering
    \includegraphics[width=\linewidth]{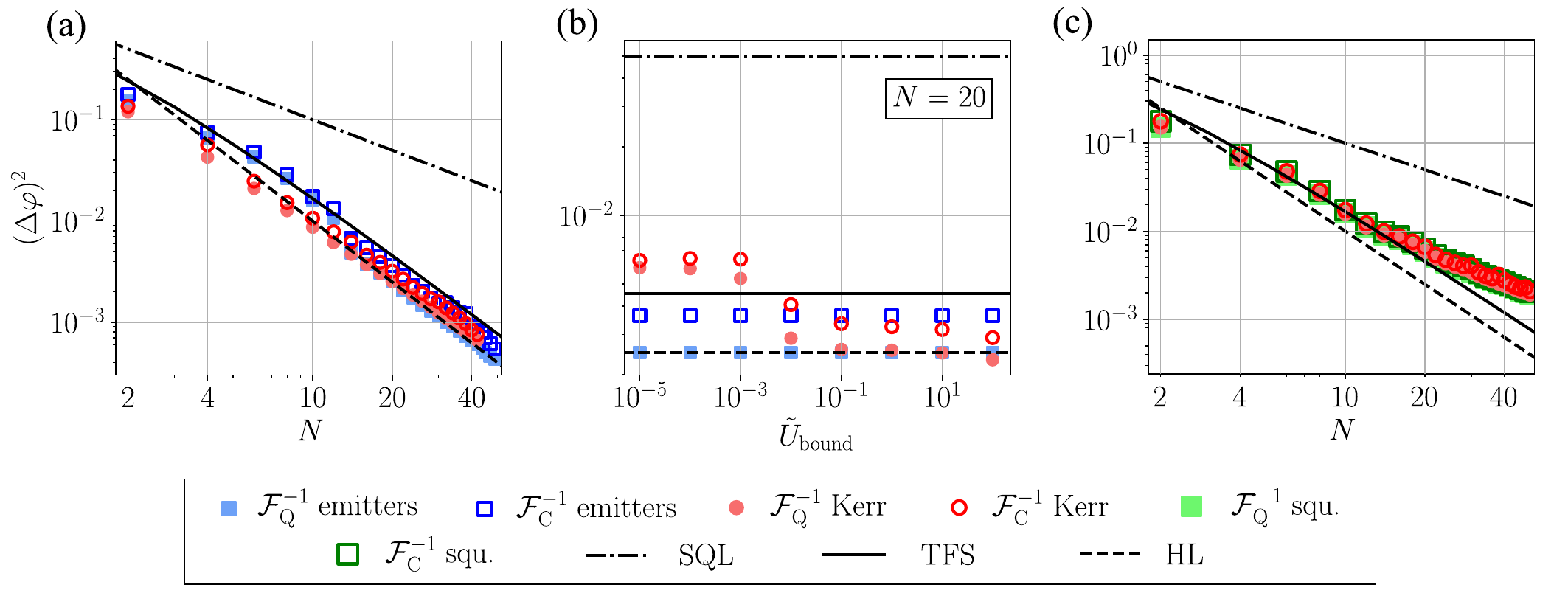}
    \caption{Estimation error $(\Delta\varphi)^2$ in the noiseless scenario starting from squeezed coherent states with $\alpha=\sqrt{N/2}$ and a squeezing factor of $10$ dB. Panel (a) shows our results as a function of the mean number of photons $N$. In panel (b), we address the effect of a bound in the Kerr non-linearity strength $\tilde{U}_{\rm bound}$ for a mean number of photons $N=20$. Panel (c) compares the results of the Kerr ansatz with $\tilde{U}_{\rm bound}=10^{-4}$ with those obtained removing the preparation and measurement PQCs. In the three panels, blue squares/red circles/green squares correspond to the results of the emitters ansatz/Kerr ansatz/squeezed states without preparation and measurement PQCs: filled (void) markers are the inverse of the QFI (CFI) $\mathcal{F}^{-1}_{\rm Q(C)}$ in the three cases. Dashed-dotted/solid/dashed lines signal the SQL/TFS/HL scaling. All calculations were made employing PQCs with depth $d=5$.}
    \label{fig:squeezed}
\end{figure*}

In this Section we perform an analog calculation to that shown in the Main Text but employing squeezed coherent states~\cite{gerry2005introductory} as initial states:
\begin{align}
    \ket{\alpha,r} &= \frac{1}{\sqrt{\cosh{r}}} \exp{\left(-\frac{1}{2}|\alpha|^2-\frac{1}{2}\alpha^{*2}\tanh{r}\right)}
    \nonumber\\
    &\times \sum^\infty_{n=0}\frac{\left(\frac{1}{2}\tanh{r}\right)^{n/2}}{\sqrt{n!}}H_n\left(\frac{\gamma}{\sqrt{\sinh{(2r)}}}\right)\ket{n},
\label{eq:squeezed}
\end{align}
where $\gamma = \alpha\cosh{r}+\alpha^{*}\sinh{r}$ and $H_n$ is the Hermite polynomial of grade $n$. The initial state is therefore $\ket{\psi_0} = \ket{\alpha, r}_1\otimes\ket{\alpha, r}_2$ for the Kerr ansatz and $\ket{\psi_0} = \ket{\alpha, r}_1\otimes\ket{\alpha, r}_2\otimes\ket{g}_1\otimes\ket{g}_2$ for the emitters one. We choose $\alpha = \sqrt{N/2}$ as in the coherent states case, and $r$ (the squeezing parameter) to be $10$ dB, which is within the reach of state-of-the-art optical technology~\cite{Schnabel2017}.
In practice, the infinite sum in Eq.~\eqref{eq:squeezed} is truncated at $N$, ensuring that the maximum number of photons in the system is $2N$. Using squeezed coherent states instead of coherent ones as initial states could be of help since the algorithm already starts from non-classical states, making it easier to obtain a quantum advantage.

Fig.~\ref{fig:squeezed}(a) shows the estimation error $(\Delta\varphi)^2$ obtained from maximizing the QFI and the CFI using our VQA as a function of $N$, and for a circuit depth $d=5$. As one can see, the difference between employing squeezed or coherent states (Fig.3a of the Main Text) as initial states is negligible: in both cases, the optimizer is able to find almost identical values for the QFI and the CFI. This is true for both ansätze.

To further explore whether squeezed states provide any advantage over coherent ones, we also plot the values of $\mathcal{F}^{-1}_{\rm Q,C}$ obtained by both ansätze as a function of a bound in the Kerr non-linearity variational parameter $\tilde{U}_{\rm bound}$. We fix the mean number of photons at $N=20$. Nevertheless, we obtain a very similar behavior to that of coherent states (shown in Fig.~\ref{fig:noiseless}(b) of the Main Text), with a threshold located at $\tilde{U}_{\rm bound}=10^{-2}$ separating below a regime where the Kerr ansatz results lie above those obtained with the emitters one, and above one where the results of both ansätze are very similar. Therefore, optical platforms only able to reach $U/\kappa\sim 10^{-2}$ at most~\cite{Delteil2019} will suffer from the same expressibility problems employing either coherent or squeezed states. Overall, the results of Fig.~\ref{fig:squeezed}(a,b) discard the possibility of obtaining better results using squeezed coherent states. 

Finally, to understand better what is happening below the Kerr non-linearity bound threshold, we plot in Fig.~\ref{fig:squeezed}(c) the values of  $\mathcal{F}^{-1}_{\rm Q,C}$ obtained with the Kerr ansatz with a bound $\tilde{U}_{\rm bound}=10^{-4}$ as a function of $N$. Here we also plot the results for the inverse of the QFI and the CFI using squeezed coherent states with $\alpha=\sqrt{N/2}$ and a squeezing factor of $10$ dB, i.e., removing the preparation and measurement PQCs (or equivalently setting $\boldsymbol{\theta}=\boldsymbol{\mu}=0$).
The results are identical in both cases, showing that the bound Kerr ansatz is not able to surpass the results of squeezed states alone due to the small value of $\tilde{U}_{\rm bound}$ preventing the optimizer to explore a larger region of the Hilbert space. In any case, these results are better than those obtained with coherent states alone, as is shown in Sec.~\ref{sec:kerr below threshold}, since the former are already non-classical. Note that, while squeezed states saturate the TFS scaling at small values of $N$, for $N\gtrsim 20$ they deviate and start following a classical $1/N$ scaling.

%%%%%%%%%%%%%%%%%%%%%%%%%%%%%%%%%%%%%
\section{Measuring the QFI}
\label{sec:measuring qfi}

In this Section, we provide more insight on the different techniques proposed to measure the QFI.
In our calculations, we compute the QFI by means of Eq.~\eqref{eq:QFI}. This procedure consists of evaluating the evolution of the quantum state through the MZI subjected to two different phase differences: $\varphi$ and $\varphi + \delta$.  Such an approximation is valid in the limit of small $\delta$. In order to obtain the QFI, one must evaluate the fidelity between the states generated by the two different evolutions. We choose to calculate the QFI this way because the alternatives (which involve either the eigendecomposition of the density matrix or evaluating the symmetric logarithmic derivative~\cite{Koczor2020,vitale2023estimation}) are computationally more expensive. Besides, in our manuscript we are dealing with a two-mode photonic system (with each of the modes coupled to a two-level emitter in the case of the emitters ansatz) in which the dimension of the total Hilbert space scales quadratically with respect to the mean number of photons present in the system $N$. We therefore expect measurements of the QFI (or, equivalently, the fidelity) to be simpler than in the most general case, in which the complexity scales exponentially with $N$. Even in that case, there exist a number of strategies aimed at mitigating the cost of brute-force full quantum state tomography to access the spectrum of eigenvalues and eigenstates of the density matrices. Here we mention the most promising ones, although the list is not complete.

\begin{enumerate}
    \item In the case of states that are well approximated by matrix product states (MPS)~\cite{Cramer2010} proposed two schemes for efficient quantum state tomography which only require a linear amount (on the system size) of local observables as well as polynomial classical post-processing of the data. Efficient quantum state tomography can be also obtained by harnessing conditional generative adversarial neural networks~\cite{Ahmed2021}, which leads to orders of magnitude fewer iterative steps than full quantum state tomography.
    \item An alternative approach to measure the fidelity between two quantum states is given by randomized measurements~\cite{Elben2023}. This strategy consists of repeatedly preparing and measuring a quantum state in a randomly chosen basis. A classical computer then processes the measurement outcome to estimate the desired property. In particular, a fidelity measurement will still feature an exponential complexity on the system size N, but better than with full tomography. Randomized measurements were also employed in~\cite{vitale2023estimation} to construct a series of polynomial lower bounds that converge to the QFI.
    \item An additional application of randomized measurements that further simplifies the measurement of several properties of the quantum system is classical shadows~\cite{Huang2020classicalshadows}. In this approach, the number of measurements to be performed is independent of the system size. In particular, it has been applied to measure the fidelity of a state preparation process~\cite{Struchalin2021} leading to higher fidelities with a number of operations orders of magnitude smaller than with maximum-likelihood estimation (which is an incomplete tomography method).
    \item One can also capitalize on the relation that exists between the quantum Fisher information and the variance of the operator generating the parameter encoding~\cite{Toth2013,toth2018lower,Toth2022} to estimate the QFI thorough a tight lower bound~\cite{Apellaniz2017}. This can be applied in large systems while requiring few operator expectation values.
    \item Further approaches to measure the fidelity include the SWAP test~\cite{Buhrman2001} and generalizing the quantum switch employed in~\cite{Abanin2012} to measure entanglement entropy.
\end{enumerate}

To sum up, the two-mode nature of our system, whose Hilbert space dimension grows quadratically with the mean photon number, combined with the growing number of techniques to experimentally estimate the fidelity (and thus the QFI) makes us confident about the possibility of implementing our scheme in experimental platforms.

%%%%%%%%%%%%%%%%%%%%%%%%%%%%%%%%%%%%%%%%%%%%%%%%%%%%
\section{Physical Implementation of Tunable Optical Non-linearities}
\label{sec:implementation}

In this Section we discuss in more detail the physical implementation of the tunable optical non-linearities required by our protocol.
The operations in which the two ansätze are based are given by the trotterization of the natural evolution of quantum states under their respective system’s Hamiltonian. This is an example of analog quantum computing~\cite{Johnson2014,Daley2022}, which is a particularly feasible way to exploit state-of-the-art quantum hardware. Nevertheless, a critical aspect of our work is the access to tunable optical non-linearities. 
In the case of the emitters ansatz, tunable emitter-photon interactions can be implemented by addressing an atom featuring a so-called lambda transition with Raman lasers~\cite{wineland1998experimental,Alton2011}. For the Kerr ansatz, tunable Kerr non-linearities have been implemented in superconducting circuits working in the microwave regime~\cite{Frattini2017,he2023fast}. However, it is not so obvious how to achieve a tunable Kerr non-linearity at optical frequencies, although there is a recent proposal in the few-photon regime based on the coupling of an infrared resonator to intersubband quantum well transition dipoles~\cite{arias2023coherent}.

However, even in the worst-case scenario in which such tunability cannot be realized, one can simulate the algorithm in a classical computer and then fabricate the desired setup in which fixed non-linearities accounting for the corresponding optimal values are applied to each mode. This is an example of the \textit{in silicon} approach that we mentioned in the Main Text. Thus, overall, we do not expect the requirement of tunable interactions to be a bottleneck for the implementation of our proposal.

\section{Results for $d=2$}
\label{sec:d=2}

\begin{figure*}[tb]
    \centering
    \includegraphics[width=\linewidth]{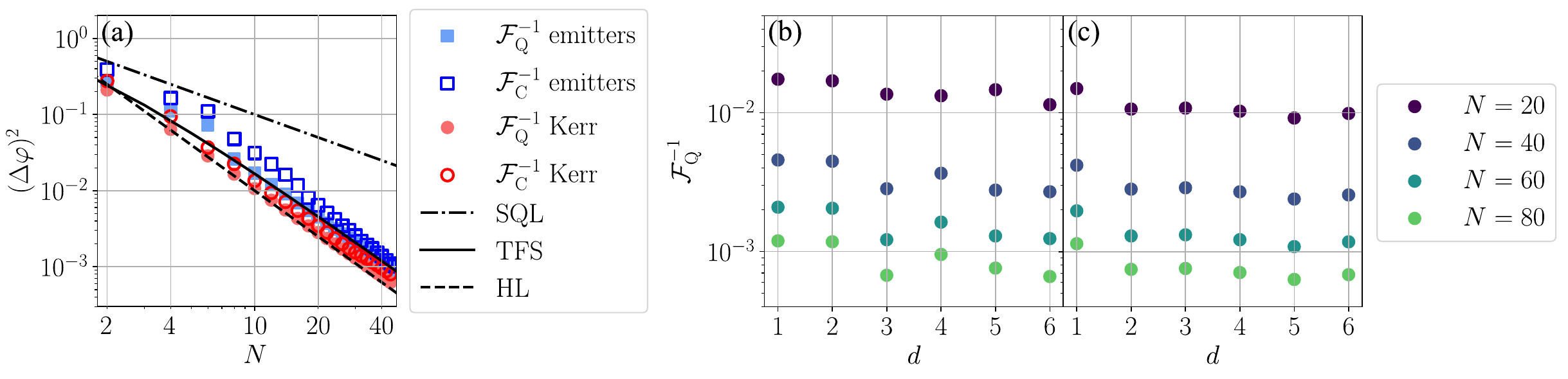}
    \caption{Results obtained by employing preparation and measurement PQCs with depth $d=2$.
    (a) Estimation error $(\Delta\varphi)^2$ in the noiseless scenario as a function of the mean number of photons $N$. Blue squares (red circles) correspond to the results of the emitters (Kerr) ansatz: filled (void) markers are the inverse of the QFI (CFI) $\mathcal{F}^{-1}_{\rm Q(C)}$ in the two cases. Dashed-dotted/solid/dashed lines signal the SQL/TFS/HL scaling. (b-c) Scaling of the inverse of the QFI $\mathcal{F}^{-1}_{\rm Q}$ as a function of the number of layers $d$ of the PQC, for different values of $N$. Contrary to Fig.~\ref{fig:scaling d} of the Main Text, these results are calculated by fixing $d$ and employing the optimal parameters for $N-1$ as the initial parameters for $N$.
    (b) Emitters ansatz.
    (c) Kerr ansatz.}
    \label{fig:d=2}
\end{figure*}

In this Section we provide more information on the results for a circuit depth $d=2$, the convergence of the QFI as a function of the number of layers $d$ of the preparation PQC, and the reason behind choosing $d=5$ in Figs.~\ref{fig:noiseless}-\ref{fig:noise} of the Main Text.

We start by providing the analog of Fig.~\ref{fig:noiseless} of the Main Text but employing a circuit depth $d=2$ for both the preparation and the measurement PQCs. The results, shown in panel (a) of Fig.~\ref{fig:d=2}, are quite similar to those obtained with $d=5$. Even though the Kerr ansatz is able to produce states reaching the HL for $N\lesssim 20$, above this value they start deviating upwards, giving slightly larger values of $\mathcal{F}^{-1}_{\rm Q}$, but still very close to the HL. As for the emitters ansatz, for $N\gtrsim 10$ they saturate the TFS scaling. A linear fit reveals that the two ansätze follow $\mathcal{F}^{-1}_{\rm Q}\sim 1/N^\beta$ with $\beta=1.90$ for the Kerr ansatz and $\beta=1.93$ for the emitters one. Regarding the CFI, it closely follows the results for the QFI. For the emitters ansatz, this behavior is very similar to what we found for $d=5$. However, for the Kerr ansatz, the agreement was even better for $d=5$. In spite of this, with $d=2$ our VQA is still able to generate states featuring a large metrological advantage, reaching the value of QFI of TFS employing the emitters ansatz, and even beating it and approaching the HL in the case of the Kerr ansatz. This supports our claim of a highly-efficient method for the generation of metrologically-relevant quantum states.

Now we further clarify why we picked $d=5$ for Figs.~\ref{fig:noiseless}-\ref{fig:noise} in the Main Text. Actually, such figures are calculated differently from Fig.~\ref{fig:scaling d}. In the latter figure, the total mean-photon number $N$ was fixed, and then the data for $d$ was computed using the optimal parameters for $d-1$ as the new initial parameters. In Figs.~\ref{fig:noiseless}-\ref{fig:noise}, $d$ was fixed and the data for $N$ was computed using the optimal parameters for $N-1$ as initial parameters. This implies that, for the same value of $d$, using the latter method more optimizations have been carried for values of $N>d$, which leads to better results than in the former case. 

This is visible in panels (b-c) of Fig.~\ref{fig:d=2}, where we show the analog of Fig.~\ref{fig:scaling d} of the Main Text but carrying the optimization using the latter method (i.e., fixing $d$ and increasing $N$). The inverse of the QFI $\mathcal{F}^{-1}_{\rm Q}$ is plotted as a function of the circuit depth $d$ for several values of the mean-photon number $N$, for the emitters (panel b) and the Kerr (panel c) ansätze. In the two cases, even with $d=1$ the VQA gives results which are close to the converged ones presented in Fig.~2 of the Main Text. As mentioned before, this is a consequence of the larger number of optimizations carried with this method before reaching the value of $N$ considered. For $d>1$, $\mathcal{F}^{-1}_{\rm Q}$ oscillates, with some values of $d$ performing better than others. After an exploration of the results obtained with different values of the circuit depth in the range $d=1-6$, we concluded that PQCs with $d=5$ attained slightly better results for both ansätze. This is why we employed this value in Figs.~\ref{fig:noiseless}-\ref{fig:noise} of the Main Text.

\section{Properties of the Output States of the VQA}
\label{sec:states}

\begin{figure}[tb]
    \centering
    \includegraphics[width=\linewidth]{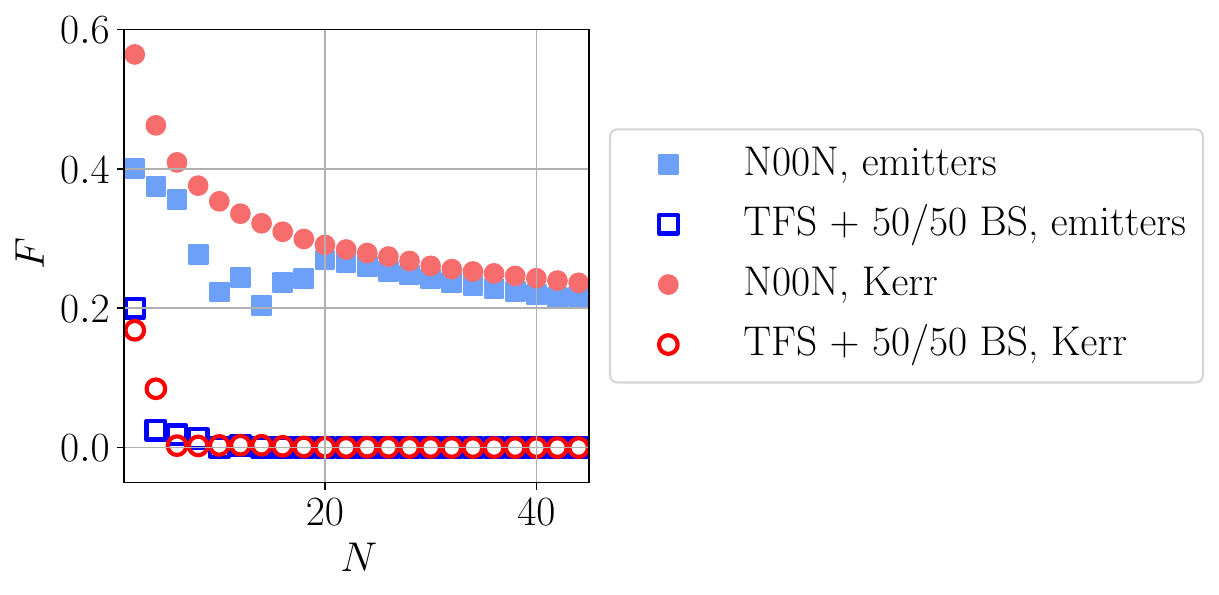}
    \caption{Fidelity $F$ of the optimal states generated by the preparation PQC after going through a 50/50 beamsplitter with respect to NOON states and TFS after passing thorough a 50/50 beamsplitter (TFS+50/50 BS), for both the emitters and the Kerr ansätze, as a function of the mean number of photons $N$. All calculations were made employing PQCs with depth $d=5$.}
    \label{fig:fidelity}
\end{figure}

In this Section we explore the nature of the optimal states prepared by the VQA, and how close they are to NOON states and TFS. In Fig.~\ref{fig:fidelity}(a) we calculated the fidelity (i.e., the complex modulus of the scalar product between two quantum states) of the states generated by the emitters and Kerr ansätze in the preparation stage of the VQA (with depth $d=5$) after going through the first symmetric (50/50) beamsplitter of the MZI with respect to TFS after passing through the same symmetric beamsplitter (TFS+50/50 BS) as well as with respect to NOON states. This is the correct comparison as TFS are defined prior to enter the first beamsplitter of the MZI interferometer, while NOON states are directly sent through the phase encoding. As one can see, as $N$ grows, the fidelities with respect to TFS+50/50 BS rapidly decay to zero. On the other hand, the fidelities with respect to NOON states are finite even for large values of $N$. This is true for both ansätze. In other words, the states generated by our VQA hold some similarity with NOON states even for mean-photon numbers $N\gtrsim 40$. The fact that they share a relatively low fidelity ($F\simeq 0.2-0.25$ for $N\simeq 40$) should not be disturbing, as metrological advantage can be obtained with a variety of different states. Besides, our VQA does not employ the fidelity with respect to a target state as cost function, but rather aims to maximize the QFI without caring about the particular state obtained.

\begin{figure}[tb]
    \centering
    \includegraphics[width=\linewidth]{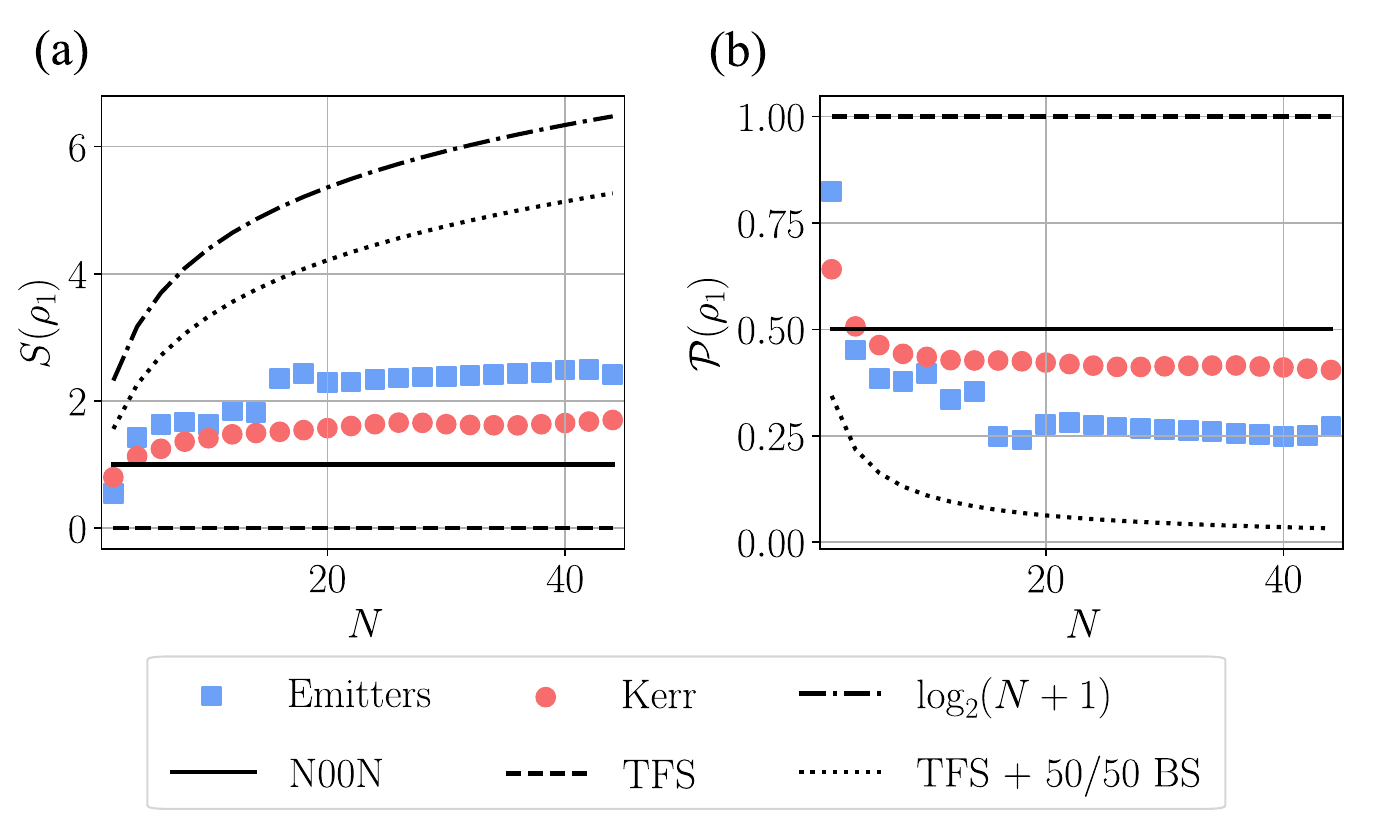}
    \caption{(a) Von Neumann entropy of entanglement of the reduced density matrix of the first photonic mode $S(\rho_1)$ as a function of the mean number of photons $N$. The dashed-dotted line shows the maximum entropy $\log_2{(N+1)}$ for a mode with $N+1$ possible states. (b) Purity of the reduced density matrix of the first photonic mode $\mathcal{P}(\rho_1)$ as a function of the mean number of photons $N$. In both panels, blue squares (red circles) are the results of the emitters (Kerr) ansatz after passing through the first symmetric (50/50) beamsplitter of the MZI. The solid (dashed) lines represent the results for NOON states (TFS), while the dotted lines are obtained with TFS sent through a symmetric beamsplitter. All calculations were made employing PQCs with depth $d=5$.}
    \label{fig:entropy}
\end{figure}

In order to further dive into the nature of the optimal states produced by our VQA, an interesting benchmark is to address whether there is entanglement between the two photonic modes (similarly to what happens in a NOON state) or if both of them are uncorrelated (like in TFS). To explore this, we calculated the Von Neumann entropy of entanglement (shown in Fig.~\ref{fig:entropy}(a)) as well as the purity of the reduced density matrix (see Fig.~\ref{fig:entropy}(b)) of the first photonic mode. 

We start from the density matrix describing the two photonic modes of our system $\rho_{\rm phot}$. In the case of the emitters ansatz, such density matrix can be obtained by tracing out the emitters degrees of freedom, i.e. $\rho_{\rm phot} = \mathrm{Tr}_{\rm emit}\{\rho\}$, where $\rho$ is the density matrix describing the total system of photons and emitters. At this point, and without loss of generality, we can trace out the second photonic mode, obtaining $\rho_1 = \mathrm{Tr}_2 \{\rho_{\rm phot}\}$. The entropy of entanglement is given by~\cite{Eisert2010}
\begin{align}
    S(\rho_1) = -\mathrm{Tr}\left(\rho_1\log_2\rho_1\right),
\end{align}
or equivalently by
\begin{align}
    S(\rho_1) = - \sum_i \lambda_i \log_2\left(\lambda_i\right),
\end{align}
where $\{\lambda_i\}$ is the set of $M$ eigenvalues of $\rho_1$ and the sum above runs from $i=1$ to $i=M$.

The results obtained with the states generated by the emitters and the Kerr ansätze are shown in Fig.~\ref{fig:entropy}(a) (again, taking such states after they have gone through the first 50/50 beamsplitter of the MZI). These are benchmarked against the values of the entropy of entanglement for TFS (which are separable and thus give $S(\rho_1)=0$), NOON states (in which only two states are entangled, and thus $S(\rho_1)=\log_2(2)=1$), TFS going through a symmetric (50/50) beamsplitter (for which all possible combinations featuring an even number of photons are entangled), and the maximum possible value $S(\rho_1)=\log_2(N+1)$ for $N+1$ available quantum states. As the mean number of photons $N$ increases, the values of $S(\rho_1)$ obtained with the two ansätze increase following a logarithmic law, surpassing the entropy of entanglement of NOON states as soon as for $N=4$, although their values lie below those obtained by TFS passing through a symmetric beamsplitter. This is telling us that both ansätze are generating entanglement between the two photonic modes. Therefore, tunneling between the two modes is a crucial resource for the ansätze. This also confirms that the optimal states of our formalism do not resemble TFS.

A similar metric addressing the entanglement between the two photonic modes is the purity of the first photonic mode, i.e.,
\begin{align}
    \mathcal{P}(\rho_1) = \mathrm{Tr}\{\rho_1^2\}.
\end{align}
Since a pure state always satisfies $\mathcal{P}=1$, if there is no entanglement between the two photonic modes we should obtain $\mathcal{P}(\rho_1)=1$. On the contrary, this value will be lower than $1$ if entanglement is present.

The results are shown in Fig.~\ref{fig:entropy}(b), and they go along the lines of Fig.~\ref{fig:entropy}(a): the purity of states generated by the two ansätze (as before, taking such states
after they have gone through the first 50/50 beamsplitter of the MZI) rapidly diminishes as $N$ increases. The obtained values of $\mathcal{P}(\rho_1)$ are below $1$, signaling that the resulting $\rho_1$ is a mixed state for both ansätze and that the two photonic modes are entangled. These results are benchmarked with the values of $\mathcal{P}(\rho_1)$ resulting from NOON states, TFS, and TFS going through a symmetric beamsplitter. As one can expect from Fig.~\ref{fig:entropy}(a), the optimal states found with our VQA provide values of the purity between those of NOON states and TFS passing through the beamsplitter.

Overall, the results presented in this Section confirm that the two ansätze are generating states that are different from the NOON and TFS, and also different between them.

\begin{figure*}[tb]
    \centering
    \includegraphics[width=\linewidth]{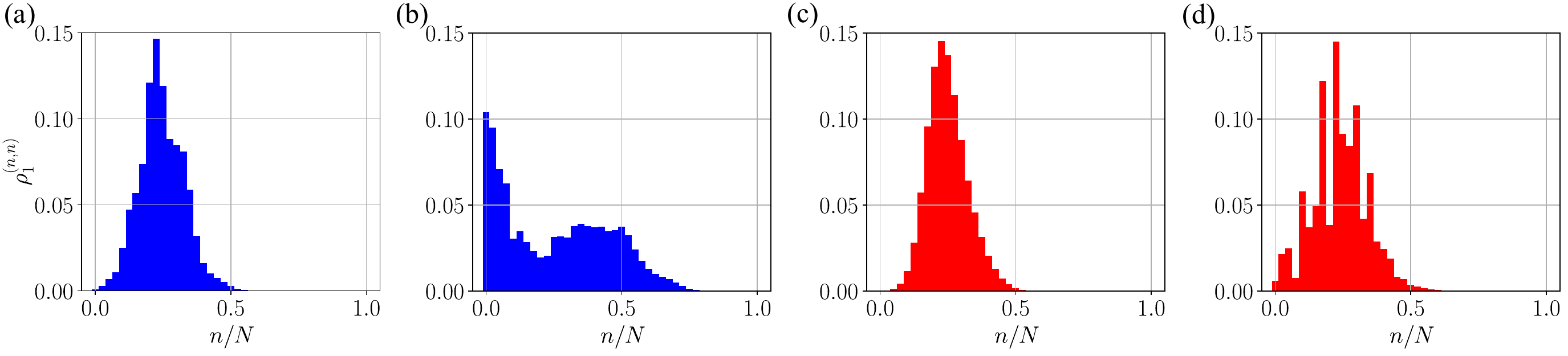}
    \caption{Diagonal elements of the reduced density matrix of the first photonic mode $\rho^{(n,n)}_1$ as a function of the number of photons in that mode $n$. The total number of photons is fixed at $N=20$. (a) Optimal state maximizing the QFI prepared using the emitters ansatz. (b) Optimal state maximizing the CFI prepared using the emitters ansatz. (c) Optimal state maximizing the QFI prepared using the Kerr ansatz. (d) Optimal state maximizing the CFI prepared using the Kerr ansatz. All calculations were made employing PQCs with depth $d=5$.}
    \label{fig:diagonal}
\end{figure*}

Finally, in Fig.~\ref{fig:diagonal} we plotted the diagonal terms of the reduced density matrix of the first photonic mode $\rho^{(n,n)}_1$ in the Fock states basis $\ket{n}$, where $n$ is the number of photons in that mode, after going through the preparation (a, c) and measurement (b, d) PQCs. We fixed the mean number of photons at $N=20$. Panels (a, b) show the results for the emitters ansatz, while panels (c, d) display those obtained with the Kerr one. As one can see, the optimal probe states given by the preparation stage are already different from coherent states, which would feature a Poisson distribution. Moreover, the measurement PQC further modifies the shape of the states.

Regarding the effect of noise, we have checked that the states prepared by our VQA still feature similar properties for finite values of $\kappa$.

\section{Optimal Parameters}
\label{sec:parameters}

\begin{figure*}[tb]
    \centering
    \includegraphics[width=\linewidth]{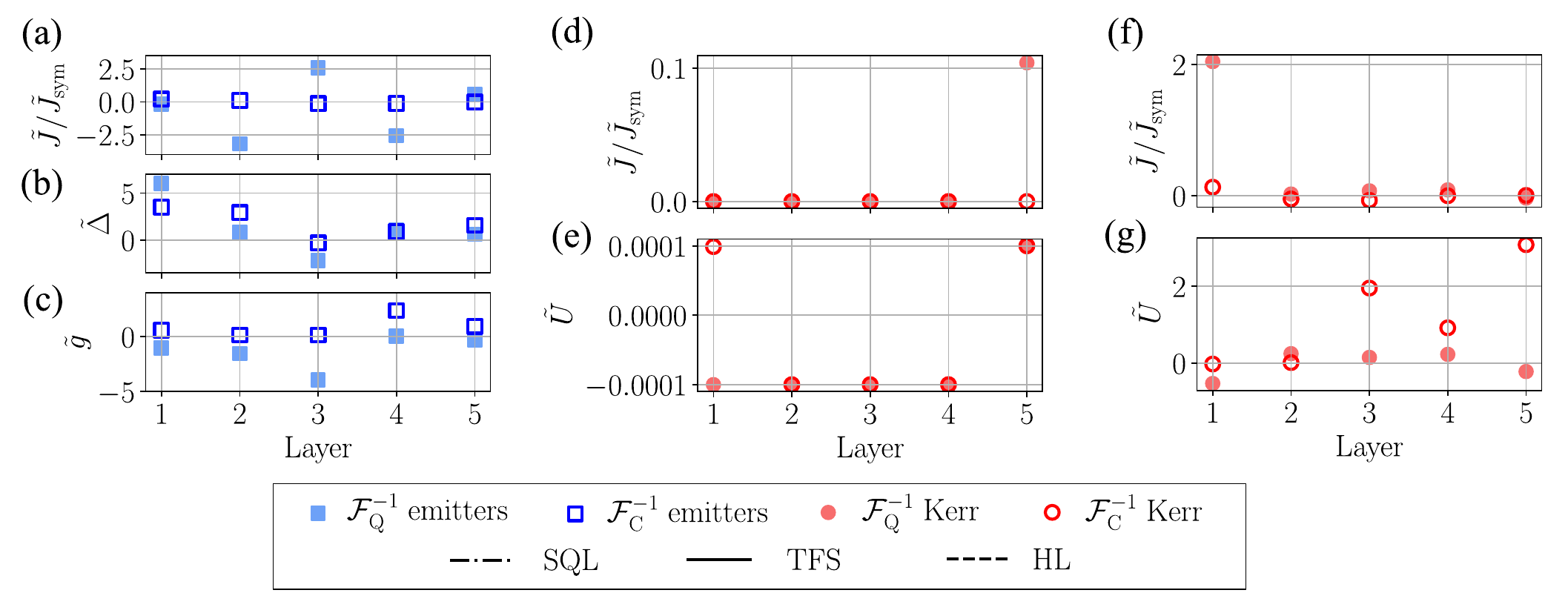}
    \caption{Optimal parameters $\boldsymbol{\theta}_{\rm opt}$ and $\boldsymbol{\mu}_{\rm opt}$ for the noiseless ($\kappa=0$) case with $N=20$ and a circuit depth $d=5$. (a) Emitters ansatz. (b) Kerr ansatz with $\tilde{U}_{\rm bound}=10^{-4}$. (c) Unbound Kerr ansatz. The tunneling parameter $\tilde{J}$ is plotted in units of that of a symmetric beamsplitter $\tilde{J}_{\rm sym}=\pi/4$.}
    \label{fig:parameters}
\end{figure*}

In this Section we examine the optimal parameters $\boldsymbol{\theta}_{\rm opt}$ and $\boldsymbol{\mu}_{\rm opt}$ obtained by the classical optimizer in the noiseless case ($\kappa=0$), for preparation and measurement PQCs with depth $d=5$, and for $N=20$ photons. The aim is to make sure that the parameters maximizing the QFI and the CFI can be attained in real experimental platforms. This is especially concerning for the variational parameters $\tilde{g}$ and $\tilde{U}$, which encode correspondingly the interaction strength in the emitters and Kerr ansatz and are of the order of $g/\kappa$ and $U/\kappa$, respectively.
Fig.~\ref{fig:parameters}(a,b,c) shows the values of $\tilde{J}$, $\tilde{\Delta}$, and $\tilde{g}$ obtained with the emitters ansatz. As one can see, the maximum value of the emitter-photon coupling required is $|g|/\kappa\simeq 5$. This can be realized in state-of-the-art cavity-QED experiments~\cite{GU2017,Meschede1985,Thompson1992,Lodahl2015,SanchezMunoz2018}.

Regarding the Kerr ansatz, we have to distinguish between the behavior below and above the expressibility threshold shown in Fig.~\ref{fig:noiseless}(b) of the Main Text. Therefore, we compare the results for a Kerr non-linearity bound $\tilde{U}_{\rm bound}=10^{-4}$ (shown in Fig.~\ref{fig:parameters}(d,e)) with those for a boundless non-linearity (shown in Fig.~\ref{fig:parameters}(f,g)). As we can see, in the first case the optimal values of $|\tilde{U}|$ given by the optimizer saturate $|\tilde{U}_{\rm bound}|$ for all layers, thus signaling that the Kerr non-linearity bound is forcing the optimizer to stay in a restricted region of the Hilbert space where it cannot find the global minimum of the cost function, and therefore limiting its expressibility. 

However, once we eliminate the Kerr non-linearity bound, we see in Fig.~\ref{fig:parameters}(g) that the optimal parameters contain values of $U/\kappa\sim 1$. In a real experiment, this would require microwave platforms, which are capable of reaching $U/\kappa\sim 10^2$~\cite{Koch2007,Kirchmair2013,Puri2017}. However, going above the expressibility threshold may be also within the reach of systems working in the optical regime, as the current limit in $U/\kappa \sim 10^{-2}$~\cite{Delteil2019} lies just in the middle of the $\tilde{U}_{\rm bound}$ threshold.

\section{Kerr Ansatz below the Non-linearity Threshold}
\label{sec:kerr below threshold}

\begin{figure}[tb]
    \centering
    \includegraphics[width=\linewidth]{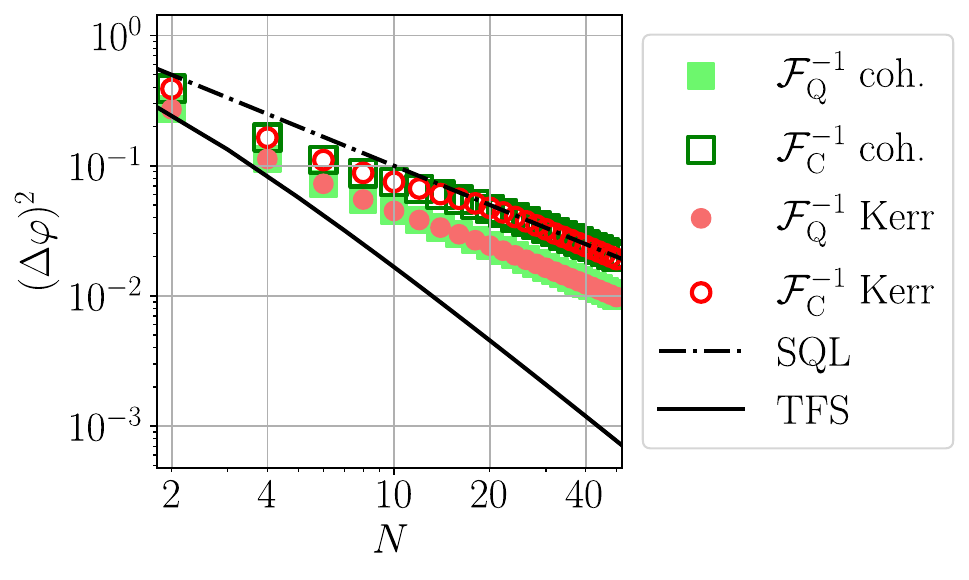}
    \caption{Estimation error $(\Delta\varphi)^2$ as a function of the mean number of photons $N$ for the Kerr ansatz with $\tilde{U}_{\rm bound}=10^{-4}$ and a circuit depth $d=5$, as well as for two input coherent states with mean photon number $|\alpha|^2=N/2$ without the preparation and measurement PQCs.}
    \label{fig:kerr below bound}
\end{figure}

In this Section we apply our VQA to a Kerr ansatz featuring a non-linearity bound $\tilde{U}_{\rm bound}=10^{-4}$, i.e., below the expressibility threshold, and in the noiseless case ($\kappa=0$). We will show that such bound ansatz is unable to provide a quantum advantage. Fig.~\ref{fig:kerr below bound} shows the results of our VQA for the inverse of the QFI ($\mathcal{F}^{-1}_{\rm Q}$) and the CFI ($\mathcal{F}^{-1}_{\rm C}$) as a function of $N$. As one can see, although the inverse of the QFI lies slightly below the SQL, it follows the same $1/N$ scaling, while the CFI does not reach the corresponding values of QFI, and follows the SQL. A circuit depth $d=5$ was employed. These are the same results that are obtained employing coherent states with $\alpha=N/2$ at the input of the MZI, and removing the preparation and measurement PQCs (i.e., taking $\boldsymbol{\theta}=\boldsymbol{\mu}=0$). This supports the results shown in Fig.~\ref{fig:parameters}(e) of Sec.~\ref{sec:parameters}, signaling that having a $\tilde{U}$ above the threshold in $\tilde{U}_{\rm bound}$ is indispensable to allow the classical optimizer to explore the region of the Hilbert space where solutions that feature a quantum advantage can be found.

\section{Critical Bound for Different Photon Numbers}
\label{sec:critical bound N}

\begin{figure}[tb]
    \centering
    \includegraphics[width=\linewidth]{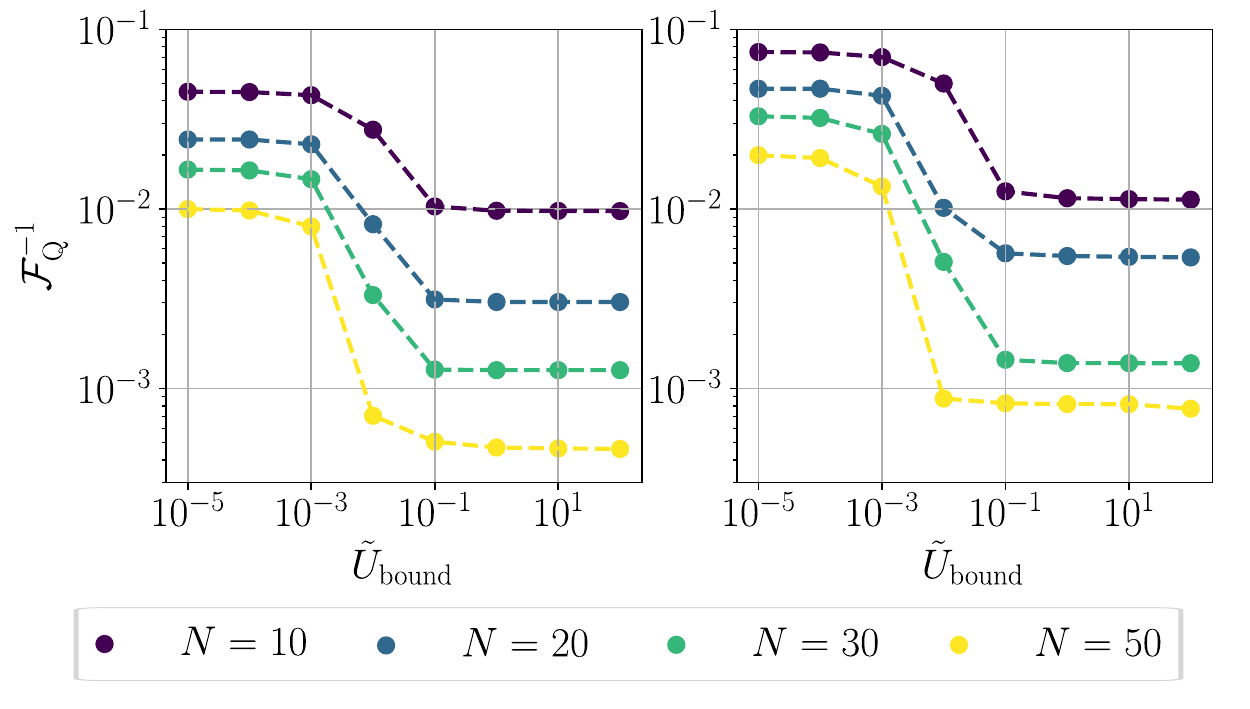}
    \caption{Inverse of the QFI $\mathcal{F}^{-1}_{\rm Q}$ (a) and CFI $\mathcal{F}^{-1}_{\rm C}$ (b) for the Kerr ansatz as a function of the bound in the Kerr non-linearity strength $\tilde{U}_{\rm bound}$ for several values of the mean number of photons $N$. All calculations were made employing PQCs with depth $d=5$.}
    \label{fig:critical bound N}
\end{figure}

In this Section we study the dependence of the critical value of $\tilde{U}_{\rm bound}$ above which the Kerr non-linearity ansatz is able to provide optimal solutions with respect to the mean number of photons $N$. To do it, in Fig.~\ref{fig:critical bound N} we plot the equivalent of Fig.~\ref{fig:noiseless}(b) of the Main Text for several values of $N$. As one can see, the crossover between the two regimes takes place around $\tilde{U}_{\rm bound}=10^{-2}$ independently of $N$ for both the QFI (panel a) and the CFI (panel b). A circuit depth $d=5$ was employed throughout all these calculations.

%%%%%%%%%%%%%%%%%%%%%%%%%%%%%%%%%%

\bibliographystyle{apsrev4-2}
\bibliography{cleaned_references}

\begin{thebibliography}{92}%
\makeatletter
\providecommand \@ifxundefined [1]{%
 \@ifx{#1\undefined}
}%
\providecommand \@ifnum [1]{%
 \ifnum #1\expandafter \@firstoftwo
 \else \expandafter \@secondoftwo
 \fi
}%
\providecommand \@ifx [1]{%
 \ifx #1\expandafter \@firstoftwo
 \else \expandafter \@secondoftwo
 \fi
}%
\providecommand \natexlab [1]{#1}%
\providecommand \enquote  [1]{``#1''}%
\providecommand \bibnamefont  [1]{#1}%
\providecommand \bibfnamefont [1]{#1}%
\providecommand \citenamefont [1]{#1}%
\providecommand \href@noop [0]{\@secondoftwo}%
\providecommand \href [0]{\begingroup \@sanitize@url \@href}%
\providecommand \@href[1]{\@@startlink{#1}\@@href}%
\providecommand \@@href[1]{\endgroup#1\@@endlink}%
\providecommand \@sanitize@url [0]{\catcode `\\12\catcode `\$12\catcode
  `\&12\catcode `\#12\catcode `\^12\catcode `\_12\catcode `\%12\relax}%
\providecommand \@@startlink[1]{}%
\providecommand \@@endlink[0]{}%
\providecommand \url  [0]{\begingroup\@sanitize@url \@url }%
\providecommand \@url [1]{\endgroup\@href {#1}{\urlprefix }}%
\providecommand \urlprefix  [0]{URL }%
\providecommand \Eprint [0]{\href }%
\providecommand \doibase [0]{https://doi.org/}%
\providecommand \selectlanguage [0]{\@gobble}%
\providecommand \bibinfo  [0]{\@secondoftwo}%
\providecommand \bibfield  [0]{\@secondoftwo}%
\providecommand \translation [1]{[#1]}%
\providecommand \BibitemOpen [0]{}%
\providecommand \bibitemStop [0]{}%
\providecommand \bibitemNoStop [0]{.\EOS\space}%
\providecommand \EOS [0]{\spacefactor3000\relax}%
\providecommand \BibitemShut  [1]{\csname bibitem#1\endcsname}%
\let\auto@bib@innerbib\@empty
%</preamble>
\bibitem [{\citenamefont {Bollinger}\ \emph {et~al.}(1996)\citenamefont
  {Bollinger}, \citenamefont {Itano}, \citenamefont {Wineland},\ and\
  \citenamefont {Heinzen}}]{Bollinger1996}%
  \BibitemOpen
  \bibfield  {author} {\bibinfo {author} {\bibfnamefont {J.~J.}\ \bibnamefont
  {Bollinger}}, \bibinfo {author} {\bibfnamefont {W.~M.}\ \bibnamefont
  {Itano}}, \bibinfo {author} {\bibfnamefont {D.~J.}\ \bibnamefont
  {Wineland}},\ and\ \bibinfo {author} {\bibfnamefont {D.~J.}\ \bibnamefont
  {Heinzen}},\ }\href {https://doi.org/10.1103/PhysRevA.54.R4649} {\bibfield
  {journal} {\bibinfo  {journal} {Phys. Rev. A}\ }\textbf {\bibinfo {volume}
  {54}},\ \bibinfo {pages} {R4649} (\bibinfo {year} {1996})}\BibitemShut
  {NoStop}%
\bibitem [{\citenamefont {Leibfried}\ \emph {et~al.}(2004)\citenamefont
  {Leibfried}, \citenamefont {Barrett}, \citenamefont {Schaetz}, \citenamefont
  {Britton}, \citenamefont {Chiaverini}, \citenamefont {Itano}, \citenamefont
  {Jost}, \citenamefont {Langer},\ and\ \citenamefont
  {Wineland}}]{Leibfried2004}%
  \BibitemOpen
  \bibfield  {author} {\bibinfo {author} {\bibfnamefont {D.}~\bibnamefont
  {Leibfried}}, \bibinfo {author} {\bibfnamefont {M.~D.}\ \bibnamefont
  {Barrett}}, \bibinfo {author} {\bibfnamefont {T.}~\bibnamefont {Schaetz}},
  \bibinfo {author} {\bibfnamefont {J.}~\bibnamefont {Britton}}, \bibinfo
  {author} {\bibfnamefont {J.}~\bibnamefont {Chiaverini}}, \bibinfo {author}
  {\bibfnamefont {W.~M.}\ \bibnamefont {Itano}}, \bibinfo {author}
  {\bibfnamefont {J.~D.}\ \bibnamefont {Jost}}, \bibinfo {author}
  {\bibfnamefont {C.}~\bibnamefont {Langer}},\ and\ \bibinfo {author}
  {\bibfnamefont {D.~J.}\ \bibnamefont {Wineland}},\ }\href
  {https://doi.org/10.1126/science.1097576} {\bibfield  {journal} {\bibinfo
  {journal} {Science}\ }\textbf {\bibinfo {volume} {304}},\ \bibinfo {pages}
  {1476} (\bibinfo {year} {2004})}\BibitemShut {NoStop}%
\bibitem [{\citenamefont {Giovannetti}\ \emph {et~al.}(2004)\citenamefont
  {Giovannetti}, \citenamefont {Lloyd},\ and\ \citenamefont
  {Maccone}}]{Giovannetti2004}%
  \BibitemOpen
  \bibfield  {author} {\bibinfo {author} {\bibfnamefont {V.}~\bibnamefont
  {Giovannetti}}, \bibinfo {author} {\bibfnamefont {S.}~\bibnamefont {Lloyd}},\
  and\ \bibinfo {author} {\bibfnamefont {L.}~\bibnamefont {Maccone}},\ }\href
  {https://doi.org/10.1126/science.1104149} {\bibfield  {journal} {\bibinfo
  {journal} {Science}\ }\textbf {\bibinfo {volume} {306}},\ \bibinfo {pages}
  {1330} (\bibinfo {year} {2004})}\BibitemShut {NoStop}%
\bibitem [{\citenamefont {Giovannetti}\ \emph {et~al.}(2006)\citenamefont
  {Giovannetti}, \citenamefont {Lloyd},\ and\ \citenamefont
  {Maccone}}]{Giovannetti2006}%
  \BibitemOpen
  \bibfield  {author} {\bibinfo {author} {\bibfnamefont {V.}~\bibnamefont
  {Giovannetti}}, \bibinfo {author} {\bibfnamefont {S.}~\bibnamefont {Lloyd}},\
  and\ \bibinfo {author} {\bibfnamefont {L.}~\bibnamefont {Maccone}},\ }\href
  {https://doi.org/10.1103/PhysRevLett.96.010401} {\bibfield  {journal}
  {\bibinfo  {journal} {Phys. Rev. Lett.}\ }\textbf {\bibinfo {volume} {96}},\
  \bibinfo {pages} {010401} (\bibinfo {year} {2006})}\BibitemShut {NoStop}%
\bibitem [{\citenamefont {Giovannetti}\ \emph {et~al.}(2011)\citenamefont
  {Giovannetti}, \citenamefont {Lloyd},\ and\ \citenamefont
  {Maccone}}]{giovannetti11a}%
  \BibitemOpen
  \bibfield  {author} {\bibinfo {author} {\bibfnamefont {V.}~\bibnamefont
  {Giovannetti}}, \bibinfo {author} {\bibfnamefont {S.}~\bibnamefont {Lloyd}},\
  and\ \bibinfo {author} {\bibfnamefont {L.}~\bibnamefont {Maccone}},\
  }\href@noop {} {\bibfield  {journal} {\bibinfo  {journal} {Nat. Photon.}\
  }\textbf {\bibinfo {volume} {5}},\ \bibinfo {pages} {222} (\bibinfo {year}
  {2011})}\BibitemShut {NoStop}%
\bibitem [{\citenamefont {Tóth}\ and\ \citenamefont
  {Apellaniz}(2014)}]{Toth_2014}%
  \BibitemOpen
  \bibfield  {author} {\bibinfo {author} {\bibfnamefont {G.}~\bibnamefont
  {Tóth}}\ and\ \bibinfo {author} {\bibfnamefont {I.}~\bibnamefont
  {Apellaniz}},\ }\href {https://doi.org/10.1088/1751-8113/47/42/424006}
  {\bibfield  {journal} {\bibinfo  {journal} {Journal of Physics A:
  Mathematical and Theoretical}\ }\textbf {\bibinfo {volume} {47}},\ \bibinfo
  {pages} {424006} (\bibinfo {year} {2014})}\BibitemShut {NoStop}%
\bibitem [{\citenamefont {Degen}\ \emph {et~al.}(2017)\citenamefont {Degen},
  \citenamefont {Reinhard},\ and\ \citenamefont {Cappellaro}}]{Degen2017}%
  \BibitemOpen
  \bibfield  {author} {\bibinfo {author} {\bibfnamefont {C.~L.}\ \bibnamefont
  {Degen}}, \bibinfo {author} {\bibfnamefont {F.}~\bibnamefont {Reinhard}},\
  and\ \bibinfo {author} {\bibfnamefont {P.}~\bibnamefont {Cappellaro}},\
  }\href {https://doi.org/10.1103/RevModPhys.89.035002} {\bibfield  {journal}
  {\bibinfo  {journal} {Rev. Mod. Phys.}\ }\textbf {\bibinfo {volume} {89}},\
  \bibinfo {pages} {035002} (\bibinfo {year} {2017})}\BibitemShut {NoStop}%
\bibitem [{\citenamefont {Dowling}\ and\ \citenamefont
  {Seshadreesan}(2015)}]{Dowling2015}%
  \BibitemOpen
  \bibfield  {author} {\bibinfo {author} {\bibfnamefont {J.~P.}\ \bibnamefont
  {Dowling}}\ and\ \bibinfo {author} {\bibfnamefont {K.~P.}\ \bibnamefont
  {Seshadreesan}},\ }\href {https://doi.org/10.1109/JLT.2014.2386795}
  {\bibfield  {journal} {\bibinfo  {journal} {Journal of Lightwave Technology}\
  }\textbf {\bibinfo {volume} {33}},\ \bibinfo {pages} {2359} (\bibinfo {year}
  {2015})}\BibitemShut {NoStop}%
\bibitem [{\citenamefont {Demkowicz-Dobrzański}\ \emph
  {et~al.}(2015)\citenamefont {Demkowicz-Dobrzański}, \citenamefont
  {Jarzyna},\ and\ \citenamefont {Kołodyński}}]{DEMKOWICZDOBRZANSKI2015}%
  \BibitemOpen
  \bibfield  {author} {\bibinfo {author} {\bibfnamefont {R.}~\bibnamefont
  {Demkowicz-Dobrzański}}, \bibinfo {author} {\bibfnamefont {M.}~\bibnamefont
  {Jarzyna}},\ and\ \bibinfo {author} {\bibfnamefont {J.}~\bibnamefont
  {Kołodyński}}\ }(\bibinfo  {publisher} {Elsevier},\ \bibinfo {year}
  {2015})\ pp.\ \bibinfo {pages} {345--435}\BibitemShut {NoStop}%
\bibitem [{\citenamefont {Pirandola}\ \emph {et~al.}(2018)\citenamefont
  {Pirandola}, \citenamefont {Bardhan}, \citenamefont {Gehring}, \citenamefont
  {Weedbrook},\ and\ \citenamefont {Lloyd}}]{Pirandola2018}%
  \BibitemOpen
  \bibfield  {author} {\bibinfo {author} {\bibfnamefont {S.}~\bibnamefont
  {Pirandola}}, \bibinfo {author} {\bibfnamefont {B.~R.}\ \bibnamefont
  {Bardhan}}, \bibinfo {author} {\bibfnamefont {T.}~\bibnamefont {Gehring}},
  \bibinfo {author} {\bibfnamefont {C.}~\bibnamefont {Weedbrook}},\ and\
  \bibinfo {author} {\bibfnamefont {S.}~\bibnamefont {Lloyd}},\ }\href
  {https://doi.org/10.1038/s41566-018-0301-6} {\bibfield  {journal} {\bibinfo
  {journal} {Nature Photonics}\ }\textbf {\bibinfo {volume} {12}},\ \bibinfo
  {pages} {724} (\bibinfo {year} {2018})}\BibitemShut {NoStop}%
\bibitem [{\citenamefont {Polino}\ \emph {et~al.}(2020)\citenamefont {Polino},
  \citenamefont {Valeri}, \citenamefont {Spagnolo},\ and\ \citenamefont
  {Sciarrino}}]{Polino2020}%
  \BibitemOpen
  \bibfield  {author} {\bibinfo {author} {\bibfnamefont {E.}~\bibnamefont
  {Polino}}, \bibinfo {author} {\bibfnamefont {M.}~\bibnamefont {Valeri}},
  \bibinfo {author} {\bibfnamefont {N.}~\bibnamefont {Spagnolo}},\ and\
  \bibinfo {author} {\bibfnamefont {F.}~\bibnamefont {Sciarrino}},\ }\href
  {https://doi.org/10.1116/5.0007577} {\bibfield  {journal} {\bibinfo
  {journal} {AVS Quantum Science}\ }\textbf {\bibinfo {volume} {2}},\ \bibinfo
  {pages} {024703} (\bibinfo {year} {2020})}\BibitemShut {NoStop}%
\bibitem [{\citenamefont {Holland}\ and\ \citenamefont
  {Burnett}(1993)}]{Holland1993}%
  \BibitemOpen
  \bibfield  {author} {\bibinfo {author} {\bibfnamefont {M.~J.}\ \bibnamefont
  {Holland}}\ and\ \bibinfo {author} {\bibfnamefont {K.}~\bibnamefont
  {Burnett}},\ }\href {https://doi.org/10.1103/PhysRevLett.71.1355} {\bibfield
  {journal} {\bibinfo  {journal} {Phys. Rev. Lett.}\ }\textbf {\bibinfo
  {volume} {71}},\ \bibinfo {pages} {1355} (\bibinfo {year}
  {1993})}\BibitemShut {NoStop}%
\bibitem [{\citenamefont {Campos}\ \emph {et~al.}(2003)\citenamefont {Campos},
  \citenamefont {Gerry},\ and\ \citenamefont {Benmoussa}}]{campos03a}%
  \BibitemOpen
  \bibfield  {author} {\bibinfo {author} {\bibfnamefont {R.~A.}\ \bibnamefont
  {Campos}}, \bibinfo {author} {\bibfnamefont {C.~C.}\ \bibnamefont {Gerry}},\
  and\ \bibinfo {author} {\bibfnamefont {A.}~\bibnamefont {Benmoussa}},\ }\href
  {https://doi.org/10.1103/PhysRevA.68.023810} {\bibfield  {journal} {\bibinfo
  {journal} {Phys. Rev. A}\ }\textbf {\bibinfo {volume} {68}},\ \bibinfo
  {pages} {23810} (\bibinfo {year} {2003})}\BibitemShut {NoStop}%
\bibitem [{\citenamefont {Afek}\ \emph {et~al.}(2010)\citenamefont {Afek},
  \citenamefont {Ambar},\ and\ \citenamefont {Silberberg}}]{Afek2010}%
  \BibitemOpen
  \bibfield  {author} {\bibinfo {author} {\bibfnamefont {I.}~\bibnamefont
  {Afek}}, \bibinfo {author} {\bibfnamefont {O.}~\bibnamefont {Ambar}},\ and\
  \bibinfo {author} {\bibfnamefont {Y.}~\bibnamefont {Silberberg}},\ }\href
  {https://doi.org/10.1126/science.1188172} {\bibfield  {journal} {\bibinfo
  {journal} {Science}\ }\textbf {\bibinfo {volume} {328}},\ \bibinfo {pages}
  {879} (\bibinfo {year} {2010})}\BibitemShut {NoStop}%
\bibitem [{\citenamefont {Israel}\ \emph {et~al.}(2012)\citenamefont {Israel},
  \citenamefont {Afek}, \citenamefont {Rosen}, \citenamefont {Ambar},\ and\
  \citenamefont {Silberberg}}]{Israel2012}%
  \BibitemOpen
  \bibfield  {author} {\bibinfo {author} {\bibfnamefont {Y.}~\bibnamefont
  {Israel}}, \bibinfo {author} {\bibfnamefont {I.}~\bibnamefont {Afek}},
  \bibinfo {author} {\bibfnamefont {S.}~\bibnamefont {Rosen}}, \bibinfo
  {author} {\bibfnamefont {O.}~\bibnamefont {Ambar}},\ and\ \bibinfo {author}
  {\bibfnamefont {Y.}~\bibnamefont {Silberberg}},\ }\href
  {https://doi.org/10.1103/PhysRevA.85.022115} {\bibfield  {journal} {\bibinfo
  {journal} {Phys. Rev. A}\ }\textbf {\bibinfo {volume} {85}},\ \bibinfo
  {pages} {022115} (\bibinfo {year} {2012})}\BibitemShut {NoStop}%
\bibitem [{\citenamefont {Rozema}\ \emph {et~al.}(2014)\citenamefont {Rozema},
  \citenamefont {Bateman}, \citenamefont {Mahler}, \citenamefont {Okamoto},
  \citenamefont {Feizpour}, \citenamefont {Hayat},\ and\ \citenamefont
  {Steinberg}}]{Rozema2014}%
  \BibitemOpen
  \bibfield  {author} {\bibinfo {author} {\bibfnamefont {L.~A.}\ \bibnamefont
  {Rozema}}, \bibinfo {author} {\bibfnamefont {J.~D.}\ \bibnamefont {Bateman}},
  \bibinfo {author} {\bibfnamefont {D.~H.}\ \bibnamefont {Mahler}}, \bibinfo
  {author} {\bibfnamefont {R.}~\bibnamefont {Okamoto}}, \bibinfo {author}
  {\bibfnamefont {A.}~\bibnamefont {Feizpour}}, \bibinfo {author}
  {\bibfnamefont {A.}~\bibnamefont {Hayat}},\ and\ \bibinfo {author}
  {\bibfnamefont {A.~M.}\ \bibnamefont {Steinberg}},\ }\href
  {https://doi.org/10.1103/PhysRevLett.112.223602} {\bibfield  {journal}
  {\bibinfo  {journal} {Phys. Rev. Lett.}\ }\textbf {\bibinfo {volume} {112}},\
  \bibinfo {pages} {223602} (\bibinfo {year} {2014})}\BibitemShut {NoStop}%
\bibitem [{\citenamefont {Vogel}\ \emph {et~al.}(1993)\citenamefont {Vogel},
  \citenamefont {Akulin},\ and\ \citenamefont {Schleich}}]{Vogel1993}%
  \BibitemOpen
  \bibfield  {author} {\bibinfo {author} {\bibfnamefont {K.}~\bibnamefont
  {Vogel}}, \bibinfo {author} {\bibfnamefont {V.~M.}\ \bibnamefont {Akulin}},\
  and\ \bibinfo {author} {\bibfnamefont {W.~P.}\ \bibnamefont {Schleich}},\
  }\href {https://doi.org/10.1103/PhysRevLett.71.1816} {\bibfield  {journal}
  {\bibinfo  {journal} {Phys. Rev. Lett.}\ }\textbf {\bibinfo {volume} {71}},\
  \bibinfo {pages} {1816} (\bibinfo {year} {1993})}\BibitemShut {NoStop}%
\bibitem [{\citenamefont {Law}\ and\ \citenamefont {Eberly}(1996)}]{Law1996}%
  \BibitemOpen
  \bibfield  {author} {\bibinfo {author} {\bibfnamefont {C.~K.}\ \bibnamefont
  {Law}}\ and\ \bibinfo {author} {\bibfnamefont {J.~H.}\ \bibnamefont
  {Eberly}},\ }\href {https://doi.org/10.1103/PhysRevLett.76.1055} {\bibfield
  {journal} {\bibinfo  {journal} {Phys. Rev. Lett.}\ }\textbf {\bibinfo
  {volume} {76}},\ \bibinfo {pages} {1055} (\bibinfo {year}
  {1996})}\BibitemShut {NoStop}%
\bibitem [{\citenamefont {Gonz\'alez-Tudela}\ \emph {et~al.}(2015)\citenamefont
  {Gonz\'alez-Tudela}, \citenamefont {Paulisch}, \citenamefont {Chang},
  \citenamefont {Kimble},\ and\ \citenamefont {Cirac}}]{Gonzalez-Tudela2015}%
  \BibitemOpen
  \bibfield  {author} {\bibinfo {author} {\bibfnamefont {A.}~\bibnamefont
  {Gonz\'alez-Tudela}}, \bibinfo {author} {\bibfnamefont {V.}~\bibnamefont
  {Paulisch}}, \bibinfo {author} {\bibfnamefont {D.~E.}\ \bibnamefont {Chang}},
  \bibinfo {author} {\bibfnamefont {H.~J.}\ \bibnamefont {Kimble}},\ and\
  \bibinfo {author} {\bibfnamefont {J.~I.}\ \bibnamefont {Cirac}},\ }\href
  {https://doi.org/10.1103/PhysRevLett.115.163603} {\bibfield  {journal}
  {\bibinfo  {journal} {Phys. Rev. Lett.}\ }\textbf {\bibinfo {volume} {115}},\
  \bibinfo {pages} {163603} (\bibinfo {year} {2015})}\BibitemShut {NoStop}%
\bibitem [{\citenamefont {Uria}\ \emph {et~al.}(2020)\citenamefont {Uria},
  \citenamefont {Solano},\ and\ \citenamefont {Hermann-Avigliano}}]{Uria2020}%
  \BibitemOpen
  \bibfield  {author} {\bibinfo {author} {\bibfnamefont {M.}~\bibnamefont
  {Uria}}, \bibinfo {author} {\bibfnamefont {P.}~\bibnamefont {Solano}},\ and\
  \bibinfo {author} {\bibfnamefont {C.}~\bibnamefont {Hermann-Avigliano}},\
  }\href {https://doi.org/10.1103/PhysRevLett.125.093603} {\bibfield  {journal}
  {\bibinfo  {journal} {Phys. Rev. Lett.}\ }\textbf {\bibinfo {volume} {125}},\
  \bibinfo {pages} {093603} (\bibinfo {year} {2020})}\BibitemShut {NoStop}%
\bibitem [{\citenamefont {Uria}\ \emph {et~al.}(2023)\citenamefont {Uria},
  \citenamefont {Maldonado-Trapp}, \citenamefont {Hermann-Avigliano},\ and\
  \citenamefont {Solano}}]{Uria2023}%
  \BibitemOpen
  \bibfield  {author} {\bibinfo {author} {\bibfnamefont {M.}~\bibnamefont
  {Uria}}, \bibinfo {author} {\bibfnamefont {A.}~\bibnamefont
  {Maldonado-Trapp}}, \bibinfo {author} {\bibfnamefont {C.}~\bibnamefont
  {Hermann-Avigliano}},\ and\ \bibinfo {author} {\bibfnamefont
  {P.}~\bibnamefont {Solano}},\ }\href
  {https://doi.org/10.1103/PhysRevResearch.5.013165} {\bibfield  {journal}
  {\bibinfo  {journal} {Phys. Rev. Res.}\ }\textbf {\bibinfo {volume} {5}},\
  \bibinfo {pages} {013165} (\bibinfo {year} {2023})}\BibitemShut {NoStop}%
\bibitem [{\citenamefont {Guerlin}\ \emph {et~al.}(2007)\citenamefont
  {Guerlin}, \citenamefont {Bernu}, \citenamefont {Del{\'e}glise},
  \citenamefont {Sayrin}, \citenamefont {Gleyzes}, \citenamefont {Kuhr},
  \citenamefont {Brune}, \citenamefont {Raimond},\ and\ \citenamefont
  {Haroche}}]{Guerlin2007}%
  \BibitemOpen
  \bibfield  {author} {\bibinfo {author} {\bibfnamefont {C.}~\bibnamefont
  {Guerlin}}, \bibinfo {author} {\bibfnamefont {J.}~\bibnamefont {Bernu}},
  \bibinfo {author} {\bibfnamefont {S.}~\bibnamefont {Del{\'e}glise}}, \bibinfo
  {author} {\bibfnamefont {C.}~\bibnamefont {Sayrin}}, \bibinfo {author}
  {\bibfnamefont {S.}~\bibnamefont {Gleyzes}}, \bibinfo {author} {\bibfnamefont
  {S.}~\bibnamefont {Kuhr}}, \bibinfo {author} {\bibfnamefont {M.}~\bibnamefont
  {Brune}}, \bibinfo {author} {\bibfnamefont {J.-M.}\ \bibnamefont {Raimond}},\
  and\ \bibinfo {author} {\bibfnamefont {S.}~\bibnamefont {Haroche}},\ }\href
  {https://doi.org/10.1038/nature06057} {\bibfield  {journal} {\bibinfo
  {journal} {Nature}\ }\textbf {\bibinfo {volume} {448}},\ \bibinfo {pages}
  {889} (\bibinfo {year} {2007})}\BibitemShut {NoStop}%
\bibitem [{\citenamefont {Del{\'e}glise}\ \emph {et~al.}(2008)\citenamefont
  {Del{\'e}glise}, \citenamefont {Dotsenko}, \citenamefont {Sayrin},
  \citenamefont {Bernu}, \citenamefont {Brune}, \citenamefont {Raimond},\ and\
  \citenamefont {Haroche}}]{Deleglise2008}%
  \BibitemOpen
  \bibfield  {author} {\bibinfo {author} {\bibfnamefont {S.}~\bibnamefont
  {Del{\'e}glise}}, \bibinfo {author} {\bibfnamefont {I.}~\bibnamefont
  {Dotsenko}}, \bibinfo {author} {\bibfnamefont {C.}~\bibnamefont {Sayrin}},
  \bibinfo {author} {\bibfnamefont {J.}~\bibnamefont {Bernu}}, \bibinfo
  {author} {\bibfnamefont {M.}~\bibnamefont {Brune}}, \bibinfo {author}
  {\bibfnamefont {J.-M.}\ \bibnamefont {Raimond}},\ and\ \bibinfo {author}
  {\bibfnamefont {S.}~\bibnamefont {Haroche}},\ }\href
  {https://doi.org/10.1038/nature07288} {\bibfield  {journal} {\bibinfo
  {journal} {Nature}\ }\textbf {\bibinfo {volume} {455}},\ \bibinfo {pages}
  {510} (\bibinfo {year} {2008})}\BibitemShut {NoStop}%
\bibitem [{\citenamefont {Sayrin}\ \emph {et~al.}(2011)\citenamefont {Sayrin},
  \citenamefont {Dotsenko}, \citenamefont {Zhou}, \citenamefont {Peaudecerf},
  \citenamefont {Rybarczyk}, \citenamefont {Gleyzes}, \citenamefont {Rouchon},
  \citenamefont {Mirrahimi}, \citenamefont {Amini}, \citenamefont {Brune},
  \citenamefont {Raimond},\ and\ \citenamefont {Haroche}}]{Sayrin2011}%
  \BibitemOpen
  \bibfield  {author} {\bibinfo {author} {\bibfnamefont {C.}~\bibnamefont
  {Sayrin}}, \bibinfo {author} {\bibfnamefont {I.}~\bibnamefont {Dotsenko}},
  \bibinfo {author} {\bibfnamefont {X.}~\bibnamefont {Zhou}}, \bibinfo {author}
  {\bibfnamefont {B.}~\bibnamefont {Peaudecerf}}, \bibinfo {author}
  {\bibfnamefont {T.}~\bibnamefont {Rybarczyk}}, \bibinfo {author}
  {\bibfnamefont {S.}~\bibnamefont {Gleyzes}}, \bibinfo {author} {\bibfnamefont
  {P.}~\bibnamefont {Rouchon}}, \bibinfo {author} {\bibfnamefont
  {M.}~\bibnamefont {Mirrahimi}}, \bibinfo {author} {\bibfnamefont
  {H.}~\bibnamefont {Amini}}, \bibinfo {author} {\bibfnamefont
  {M.}~\bibnamefont {Brune}}, \bibinfo {author} {\bibfnamefont {J.-M.}\
  \bibnamefont {Raimond}},\ and\ \bibinfo {author} {\bibfnamefont
  {S.}~\bibnamefont {Haroche}},\ }\href {https://doi.org/10.1038/nature10376}
  {\bibfield  {journal} {\bibinfo  {journal} {Nature}\ }\textbf {\bibinfo
  {volume} {477}},\ \bibinfo {pages} {73} (\bibinfo {year} {2011})}\BibitemShut
  {NoStop}%
\bibitem [{\citenamefont {Deng}\ \emph {et~al.}()\citenamefont {Deng},
  \citenamefont {Li}, \citenamefont {Chen}, \citenamefont {Ni}, \citenamefont
  {Cai}, \citenamefont {Mai}, \citenamefont {Zhang}, \citenamefont {Zheng},
  \citenamefont {Yu}, \citenamefont {Zou}, \citenamefont {Liu}, \citenamefont
  {Yan}, \citenamefont {Xu},\ and\ \citenamefont
  {Yu}}]{deng2023heisenberglimited}%
  \BibitemOpen
  \bibfield  {author} {\bibinfo {author} {\bibfnamefont {X.}~\bibnamefont
  {Deng}}, \bibinfo {author} {\bibfnamefont {S.}~\bibnamefont {Li}}, \bibinfo
  {author} {\bibfnamefont {Z.-J.}\ \bibnamefont {Chen}}, \bibinfo {author}
  {\bibfnamefont {Z.}~\bibnamefont {Ni}}, \bibinfo {author} {\bibfnamefont
  {Y.}~\bibnamefont {Cai}}, \bibinfo {author} {\bibfnamefont {J.}~\bibnamefont
  {Mai}}, \bibinfo {author} {\bibfnamefont {L.}~\bibnamefont {Zhang}}, \bibinfo
  {author} {\bibfnamefont {P.}~\bibnamefont {Zheng}}, \bibinfo {author}
  {\bibfnamefont {H.}~\bibnamefont {Yu}}, \bibinfo {author} {\bibfnamefont
  {C.-L.}\ \bibnamefont {Zou}}, \bibinfo {author} {\bibfnamefont
  {S.}~\bibnamefont {Liu}}, \bibinfo {author} {\bibfnamefont {F.}~\bibnamefont
  {Yan}}, \bibinfo {author} {\bibfnamefont {Y.}~\bibnamefont {Xu}},\ and\
  \bibinfo {author} {\bibfnamefont {D.}~\bibnamefont {Yu}},\ }\href@noop {}
  {}\Eprint {https://arxiv.org/abs/2306.16919} {arXiv:2306.16919} \BibitemShut
  {NoStop}%
\bibitem [{\citenamefont {Kok}\ \emph {et~al.}(2002)\citenamefont {Kok},
  \citenamefont {Lee},\ and\ \citenamefont {Dowling}}]{Kok2002}%
  \BibitemOpen
  \bibfield  {author} {\bibinfo {author} {\bibfnamefont {P.}~\bibnamefont
  {Kok}}, \bibinfo {author} {\bibfnamefont {H.}~\bibnamefont {Lee}},\ and\
  \bibinfo {author} {\bibfnamefont {J.~P.}\ \bibnamefont {Dowling}},\ }\href
  {https://doi.org/10.1103/PhysRevA.65.052104} {\bibfield  {journal} {\bibinfo
  {journal} {Phys. Rev. A}\ }\textbf {\bibinfo {volume} {65}},\ \bibinfo
  {pages} {052104} (\bibinfo {year} {2002})}\BibitemShut {NoStop}%
\bibitem [{\citenamefont {Walther}\ \emph {et~al.}(2004)\citenamefont
  {Walther}, \citenamefont {Pan}, \citenamefont {Aspelmeyer}, \citenamefont
  {Ursin}, \citenamefont {Gasparoni},\ and\ \citenamefont
  {Zeilinger}}]{Walther2004}%
  \BibitemOpen
  \bibfield  {author} {\bibinfo {author} {\bibfnamefont {P.}~\bibnamefont
  {Walther}}, \bibinfo {author} {\bibfnamefont {J.-W.}\ \bibnamefont {Pan}},
  \bibinfo {author} {\bibfnamefont {M.}~\bibnamefont {Aspelmeyer}}, \bibinfo
  {author} {\bibfnamefont {R.}~\bibnamefont {Ursin}}, \bibinfo {author}
  {\bibfnamefont {S.}~\bibnamefont {Gasparoni}},\ and\ \bibinfo {author}
  {\bibfnamefont {A.}~\bibnamefont {Zeilinger}},\ }\href
  {https://doi.org/10.1038/nature02552} {\bibfield  {journal} {\bibinfo
  {journal} {Nature}\ }\textbf {\bibinfo {volume} {429}},\ \bibinfo {pages}
  {158} (\bibinfo {year} {2004})}\BibitemShut {NoStop}%
\bibitem [{\citenamefont {Wang}\ \emph {et~al.}(2016)\citenamefont {Wang},
  \citenamefont {Chen}, \citenamefont {Li}, \citenamefont {Huang},
  \citenamefont {Liu}, \citenamefont {Chen}, \citenamefont {Luo}, \citenamefont
  {Su}, \citenamefont {Wu}, \citenamefont {Li}, \citenamefont {Lu},
  \citenamefont {Hu}, \citenamefont {Jiang}, \citenamefont {Peng},
  \citenamefont {Li}, \citenamefont {Liu}, \citenamefont {Chen}, \citenamefont
  {Lu},\ and\ \citenamefont {Pan}}]{Wang2016b}%
  \BibitemOpen
  \bibfield  {author} {\bibinfo {author} {\bibfnamefont {X.~L.}\ \bibnamefont
  {Wang}}, \bibinfo {author} {\bibfnamefont {L.~K.}\ \bibnamefont {Chen}},
  \bibinfo {author} {\bibfnamefont {W.}~\bibnamefont {Li}}, \bibinfo {author}
  {\bibfnamefont {H.~L.}\ \bibnamefont {Huang}}, \bibinfo {author}
  {\bibfnamefont {C.}~\bibnamefont {Liu}}, \bibinfo {author} {\bibfnamefont
  {C.}~\bibnamefont {Chen}}, \bibinfo {author} {\bibfnamefont {Y.~H.}\
  \bibnamefont {Luo}}, \bibinfo {author} {\bibfnamefont {Z.~E.}\ \bibnamefont
  {Su}}, \bibinfo {author} {\bibfnamefont {D.}~\bibnamefont {Wu}}, \bibinfo
  {author} {\bibfnamefont {Z.~D.}\ \bibnamefont {Li}}, \bibinfo {author}
  {\bibfnamefont {H.}~\bibnamefont {Lu}}, \bibinfo {author} {\bibfnamefont
  {Y.}~\bibnamefont {Hu}}, \bibinfo {author} {\bibfnamefont {X.}~\bibnamefont
  {Jiang}}, \bibinfo {author} {\bibfnamefont {C.~Z.}\ \bibnamefont {Peng}},
  \bibinfo {author} {\bibfnamefont {L.}~\bibnamefont {Li}}, \bibinfo {author}
  {\bibfnamefont {N.~L.}\ \bibnamefont {Liu}}, \bibinfo {author} {\bibfnamefont
  {Y.~A.}\ \bibnamefont {Chen}}, \bibinfo {author} {\bibfnamefont {C.~Y.}\
  \bibnamefont {Lu}},\ and\ \bibinfo {author} {\bibfnamefont {J.~W.}\
  \bibnamefont {Pan}},\ }\href {https://doi.org/10.1103/PHYSREVLETT.117.210502}
  {\bibfield  {journal} {\bibinfo  {journal} {Phys. Rev. Lett.}\ }\textbf
  {\bibinfo {volume} {117}},\ \bibinfo {pages} {210502} (\bibinfo {year}
  {2016})}\BibitemShut {NoStop}%
\bibitem [{\citenamefont {Wang}\ \emph {et~al.}(2019)\citenamefont {Wang},
  \citenamefont {Qin}, \citenamefont {Ding}, \citenamefont {Chen},
  \citenamefont {Chen}, \citenamefont {You}, \citenamefont {He}, \citenamefont
  {Jiang}, \citenamefont {You}, \citenamefont {Wang}, \citenamefont
  {Schneider}, \citenamefont {Renema}, \citenamefont {H{\"{o}}fling},
  \citenamefont {Lu},\ and\ \citenamefont {Pan}}]{Wang2019a}%
  \BibitemOpen
  \bibfield  {author} {\bibinfo {author} {\bibfnamefont {H.}~\bibnamefont
  {Wang}}, \bibinfo {author} {\bibfnamefont {J.}~\bibnamefont {Qin}}, \bibinfo
  {author} {\bibfnamefont {X.}~\bibnamefont {Ding}}, \bibinfo {author}
  {\bibfnamefont {M.~C.}\ \bibnamefont {Chen}}, \bibinfo {author}
  {\bibfnamefont {S.}~\bibnamefont {Chen}}, \bibinfo {author} {\bibfnamefont
  {X.}~\bibnamefont {You}}, \bibinfo {author} {\bibfnamefont {Y.~M.}\
  \bibnamefont {He}}, \bibinfo {author} {\bibfnamefont {X.}~\bibnamefont
  {Jiang}}, \bibinfo {author} {\bibfnamefont {L.}~\bibnamefont {You}}, \bibinfo
  {author} {\bibfnamefont {Z.}~\bibnamefont {Wang}}, \bibinfo {author}
  {\bibfnamefont {C.}~\bibnamefont {Schneider}}, \bibinfo {author}
  {\bibfnamefont {J.~J.}\ \bibnamefont {Renema}}, \bibinfo {author}
  {\bibfnamefont {S.}~\bibnamefont {H{\"{o}}fling}}, \bibinfo {author}
  {\bibfnamefont {C.~Y.}\ \bibnamefont {Lu}},\ and\ \bibinfo {author}
  {\bibfnamefont {J.~W.}\ \bibnamefont {Pan}},\ }\href
  {https://doi.org/10.1103/PHYSREVLETT.123.250503} {\bibfield  {journal}
  {\bibinfo  {journal} {Phys. Rev. Lett.}\ }\textbf {\bibinfo {volume} {123}},\
  \bibinfo {pages} {250503} (\bibinfo {year} {2019})}\BibitemShut {NoStop}%
\bibitem [{\citenamefont {Cerezo}\ \emph {et~al.}(2021)\citenamefont {Cerezo},
  \citenamefont {Arrasmith}, \citenamefont {Babbush}, \citenamefont {Benjamin},
  \citenamefont {Endo}, \citenamefont {Fujii}, \citenamefont {McClean},
  \citenamefont {Mitarai}, \citenamefont {Yuan}, \citenamefont {Cincio},\ and\
  \citenamefont {Coles}}]{Cerezo2021}%
  \BibitemOpen
  \bibfield  {author} {\bibinfo {author} {\bibfnamefont {M.}~\bibnamefont
  {Cerezo}}, \bibinfo {author} {\bibfnamefont {A.}~\bibnamefont {Arrasmith}},
  \bibinfo {author} {\bibfnamefont {R.}~\bibnamefont {Babbush}}, \bibinfo
  {author} {\bibfnamefont {S.~C.}\ \bibnamefont {Benjamin}}, \bibinfo {author}
  {\bibfnamefont {S.}~\bibnamefont {Endo}}, \bibinfo {author} {\bibfnamefont
  {K.}~\bibnamefont {Fujii}}, \bibinfo {author} {\bibfnamefont {J.~R.}\
  \bibnamefont {McClean}}, \bibinfo {author} {\bibfnamefont {K.}~\bibnamefont
  {Mitarai}}, \bibinfo {author} {\bibfnamefont {X.}~\bibnamefont {Yuan}},
  \bibinfo {author} {\bibfnamefont {L.}~\bibnamefont {Cincio}},\ and\ \bibinfo
  {author} {\bibfnamefont {P.~J.}\ \bibnamefont {Coles}},\ }\href
  {https://doi.org/10.1038/s42254-021-00348-9} {\bibfield  {journal} {\bibinfo
  {journal} {Nat. Phys. Rev.}\ }\textbf {\bibinfo {volume} {3}},\ \bibinfo
  {pages} {625} (\bibinfo {year} {2021})}\BibitemShut {NoStop}%
\bibitem [{\citenamefont {Bharti}\ \emph {et~al.}(2022)\citenamefont {Bharti},
  \citenamefont {Cervera-Lierta}, \citenamefont {Kyaw}, \citenamefont {Haug},
  \citenamefont {Alperin-Lea}, \citenamefont {Anand}, \citenamefont {Degroote},
  \citenamefont {Heimonen}, \citenamefont {Kottmann}, \citenamefont {Menke},
  \citenamefont {Mok}, \citenamefont {Sim}, \citenamefont {Kwek},\ and\
  \citenamefont {Aspuru-Guzik}}]{Bharti2022}%
  \BibitemOpen
  \bibfield  {author} {\bibinfo {author} {\bibfnamefont {K.}~\bibnamefont
  {Bharti}}, \bibinfo {author} {\bibfnamefont {A.}~\bibnamefont
  {Cervera-Lierta}}, \bibinfo {author} {\bibfnamefont {T.~H.}\ \bibnamefont
  {Kyaw}}, \bibinfo {author} {\bibfnamefont {T.}~\bibnamefont {Haug}}, \bibinfo
  {author} {\bibfnamefont {S.}~\bibnamefont {Alperin-Lea}}, \bibinfo {author}
  {\bibfnamefont {A.}~\bibnamefont {Anand}}, \bibinfo {author} {\bibfnamefont
  {M.}~\bibnamefont {Degroote}}, \bibinfo {author} {\bibfnamefont
  {H.}~\bibnamefont {Heimonen}}, \bibinfo {author} {\bibfnamefont {J.~S.}\
  \bibnamefont {Kottmann}}, \bibinfo {author} {\bibfnamefont {T.}~\bibnamefont
  {Menke}}, \bibinfo {author} {\bibfnamefont {W.~K.}\ \bibnamefont {Mok}},
  \bibinfo {author} {\bibfnamefont {S.}~\bibnamefont {Sim}}, \bibinfo {author}
  {\bibfnamefont {L.~C.}\ \bibnamefont {Kwek}},\ and\ \bibinfo {author}
  {\bibfnamefont {A.}~\bibnamefont {Aspuru-Guzik}},\ }\href
  {https://doi.org/10.1103/REVMODPHYS.94.015004/FIGURES/7/MEDIUM} {\bibfield
  {journal} {\bibinfo  {journal} {Rev. Mod. Phys.}\ }\textbf {\bibinfo {volume}
  {94}},\ \bibinfo {pages} {015004} (\bibinfo {year} {2022})}\BibitemShut
  {NoStop}%
\bibitem [{\citenamefont {Kaubruegger}\ \emph {et~al.}(2019)\citenamefont
  {Kaubruegger}, \citenamefont {Silvi}, \citenamefont {Kokail}, \citenamefont
  {van Bijnen}, \citenamefont {Rey}, \citenamefont {Ye}, \citenamefont
  {Kaufman},\ and\ \citenamefont {Zoller}}]{Kaubruegger2019}%
  \BibitemOpen
  \bibfield  {author} {\bibinfo {author} {\bibfnamefont {R.}~\bibnamefont
  {Kaubruegger}}, \bibinfo {author} {\bibfnamefont {P.}~\bibnamefont {Silvi}},
  \bibinfo {author} {\bibfnamefont {C.}~\bibnamefont {Kokail}}, \bibinfo
  {author} {\bibfnamefont {R.}~\bibnamefont {van Bijnen}}, \bibinfo {author}
  {\bibfnamefont {A.~M.}\ \bibnamefont {Rey}}, \bibinfo {author} {\bibfnamefont
  {J.}~\bibnamefont {Ye}}, \bibinfo {author} {\bibfnamefont {A.~M.}\
  \bibnamefont {Kaufman}},\ and\ \bibinfo {author} {\bibfnamefont
  {P.}~\bibnamefont {Zoller}},\ }\href
  {https://doi.org/10.1103/PhysRevLett.123.260505} {\bibfield  {journal}
  {\bibinfo  {journal} {Phys. Rev. Lett.}\ }\textbf {\bibinfo {volume} {123}},\
  \bibinfo {pages} {260505} (\bibinfo {year} {2019})}\BibitemShut {NoStop}%
\bibitem [{\citenamefont {Sun}\ \emph {et~al.}()\citenamefont {Sun},
  \citenamefont {Wang}, \citenamefont {Zhang}, \citenamefont {Nori},\ and\
  \citenamefont {Fan}}]{sun2023variational}%
  \BibitemOpen
  \bibfield  {author} {\bibinfo {author} {\bibfnamefont {Z.-H.}\ \bibnamefont
  {Sun}}, \bibinfo {author} {\bibfnamefont {Y.-Y.}\ \bibnamefont {Wang}},
  \bibinfo {author} {\bibfnamefont {Y.-R.}\ \bibnamefont {Zhang}}, \bibinfo
  {author} {\bibfnamefont {F.}~\bibnamefont {Nori}},\ and\ \bibinfo {author}
  {\bibfnamefont {H.}~\bibnamefont {Fan}},\ }\href@noop {} {}\Eprint
  {https://arxiv.org/abs/2306.16194} {arXiv:2306.16194} \BibitemShut {NoStop}%
\bibitem [{\citenamefont {Meyer}\ \emph {et~al.}(2021)\citenamefont {Meyer},
  \citenamefont {Borregaard},\ and\ \citenamefont {Eisert}}]{Meyer2021}%
  \BibitemOpen
  \bibfield  {author} {\bibinfo {author} {\bibfnamefont {J.~J.}\ \bibnamefont
  {Meyer}}, \bibinfo {author} {\bibfnamefont {J.}~\bibnamefont {Borregaard}},\
  and\ \bibinfo {author} {\bibfnamefont {J.}~\bibnamefont {Eisert}},\ }\href
  {https://doi.org/10.1038/s41534-021-00425-y} {\bibfield  {journal} {\bibinfo
  {journal} {npj Quantum Information}\ }\textbf {\bibinfo {volume} {7}},\
  \bibinfo {pages} {89} (\bibinfo {year} {2021})}\BibitemShut {NoStop}%
\bibitem [{\citenamefont {Ma}\ \emph {et~al.}(2021)\citenamefont {Ma},
  \citenamefont {Gokhale}, \citenamefont {Zheng}, \citenamefont {Zhou},
  \citenamefont {Yu}, \citenamefont {Jiang}, \citenamefont {Maurer},\ and\
  \citenamefont {Chong}}]{Ma2021}%
  \BibitemOpen
  \bibfield  {author} {\bibinfo {author} {\bibfnamefont {Z.}~\bibnamefont
  {Ma}}, \bibinfo {author} {\bibfnamefont {P.}~\bibnamefont {Gokhale}},
  \bibinfo {author} {\bibfnamefont {T.~X.}\ \bibnamefont {Zheng}}, \bibinfo
  {author} {\bibfnamefont {S.}~\bibnamefont {Zhou}}, \bibinfo {author}
  {\bibfnamefont {X.}~\bibnamefont {Yu}}, \bibinfo {author} {\bibfnamefont
  {L.}~\bibnamefont {Jiang}}, \bibinfo {author} {\bibfnamefont
  {P.}~\bibnamefont {Maurer}},\ and\ \bibinfo {author} {\bibfnamefont {F.~T.}\
  \bibnamefont {Chong}},\ }\href {https://doi.org/10.1109/QCE52317.2021.00063}
  {\bibfield  {journal} {\bibinfo  {journal} {Proceedings - 2021 IEEE
  International Conference on Quantum Computing and Engineering}\ ,\ \bibinfo
  {pages} {419}} (\bibinfo {year} {2021})}\BibitemShut {NoStop}%
\bibitem [{\citenamefont {Le}\ \emph {et~al.}()\citenamefont {Le},
  \citenamefont {Nguyen},\ and\ \citenamefont {Ho}}]{le2023variational}%
  \BibitemOpen
  \bibfield  {author} {\bibinfo {author} {\bibfnamefont {T.~K.}\ \bibnamefont
  {Le}}, \bibinfo {author} {\bibfnamefont {H.~Q.}\ \bibnamefont {Nguyen}},\
  and\ \bibinfo {author} {\bibfnamefont {L.~B.}\ \bibnamefont {Ho}},\
  }\href@noop {} {}\Eprint {https://arxiv.org/abs/2305.08289}
  {arXiv:2305.08289} \BibitemShut {NoStop}%
\bibitem [{\citenamefont {Koczor}\ \emph {et~al.}(2020)\citenamefont {Koczor},
  \citenamefont {Endo}, \citenamefont {Jones}, \citenamefont {Matsuzaki},\ and\
  \citenamefont {Benjamin}}]{Koczor2020}%
  \BibitemOpen
  \bibfield  {author} {\bibinfo {author} {\bibfnamefont {B.}~\bibnamefont
  {Koczor}}, \bibinfo {author} {\bibfnamefont {S.}~\bibnamefont {Endo}},
  \bibinfo {author} {\bibfnamefont {T.}~\bibnamefont {Jones}}, \bibinfo
  {author} {\bibfnamefont {Y.}~\bibnamefont {Matsuzaki}},\ and\ \bibinfo
  {author} {\bibfnamefont {S.~C.}\ \bibnamefont {Benjamin}},\ }\href
  {https://doi.org/10.1088/1367-2630/AB965E} {\bibfield  {journal} {\bibinfo
  {journal} {New Journal of Physics}\ }\textbf {\bibinfo {volume} {22}},\
  \bibinfo {pages} {083038} (\bibinfo {year} {2020})}\BibitemShut {NoStop}%
\bibitem [{\citenamefont {Beckey}\ \emph {et~al.}(2022)\citenamefont {Beckey},
  \citenamefont {Cerezo}, \citenamefont {Sone},\ and\ \citenamefont
  {Coles}}]{Beckey2022}%
  \BibitemOpen
  \bibfield  {author} {\bibinfo {author} {\bibfnamefont {J.~L.}\ \bibnamefont
  {Beckey}}, \bibinfo {author} {\bibfnamefont {M.}~\bibnamefont {Cerezo}},
  \bibinfo {author} {\bibfnamefont {A.}~\bibnamefont {Sone}},\ and\ \bibinfo
  {author} {\bibfnamefont {P.~J.}\ \bibnamefont {Coles}},\ }\href
  {https://doi.org/10.1103/PhysRevResearch.4.013083} {\bibfield  {journal}
  {\bibinfo  {journal} {Phys. Rev. Res.}\ }\textbf {\bibinfo {volume} {4}},\
  \bibinfo {pages} {013083} (\bibinfo {year} {2022})}\BibitemShut {NoStop}%
\bibitem [{\citenamefont {Yang}\ \emph {et~al.}(2022)\citenamefont {Yang},
  \citenamefont {Pang}, \citenamefont {Chen}, \citenamefont {Jordan},\ and\
  \citenamefont {del Campo}}]{Yang2022}%
  \BibitemOpen
  \bibfield  {author} {\bibinfo {author} {\bibfnamefont {J.}~\bibnamefont
  {Yang}}, \bibinfo {author} {\bibfnamefont {S.}~\bibnamefont {Pang}}, \bibinfo
  {author} {\bibfnamefont {Z.}~\bibnamefont {Chen}}, \bibinfo {author}
  {\bibfnamefont {A.~N.}\ \bibnamefont {Jordan}},\ and\ \bibinfo {author}
  {\bibfnamefont {A.}~\bibnamefont {del Campo}},\ }\href
  {https://doi.org/10.1103/PhysRevLett.128.160505} {\bibfield  {journal}
  {\bibinfo  {journal} {Phys. Rev. Lett.}\ }\textbf {\bibinfo {volume} {128}},\
  \bibinfo {pages} {160505} (\bibinfo {year} {2022})}\BibitemShut {NoStop}%
\bibitem [{\citenamefont {Kaubruegger}\ \emph {et~al.}(2023)\citenamefont
  {Kaubruegger}, \citenamefont {Shankar}, \citenamefont {Vasilyev},\ and\
  \citenamefont {Zoller}}]{Kaubruegger2023}%
  \BibitemOpen
  \bibfield  {author} {\bibinfo {author} {\bibfnamefont {R.}~\bibnamefont
  {Kaubruegger}}, \bibinfo {author} {\bibfnamefont {A.}~\bibnamefont
  {Shankar}}, \bibinfo {author} {\bibfnamefont {D.~V.}\ \bibnamefont
  {Vasilyev}},\ and\ \bibinfo {author} {\bibfnamefont {P.}~\bibnamefont
  {Zoller}},\ }\href {https://doi.org/10.1103/PRXQuantum.4.020333} {\bibfield
  {journal} {\bibinfo  {journal} {PRX Quantum}\ }\textbf {\bibinfo {volume}
  {4}},\ \bibinfo {pages} {020333} (\bibinfo {year} {2023})}\BibitemShut
  {NoStop}%
\bibitem [{\citenamefont {Krisnanda}\ \emph {et~al.}(2021)\citenamefont
  {Krisnanda}, \citenamefont {Ghosh}, \citenamefont {Paterek},\ and\
  \citenamefont {Liew}}]{KRISNANDA2021}%
  \BibitemOpen
  \bibfield  {author} {\bibinfo {author} {\bibfnamefont {T.}~\bibnamefont
  {Krisnanda}}, \bibinfo {author} {\bibfnamefont {S.}~\bibnamefont {Ghosh}},
  \bibinfo {author} {\bibfnamefont {T.}~\bibnamefont {Paterek}},\ and\ \bibinfo
  {author} {\bibfnamefont {T.~C.}\ \bibnamefont {Liew}},\ }\href
  {https://doi.org/https://doi.org/10.1016/j.neunet.2021.01.003} {\bibfield
  {journal} {\bibinfo  {journal} {Neural Networks}\ }\textbf {\bibinfo {volume}
  {136}},\ \bibinfo {pages} {141} (\bibinfo {year} {2021})}\BibitemShut
  {NoStop}%
\bibitem [{\citenamefont {Cimini}\ \emph {et~al.}()\citenamefont {Cimini},
  \citenamefont {Valeri}, \citenamefont {Piacentini}, \citenamefont
  {Ceccarelli}, \citenamefont {Corrielli}, \citenamefont {Osellame},
  \citenamefont {Spagnolo},\ and\ \citenamefont
  {Sciarrino}}]{cimini2023variational}%
  \BibitemOpen
  \bibfield  {author} {\bibinfo {author} {\bibfnamefont {V.}~\bibnamefont
  {Cimini}}, \bibinfo {author} {\bibfnamefont {M.}~\bibnamefont {Valeri}},
  \bibinfo {author} {\bibfnamefont {S.}~\bibnamefont {Piacentini}}, \bibinfo
  {author} {\bibfnamefont {F.}~\bibnamefont {Ceccarelli}}, \bibinfo {author}
  {\bibfnamefont {G.}~\bibnamefont {Corrielli}}, \bibinfo {author}
  {\bibfnamefont {R.}~\bibnamefont {Osellame}}, \bibinfo {author}
  {\bibfnamefont {N.}~\bibnamefont {Spagnolo}},\ and\ \bibinfo {author}
  {\bibfnamefont {F.}~\bibnamefont {Sciarrino}},\ }\href@noop {} {}\Eprint
  {https://arxiv.org/abs/2308.02643} {arXiv:2308.02643} \BibitemShut {NoStop}%
\bibitem [{\citenamefont {Steinbrecher}\ \emph {et~al.}(2019)\citenamefont
  {Steinbrecher}, \citenamefont {Olson}, \citenamefont {Englund},\ and\
  \citenamefont {Carolan}}]{Steinbrecher2019}%
  \BibitemOpen
  \bibfield  {author} {\bibinfo {author} {\bibfnamefont {G.~R.}\ \bibnamefont
  {Steinbrecher}}, \bibinfo {author} {\bibfnamefont {J.~P.}\ \bibnamefont
  {Olson}}, \bibinfo {author} {\bibfnamefont {D.}~\bibnamefont {Englund}},\
  and\ \bibinfo {author} {\bibfnamefont {J.}~\bibnamefont {Carolan}},\ }\href
  {https://doi.org/10.1038/s41534-019-0174-7} {\bibfield  {journal} {\bibinfo
  {journal} {npj Quantum Information}\ }\textbf {\bibinfo {volume} {5}},\
  \bibinfo {pages} {60} (\bibinfo {year} {2019})}\BibitemShut {NoStop}%
\bibitem [{\citenamefont {Kimble}(1998)}]{Kimble1998}%
  \BibitemOpen
  \bibfield  {author} {\bibinfo {author} {\bibfnamefont {H.~J.}\ \bibnamefont
  {Kimble}},\ }\href {https://doi.org/10.1238/Physica.Topical.076a00127}
  {\bibfield  {journal} {\bibinfo  {journal} {Physica Scripta}\ }\textbf
  {\bibinfo {volume} {1998}},\ \bibinfo {pages} {127} (\bibinfo {year}
  {1998})}\BibitemShut {NoStop}%
\bibitem [{\citenamefont {On}\ \emph {et~al.}()\citenamefont {On},
  \citenamefont {Ashtiani}, \citenamefont {Sanchez-Jacome}, \citenamefont
  {Perez-Lopez}, \citenamefont {Yoo},\ and\ \citenamefont
  {Blanco-Redondo}}]{on2023programmable}%
  \BibitemOpen
  \bibfield  {author} {\bibinfo {author} {\bibfnamefont {M.~B.}\ \bibnamefont
  {On}}, \bibinfo {author} {\bibfnamefont {F.}~\bibnamefont {Ashtiani}},
  \bibinfo {author} {\bibfnamefont {D.}~\bibnamefont {Sanchez-Jacome}},
  \bibinfo {author} {\bibfnamefont {D.}~\bibnamefont {Perez-Lopez}}, \bibinfo
  {author} {\bibfnamefont {S.~J.~B.}\ \bibnamefont {Yoo}},\ and\ \bibinfo
  {author} {\bibfnamefont {A.}~\bibnamefont {Blanco-Redondo}},\ }\href@noop {}
  {}\Eprint {https://arxiv.org/abs/2307.05003} {arXiv:2307.05003} \BibitemShut
  {NoStop}%
\bibitem [{\citenamefont {Sun}\ \emph {et~al.}(2021)\citenamefont {Sun},
  \citenamefont {Motta}, \citenamefont {Tazhigulov}, \citenamefont {Tan},
  \citenamefont {Chan},\ and\ \citenamefont {Minnich}}]{Sun2021}%
  \BibitemOpen
  \bibfield  {author} {\bibinfo {author} {\bibfnamefont {S.-N.}\ \bibnamefont
  {Sun}}, \bibinfo {author} {\bibfnamefont {M.}~\bibnamefont {Motta}}, \bibinfo
  {author} {\bibfnamefont {R.~N.}\ \bibnamefont {Tazhigulov}}, \bibinfo
  {author} {\bibfnamefont {A.~T.}\ \bibnamefont {Tan}}, \bibinfo {author}
  {\bibfnamefont {G.~K.-L.}\ \bibnamefont {Chan}},\ and\ \bibinfo {author}
  {\bibfnamefont {A.~J.}\ \bibnamefont {Minnich}},\ }\href
  {https://doi.org/10.1103/PRXQuantum.2.010317} {\bibfield  {journal} {\bibinfo
   {journal} {PRX Quantum}\ }\textbf {\bibinfo {volume} {2}},\ \bibinfo {pages}
  {010317} (\bibinfo {year} {2021})}\BibitemShut {NoStop}%
\bibitem [{Git()}]{GitHub}%
  \BibitemOpen
  \href@noop {} {}\bibinfo {howpublished} {Codes to reproduce the results of
  this manuscript are available at:
  \url{https://github.com/albertomdlh/quantum-metrology-variational}}\BibitemShut
  {NoStop}%
\bibitem [{\citenamefont {Gerry}\ and\ \citenamefont
  {Knight}(2005)}]{gerry2005introductory}%
  \BibitemOpen
  \bibfield  {author} {\bibinfo {author} {\bibfnamefont {C.}~\bibnamefont
  {Gerry}}\ and\ \bibinfo {author} {\bibfnamefont {P.}~\bibnamefont {Knight}},\
  }\href {https://books.google.es/books?id=CgByyoBJJwgC} {\emph {\bibinfo
  {title} {Introductory Quantum Optics}}}\ (\bibinfo  {publisher} {Cambridge
  University Press},\ \bibinfo {year} {2005})\BibitemShut {NoStop}%
\bibitem [{Note1()}]{Note1}%
  \BibitemOpen
  \bibinfo {note} {This choice of initial states is motivated by the ease with
  which coherent states can be experimentally prepared. In Appendix~\ref
  {sec:squeezed} we also consider squeezed coherent states as inputs, but there
  was no significant improvement over the results}\BibitemShut {NoStop}%
\bibitem [{\citenamefont {Cramer}\ \emph {et~al.}(2010)\citenamefont {Cramer},
  \citenamefont {Plenio}, \citenamefont {Flammia}, \citenamefont {Somma},
  \citenamefont {Gross}, \citenamefont {Bartlett}, \citenamefont
  {Landon-Cardinal}, \citenamefont {Poulin},\ and\ \citenamefont
  {Liu}}]{Cramer2010}%
  \BibitemOpen
  \bibfield  {author} {\bibinfo {author} {\bibfnamefont {M.}~\bibnamefont
  {Cramer}}, \bibinfo {author} {\bibfnamefont {M.~B.}\ \bibnamefont {Plenio}},
  \bibinfo {author} {\bibfnamefont {S.~T.}\ \bibnamefont {Flammia}}, \bibinfo
  {author} {\bibfnamefont {R.}~\bibnamefont {Somma}}, \bibinfo {author}
  {\bibfnamefont {D.}~\bibnamefont {Gross}}, \bibinfo {author} {\bibfnamefont
  {S.~D.}\ \bibnamefont {Bartlett}}, \bibinfo {author} {\bibfnamefont
  {O.}~\bibnamefont {Landon-Cardinal}}, \bibinfo {author} {\bibfnamefont
  {D.}~\bibnamefont {Poulin}},\ and\ \bibinfo {author} {\bibfnamefont {Y.-K.}\
  \bibnamefont {Liu}},\ }\href {https://doi.org/10.1038/ncomms1147} {\bibfield
  {journal} {\bibinfo  {journal} {Nature Communications}\ }\textbf {\bibinfo
  {volume} {1}},\ \bibinfo {pages} {149} (\bibinfo {year} {2010})}\BibitemShut
  {NoStop}%
\bibitem [{\citenamefont {Walther}\ \emph {et~al.}(2006)\citenamefont
  {Walther}, \citenamefont {Varcoe}, \citenamefont {Englert},\ and\
  \citenamefont {Becker}}]{Walther_2006}%
  \BibitemOpen
  \bibfield  {author} {\bibinfo {author} {\bibfnamefont {H.}~\bibnamefont
  {Walther}}, \bibinfo {author} {\bibfnamefont {B.~T.~H.}\ \bibnamefont
  {Varcoe}}, \bibinfo {author} {\bibfnamefont {B.-G.}\ \bibnamefont
  {Englert}},\ and\ \bibinfo {author} {\bibfnamefont {T.}~\bibnamefont
  {Becker}},\ }\href {https://doi.org/10.1088/0034-4885/69/5/R02} {\bibfield
  {journal} {\bibinfo  {journal} {Reports on Progress in Physics}\ }\textbf
  {\bibinfo {volume} {69}},\ \bibinfo {pages} {1325} (\bibinfo {year}
  {2006})}\BibitemShut {NoStop}%
\bibitem [{\citenamefont {Reiserer}\ and\ \citenamefont
  {Rempe}(2015)}]{Reiserer2015}%
  \BibitemOpen
  \bibfield  {author} {\bibinfo {author} {\bibfnamefont {A.}~\bibnamefont
  {Reiserer}}\ and\ \bibinfo {author} {\bibfnamefont {G.}~\bibnamefont
  {Rempe}},\ }\href {https://doi.org/10.1103/REVMODPHYS.87.1379} {\bibfield
  {journal} {\bibinfo  {journal} {Rev. Mod. Phys.}\ }\textbf {\bibinfo {volume}
  {87}},\ \bibinfo {pages} {1379} (\bibinfo {year} {2015})}\BibitemShut
  {NoStop}%
\bibitem [{\citenamefont {Blais}\ \emph {et~al.}(2021)\citenamefont {Blais},
  \citenamefont {Grimsmo}, \citenamefont {Girvin},\ and\ \citenamefont
  {Wallraff}}]{Blais2021}%
  \BibitemOpen
  \bibfield  {author} {\bibinfo {author} {\bibfnamefont {A.}~\bibnamefont
  {Blais}}, \bibinfo {author} {\bibfnamefont {A.~L.}\ \bibnamefont {Grimsmo}},
  \bibinfo {author} {\bibfnamefont {S.~M.}\ \bibnamefont {Girvin}},\ and\
  \bibinfo {author} {\bibfnamefont {A.}~\bibnamefont {Wallraff}},\ }\href
  {https://doi.org/10.1103/RevModPhys.93.025005} {\bibfield  {journal}
  {\bibinfo  {journal} {Rev. Mod. Phys.}\ }\textbf {\bibinfo {volume} {93}},\
  \bibinfo {pages} {025005} (\bibinfo {year} {2021})}\BibitemShut {NoStop}%
\bibitem [{\citenamefont {Butcher}\ and\ \citenamefont
  {Cotter}(1990)}]{butcher_cotter_1990}%
  \BibitemOpen
  \bibfield  {author} {\bibinfo {author} {\bibfnamefont {P.~N.}\ \bibnamefont
  {Butcher}}\ and\ \bibinfo {author} {\bibfnamefont {D.}~\bibnamefont
  {Cotter}},\ }\href {https://doi.org/10.1017/CBO9781139167994} {\emph
  {\bibinfo {title} {The Elements of Nonlinear Optics}}},\ Cambridge Studies in
  Modern Optics\ (\bibinfo  {publisher} {Cambridge University Press},\ \bibinfo
  {year} {1990})\BibitemShut {NoStop}%
\bibitem [{Note2()}]{Note2}%
  \BibitemOpen
  \bibinfo {note} {In Figs.~\ref {fig:noiseless}-\ref {fig:noise} we employ a
  circuit depth $d=5$ larger than the number of layers $d=2$ necessary to
  attain convergence because we found a slight improvement with respect to the
  latter case. Results for $d=2$ can be found in Appendix~\ref
  {sec:d=2}.}\BibitemShut {Stop}%
\bibitem [{\citenamefont {Gu}\ \emph {et~al.}(2017)\citenamefont {Gu},
  \citenamefont {Kockum}, \citenamefont {Miranowicz}, \citenamefont {xi~Liu},\
  and\ \citenamefont {Nori}}]{GU2017}%
  \BibitemOpen
  \bibfield  {author} {\bibinfo {author} {\bibfnamefont {X.}~\bibnamefont
  {Gu}}, \bibinfo {author} {\bibfnamefont {A.~F.}\ \bibnamefont {Kockum}},
  \bibinfo {author} {\bibfnamefont {A.}~\bibnamefont {Miranowicz}}, \bibinfo
  {author} {\bibfnamefont {Y.}~\bibnamefont {xi~Liu}},\ and\ \bibinfo {author}
  {\bibfnamefont {F.}~\bibnamefont {Nori}},\ }\href
  {https://doi.org/https://doi.org/10.1016/j.physrep.2017.10.002} {\bibfield
  {journal} {\bibinfo  {journal} {Physics Reports}\ }\textbf {\bibinfo {volume}
  {718-719}},\ \bibinfo {pages} {1} (\bibinfo {year} {2017})}\BibitemShut
  {NoStop}%
\bibitem [{\citenamefont {Meschede}\ \emph {et~al.}(1985)\citenamefont
  {Meschede}, \citenamefont {Walther},\ and\ \citenamefont
  {M\"uller}}]{Meschede1985}%
  \BibitemOpen
  \bibfield  {author} {\bibinfo {author} {\bibfnamefont {D.}~\bibnamefont
  {Meschede}}, \bibinfo {author} {\bibfnamefont {H.}~\bibnamefont {Walther}},\
  and\ \bibinfo {author} {\bibfnamefont {G.}~\bibnamefont {M\"uller}},\ }\href
  {https://doi.org/10.1103/PhysRevLett.54.551} {\bibfield  {journal} {\bibinfo
  {journal} {Phys. Rev. Lett.}\ }\textbf {\bibinfo {volume} {54}},\ \bibinfo
  {pages} {551} (\bibinfo {year} {1985})}\BibitemShut {NoStop}%
\bibitem [{\citenamefont {Thompson}\ \emph {et~al.}(1992)\citenamefont
  {Thompson}, \citenamefont {Rempe},\ and\ \citenamefont
  {Kimble}}]{Thompson1992}%
  \BibitemOpen
  \bibfield  {author} {\bibinfo {author} {\bibfnamefont {R.~J.}\ \bibnamefont
  {Thompson}}, \bibinfo {author} {\bibfnamefont {G.}~\bibnamefont {Rempe}},\
  and\ \bibinfo {author} {\bibfnamefont {H.~J.}\ \bibnamefont {Kimble}},\
  }\href {https://doi.org/10.1103/PhysRevLett.68.1132} {\bibfield  {journal}
  {\bibinfo  {journal} {Phys. Rev. Lett.}\ }\textbf {\bibinfo {volume} {68}},\
  \bibinfo {pages} {1132} (\bibinfo {year} {1992})}\BibitemShut {NoStop}%
\bibitem [{\citenamefont {Lodahl}\ \emph {et~al.}(2015)\citenamefont {Lodahl},
  \citenamefont {Mahmoodian},\ and\ \citenamefont {Stobbe}}]{Lodahl2015}%
  \BibitemOpen
  \bibfield  {author} {\bibinfo {author} {\bibfnamefont {P.}~\bibnamefont
  {Lodahl}}, \bibinfo {author} {\bibfnamefont {S.}~\bibnamefont {Mahmoodian}},\
  and\ \bibinfo {author} {\bibfnamefont {S.}~\bibnamefont {Stobbe}},\ }\href
  {https://doi.org/10.1103/RevModPhys.87.347} {\bibfield  {journal} {\bibinfo
  {journal} {Rev. Mod. Phys.}\ }\textbf {\bibinfo {volume} {87}},\ \bibinfo
  {pages} {347} (\bibinfo {year} {2015})}\BibitemShut {NoStop}%
\bibitem [{\citenamefont {{n}oz}\ \emph {et~al.}(2018)\citenamefont {{n}oz},
  \citenamefont {Laussy}, \citenamefont {del Valle}, \citenamefont {Tejedor},\
  and\ \citenamefont {Gonz\'{a}lez-Tudela}}]{SanchezMunoz2018}%
  \BibitemOpen
  \bibfield  {author} {\bibinfo {author} {\bibfnamefont {C.~S.~M.}\
  \bibnamefont {{n}oz}}, \bibinfo {author} {\bibfnamefont {F.~P.}\ \bibnamefont
  {Laussy}}, \bibinfo {author} {\bibfnamefont {E.}~\bibnamefont {del Valle}},
  \bibinfo {author} {\bibfnamefont {C.}~\bibnamefont {Tejedor}},\ and\ \bibinfo
  {author} {\bibfnamefont {A.}~\bibnamefont {Gonz\'{a}lez-Tudela}},\ }\href
  {https://doi.org/10.1364/OPTICA.5.000014} {\bibfield  {journal} {\bibinfo
  {journal} {Optica}\ }\textbf {\bibinfo {volume} {5}},\ \bibinfo {pages} {14}
  (\bibinfo {year} {2018})}\BibitemShut {NoStop}%
\bibitem [{\citenamefont {Delteil}\ \emph {et~al.}(2019)\citenamefont
  {Delteil}, \citenamefont {Fink}, \citenamefont {Schade}, \citenamefont
  {H{\"{o}}fling}, \citenamefont {Schneider},\ and\ \citenamefont
  {İmamoglu}}]{Delteil2019}%
  \BibitemOpen
  \bibfield  {author} {\bibinfo {author} {\bibfnamefont {A.}~\bibnamefont
  {Delteil}}, \bibinfo {author} {\bibfnamefont {T.}~\bibnamefont {Fink}},
  \bibinfo {author} {\bibfnamefont {A.}~\bibnamefont {Schade}}, \bibinfo
  {author} {\bibfnamefont {S.}~\bibnamefont {H{\"{o}}fling}}, \bibinfo {author}
  {\bibfnamefont {C.}~\bibnamefont {Schneider}},\ and\ \bibinfo {author}
  {\bibfnamefont {A.}~\bibnamefont {İmamoglu}},\ }\href
  {https://doi.org/10.1038/s41563-019-0282-y} {\bibfield  {journal} {\bibinfo
  {journal} {Nat. Mater.}\ }\textbf {\bibinfo {volume} {18}},\ \bibinfo {pages}
  {219} (\bibinfo {year} {2019})}\BibitemShut {NoStop}%
\bibitem [{\citenamefont {Mu{\~{n}}oz-Matutano}\ \emph
  {et~al.}(2019)\citenamefont {Mu{\~{n}}oz-Matutano}, \citenamefont {Wood},
  \citenamefont {Johnsson}, \citenamefont {Vidal}, \citenamefont {Baragiola},
  \citenamefont {Reinhard}, \citenamefont {Lema{\^{i}}tre}, \citenamefont
  {Bloch}, \citenamefont {Amo}, \citenamefont {Nogues}, \citenamefont {Besga},
  \citenamefont {Richard},\ and\ \citenamefont {Volz}}]{Munoz-Matutano2019}%
  \BibitemOpen
  \bibfield  {author} {\bibinfo {author} {\bibfnamefont {G.}~\bibnamefont
  {Mu{\~{n}}oz-Matutano}}, \bibinfo {author} {\bibfnamefont {A.}~\bibnamefont
  {Wood}}, \bibinfo {author} {\bibfnamefont {M.}~\bibnamefont {Johnsson}},
  \bibinfo {author} {\bibfnamefont {X.}~\bibnamefont {Vidal}}, \bibinfo
  {author} {\bibfnamefont {B.~Q.}\ \bibnamefont {Baragiola}}, \bibinfo {author}
  {\bibfnamefont {A.}~\bibnamefont {Reinhard}}, \bibinfo {author}
  {\bibfnamefont {A.}~\bibnamefont {Lema{\^{i}}tre}}, \bibinfo {author}
  {\bibfnamefont {J.}~\bibnamefont {Bloch}}, \bibinfo {author} {\bibfnamefont
  {A.}~\bibnamefont {Amo}}, \bibinfo {author} {\bibfnamefont {G.}~\bibnamefont
  {Nogues}}, \bibinfo {author} {\bibfnamefont {B.}~\bibnamefont {Besga}},
  \bibinfo {author} {\bibfnamefont {M.}~\bibnamefont {Richard}},\ and\ \bibinfo
  {author} {\bibfnamefont {T.}~\bibnamefont {Volz}},\ }\href
  {https://doi.org/10.1038/s41563-019-0281-z} {\bibfield  {journal} {\bibinfo
  {journal} {Nat. Mater.}\ }\textbf {\bibinfo {volume} {18}},\ \bibinfo {pages}
  {213} (\bibinfo {year} {2019})}\BibitemShut {NoStop}%
\bibitem [{\citenamefont {Koch}\ \emph {et~al.}(2007)\citenamefont {Koch},
  \citenamefont {Yu}, \citenamefont {Gambetta}, \citenamefont {Houck},
  \citenamefont {Schuster}, \citenamefont {Majer}, \citenamefont {Blais},
  \citenamefont {Devoret}, \citenamefont {Girvin},\ and\ \citenamefont
  {Schoelkopf}}]{Koch2007}%
  \BibitemOpen
  \bibfield  {author} {\bibinfo {author} {\bibfnamefont {J.}~\bibnamefont
  {Koch}}, \bibinfo {author} {\bibfnamefont {T.~M.}\ \bibnamefont {Yu}},
  \bibinfo {author} {\bibfnamefont {J.}~\bibnamefont {Gambetta}}, \bibinfo
  {author} {\bibfnamefont {A.~A.}\ \bibnamefont {Houck}}, \bibinfo {author}
  {\bibfnamefont {D.~I.}\ \bibnamefont {Schuster}}, \bibinfo {author}
  {\bibfnamefont {J.}~\bibnamefont {Majer}}, \bibinfo {author} {\bibfnamefont
  {A.}~\bibnamefont {Blais}}, \bibinfo {author} {\bibfnamefont {M.~H.}\
  \bibnamefont {Devoret}}, \bibinfo {author} {\bibfnamefont {S.~M.}\
  \bibnamefont {Girvin}},\ and\ \bibinfo {author} {\bibfnamefont {R.~J.}\
  \bibnamefont {Schoelkopf}},\ }\href
  {https://doi.org/10.1103/PhysRevA.76.042319} {\bibfield  {journal} {\bibinfo
  {journal} {Phys. Rev. A}\ }\textbf {\bibinfo {volume} {76}},\ \bibinfo
  {pages} {042319} (\bibinfo {year} {2007})}\BibitemShut {NoStop}%
\bibitem [{\citenamefont {Kirchmair}\ \emph {et~al.}(2013)\citenamefont
  {Kirchmair}, \citenamefont {Vlastakis}, \citenamefont {Leghtas},
  \citenamefont {Nigg}, \citenamefont {Paik}, \citenamefont {Ginossar},
  \citenamefont {Mirrahimi}, \citenamefont {Frunzio}, \citenamefont {Girvin},\
  and\ \citenamefont {Schoelkopf}}]{Kirchmair2013}%
  \BibitemOpen
  \bibfield  {author} {\bibinfo {author} {\bibfnamefont {G.}~\bibnamefont
  {Kirchmair}}, \bibinfo {author} {\bibfnamefont {B.}~\bibnamefont
  {Vlastakis}}, \bibinfo {author} {\bibfnamefont {Z.}~\bibnamefont {Leghtas}},
  \bibinfo {author} {\bibfnamefont {S.~E.}\ \bibnamefont {Nigg}}, \bibinfo
  {author} {\bibfnamefont {H.}~\bibnamefont {Paik}}, \bibinfo {author}
  {\bibfnamefont {E.}~\bibnamefont {Ginossar}}, \bibinfo {author}
  {\bibfnamefont {M.}~\bibnamefont {Mirrahimi}}, \bibinfo {author}
  {\bibfnamefont {L.}~\bibnamefont {Frunzio}}, \bibinfo {author} {\bibfnamefont
  {S.~M.}\ \bibnamefont {Girvin}},\ and\ \bibinfo {author} {\bibfnamefont
  {R.~J.}\ \bibnamefont {Schoelkopf}},\ }\href
  {https://doi.org/10.1038/nature11902} {\bibfield  {journal} {\bibinfo
  {journal} {Nature}\ }\textbf {\bibinfo {volume} {495}},\ \bibinfo {pages}
  {205} (\bibinfo {year} {2013})}\BibitemShut {NoStop}%
\bibitem [{\citenamefont {Puri}\ \emph {et~al.}(2017)\citenamefont {Puri},
  \citenamefont {Andersen}, \citenamefont {Grimsmo},\ and\ \citenamefont
  {Blais}}]{Puri2017}%
  \BibitemOpen
  \bibfield  {author} {\bibinfo {author} {\bibfnamefont {S.}~\bibnamefont
  {Puri}}, \bibinfo {author} {\bibfnamefont {C.~K.}\ \bibnamefont {Andersen}},
  \bibinfo {author} {\bibfnamefont {A.~L.}\ \bibnamefont {Grimsmo}},\ and\
  \bibinfo {author} {\bibfnamefont {A.}~\bibnamefont {Blais}},\ }\href
  {https://doi.org/10.1038/ncomms15785} {\bibfield  {journal} {\bibinfo
  {journal} {Nature Comm.}\ }\textbf {\bibinfo {volume} {8}},\ \bibinfo {pages}
  {1} (\bibinfo {year} {2017})}\BibitemShut {NoStop}%
\bibitem [{\citenamefont {Gardiner}\ and\ \citenamefont
  {Zoller}(2004)}]{gardiner2004quantum}%
  \BibitemOpen
  \bibfield  {author} {\bibinfo {author} {\bibfnamefont {C.}~\bibnamefont
  {Gardiner}}\ and\ \bibinfo {author} {\bibfnamefont {P.}~\bibnamefont
  {Zoller}},\ }\href {https://books.google.es/books?id=a\_xsT8oGhdgC} {\emph
  {\bibinfo {title} {Quantum Noise: A Handbook of Markovian and Non-Markovian
  Quantum Stochastic Methods with Applications to Quantum Optics}}},\ Springer
  Series in Synergetics\ (\bibinfo  {publisher} {Springer},\ \bibinfo {year}
  {2004})\BibitemShut {NoStop}%
\bibitem [{\citenamefont {Demkowicz-Dobrza{\'{n}}ski}\ \emph
  {et~al.}(2012)\citenamefont {Demkowicz-Dobrza{\'{n}}ski}, \citenamefont
  {Ko{\l}ody{\'{n}}ski},\ and\ \citenamefont {Gu{\c{T}}{\u{a}}}}]{DD2012}%
  \BibitemOpen
  \bibfield  {author} {\bibinfo {author} {\bibfnamefont {R.}~\bibnamefont
  {Demkowicz-Dobrza{\'{n}}ski}}, \bibinfo {author} {\bibfnamefont
  {J.}~\bibnamefont {Ko{\l}ody{\'{n}}ski}},\ and\ \bibinfo {author}
  {\bibfnamefont {M.}~\bibnamefont {Gu{\c{T}}{\u{a}}}},\ }\href
  {https://doi.org/10.1038/ncomms2067} {\bibfield  {journal} {\bibinfo
  {journal} {Nature Communications}\ }\textbf {\bibinfo {volume} {3}},\
  \bibinfo {pages} {1063} (\bibinfo {year} {2012})}\BibitemShut {NoStop}%
\bibitem [{\citenamefont {Dorner}\ \emph {et~al.}(2009)\citenamefont {Dorner},
  \citenamefont {Demkowicz-Dobrzanski}, \citenamefont {Smith}, \citenamefont
  {Lundeen}, \citenamefont {Wasilewski}, \citenamefont {Banaszek},\ and\
  \citenamefont {Walmsley}}]{Dorner2009}%
  \BibitemOpen
  \bibfield  {author} {\bibinfo {author} {\bibfnamefont {U.}~\bibnamefont
  {Dorner}}, \bibinfo {author} {\bibfnamefont {R.}~\bibnamefont
  {Demkowicz-Dobrzanski}}, \bibinfo {author} {\bibfnamefont {B.~J.}\
  \bibnamefont {Smith}}, \bibinfo {author} {\bibfnamefont {J.~S.}\ \bibnamefont
  {Lundeen}}, \bibinfo {author} {\bibfnamefont {W.}~\bibnamefont {Wasilewski}},
  \bibinfo {author} {\bibfnamefont {K.}~\bibnamefont {Banaszek}},\ and\
  \bibinfo {author} {\bibfnamefont {I.~A.}\ \bibnamefont {Walmsley}},\ }\href
  {https://doi.org/10.1103/PhysRevLett.102.040403} {\bibfield  {journal}
  {\bibinfo  {journal} {Phys. Rev. Lett.}\ }\textbf {\bibinfo {volume} {102}},\
  \bibinfo {pages} {040403} (\bibinfo {year} {2009})}\BibitemShut {NoStop}%
\bibitem [{\citenamefont {Brownnutt}\ \emph {et~al.}(2015)\citenamefont
  {Brownnutt}, \citenamefont {Kumph}, \citenamefont {Rabl},\ and\ \citenamefont
  {Blatt}}]{Brownnutt2015}%
  \BibitemOpen
  \bibfield  {author} {\bibinfo {author} {\bibfnamefont {M.}~\bibnamefont
  {Brownnutt}}, \bibinfo {author} {\bibfnamefont {M.}~\bibnamefont {Kumph}},
  \bibinfo {author} {\bibfnamefont {P.}~\bibnamefont {Rabl}},\ and\ \bibinfo
  {author} {\bibfnamefont {R.}~\bibnamefont {Blatt}},\ }\href
  {https://doi.org/10.1103/RevModPhys.87.1419} {\bibfield  {journal} {\bibinfo
  {journal} {Rev. Mod. Phys.}\ }\textbf {\bibinfo {volume} {87}},\ \bibinfo
  {pages} {1419} (\bibinfo {year} {2015})}\BibitemShut {NoStop}%
\bibitem [{\citenamefont {Choi}(1975)}]{CHOI1975}%
  \BibitemOpen
  \bibfield  {author} {\bibinfo {author} {\bibfnamefont {M.-D.}\ \bibnamefont
  {Choi}},\ }\href
  {https://doi.org/https://doi.org/10.1016/0024-3795(75)90075-0} {\bibfield
  {journal} {\bibinfo  {journal} {Linear Algebra and its Applications}\
  }\textbf {\bibinfo {volume} {10}},\ \bibinfo {pages} {285} (\bibinfo {year}
  {1975})}\BibitemShut {NoStop}%
\bibitem [{\citenamefont {Jozsa}(1994)}]{Jozsa1994}%
  \BibitemOpen
  \bibfield  {author} {\bibinfo {author} {\bibfnamefont {R.}~\bibnamefont
  {Jozsa}},\ }\href {https://doi.org/10.1080/09500349414552171} {\bibfield
  {journal} {\bibinfo  {journal} {Journal of Modern Optics}\ }\textbf {\bibinfo
  {volume} {41}},\ \bibinfo {pages} {2315} (\bibinfo {year}
  {1994})}\BibitemShut {NoStop}%
\bibitem [{\citenamefont {Powell}(1994)}]{Powell1994}%
  \BibitemOpen
  \bibfield  {author} {\bibinfo {author} {\bibfnamefont {M.~J.~D.}\
  \bibnamefont {Powell}},\ }\bibinfo {title} {A direct search optimization
  method that models the objective and constraint functions by linear
  interpolation},\ in\ \href {https://doi.org/10.1007/978-94-015-8330-5_4}
  {\emph {\bibinfo {booktitle} {Advances in Optimization and Numerical
  Analysis}}}\ (\bibinfo  {publisher} {Springer Netherlands},\ \bibinfo
  {address} {Dordrecht},\ \bibinfo {year} {1994})\ pp.\ \bibinfo {pages}
  {51--67}\BibitemShut {NoStop}%
\bibitem [{\citenamefont {Schnabel}(2017)}]{Schnabel2017}%
  \BibitemOpen
  \bibfield  {author} {\bibinfo {author} {\bibfnamefont {R.}~\bibnamefont
  {Schnabel}},\ }\href
  {https://doi.org/https://doi.org/10.1016/j.physrep.2017.04.001} {\bibfield
  {journal} {\bibinfo  {journal} {Physics Reports}\ }\textbf {\bibinfo {volume}
  {684}},\ \bibinfo {pages} {1} (\bibinfo {year} {2017})}\BibitemShut {NoStop}%
\bibitem [{\citenamefont {Vitale}\ \emph {et~al.}(2023)\citenamefont {Vitale},
  \citenamefont {Rath}, \citenamefont {Jurcevic}, \citenamefont {Elben},
  \citenamefont {Branciard},\ and\ \citenamefont
  {Vermersch}}]{vitale2023estimation}%
  \BibitemOpen
  \bibfield  {author} {\bibinfo {author} {\bibfnamefont {V.}~\bibnamefont
  {Vitale}}, \bibinfo {author} {\bibfnamefont {A.}~\bibnamefont {Rath}},
  \bibinfo {author} {\bibfnamefont {P.}~\bibnamefont {Jurcevic}}, \bibinfo
  {author} {\bibfnamefont {A.}~\bibnamefont {Elben}}, \bibinfo {author}
  {\bibfnamefont {C.}~\bibnamefont {Branciard}},\ and\ \bibinfo {author}
  {\bibfnamefont {B.}~\bibnamefont {Vermersch}},\ }\href@noop {} {\bibinfo
  {title} {Estimation of the quantum fisher information on a quantum
  processor}} (\bibinfo {year} {2023}),\ \Eprint
  {https://arxiv.org/abs/2307.16882} {arXiv:2307.16882 [quant-ph]} \BibitemShut
  {NoStop}%
\bibitem [{\citenamefont {Ahmed}\ \emph {et~al.}(2021)\citenamefont {Ahmed},
  \citenamefont {S\'anchez Mu\~noz}, \citenamefont {Nori},\ and\ \citenamefont
  {Kockum}}]{Ahmed2021}%
  \BibitemOpen
  \bibfield  {author} {\bibinfo {author} {\bibfnamefont {S.}~\bibnamefont
  {Ahmed}}, \bibinfo {author} {\bibfnamefont {C.}~\bibnamefont {S\'anchez
  Mu\~noz}}, \bibinfo {author} {\bibfnamefont {F.}~\bibnamefont {Nori}},\ and\
  \bibinfo {author} {\bibfnamefont {A.~F.}\ \bibnamefont {Kockum}},\ }\href
  {https://doi.org/10.1103/PhysRevLett.127.140502} {\bibfield  {journal}
  {\bibinfo  {journal} {Phys. Rev. Lett.}\ }\textbf {\bibinfo {volume} {127}},\
  \bibinfo {pages} {140502} (\bibinfo {year} {2021})}\BibitemShut {NoStop}%
\bibitem [{\citenamefont {Elben}\ \emph {et~al.}(2023)\citenamefont {Elben},
  \citenamefont {Flammia}, \citenamefont {Huang}, \citenamefont {Kueng},
  \citenamefont {Preskill}, \citenamefont {Vermersch},\ and\ \citenamefont
  {Zoller}}]{Elben2023}%
  \BibitemOpen
  \bibfield  {author} {\bibinfo {author} {\bibfnamefont {A.}~\bibnamefont
  {Elben}}, \bibinfo {author} {\bibfnamefont {S.~T.}\ \bibnamefont {Flammia}},
  \bibinfo {author} {\bibfnamefont {H.-Y.}\ \bibnamefont {Huang}}, \bibinfo
  {author} {\bibfnamefont {R.}~\bibnamefont {Kueng}}, \bibinfo {author}
  {\bibfnamefont {J.}~\bibnamefont {Preskill}}, \bibinfo {author}
  {\bibfnamefont {B.}~\bibnamefont {Vermersch}},\ and\ \bibinfo {author}
  {\bibfnamefont {P.}~\bibnamefont {Zoller}},\ }\href
  {https://doi.org/10.1038/s42254-022-00535-2} {\bibfield  {journal} {\bibinfo
  {journal} {Nature Reviews Physics}\ }\textbf {\bibinfo {volume} {5}},\
  \bibinfo {pages} {9} (\bibinfo {year} {2023})}\BibitemShut {NoStop}%
\bibitem [{\citenamefont {Huang}\ \emph {et~al.}(2020)\citenamefont {Huang},
  \citenamefont {Kueng},\ and\ \citenamefont
  {Preskill}}]{Huang2020classicalshadows}%
  \BibitemOpen
  \bibfield  {author} {\bibinfo {author} {\bibfnamefont {H.-Y.}\ \bibnamefont
  {Huang}}, \bibinfo {author} {\bibfnamefont {R.}~\bibnamefont {Kueng}},\ and\
  \bibinfo {author} {\bibfnamefont {J.}~\bibnamefont {Preskill}},\ }\href
  {https://doi.org/10.1038/s41567-020-0932-7} {\bibfield  {journal} {\bibinfo
  {journal} {Nature Physics}\ }\textbf {\bibinfo {volume} {16}},\ \bibinfo
  {pages} {1050} (\bibinfo {year} {2020})}\BibitemShut {NoStop}%
\bibitem [{\citenamefont {Struchalin}\ \emph {et~al.}(2021)\citenamefont
  {Struchalin}, \citenamefont {Zagorovskii}, \citenamefont {Kovlakov},
  \citenamefont {Straupe},\ and\ \citenamefont {Kulik}}]{Struchalin2021}%
  \BibitemOpen
  \bibfield  {author} {\bibinfo {author} {\bibfnamefont {G.}~\bibnamefont
  {Struchalin}}, \bibinfo {author} {\bibfnamefont {Y.~A.}\ \bibnamefont
  {Zagorovskii}}, \bibinfo {author} {\bibfnamefont {E.}~\bibnamefont
  {Kovlakov}}, \bibinfo {author} {\bibfnamefont {S.}~\bibnamefont {Straupe}},\
  and\ \bibinfo {author} {\bibfnamefont {S.}~\bibnamefont {Kulik}},\ }\href
  {https://doi.org/10.1103/PRXQuantum.2.010307} {\bibfield  {journal} {\bibinfo
   {journal} {PRX Quantum}\ }\textbf {\bibinfo {volume} {2}},\ \bibinfo {pages}
  {010307} (\bibinfo {year} {2021})}\BibitemShut {NoStop}%
\bibitem [{\citenamefont {T\'oth}\ and\ \citenamefont {Petz}(2013)}]{Toth2013}%
  \BibitemOpen
  \bibfield  {author} {\bibinfo {author} {\bibfnamefont {G.}~\bibnamefont
  {T\'oth}}\ and\ \bibinfo {author} {\bibfnamefont {D.}~\bibnamefont {Petz}},\
  }\href {https://doi.org/10.1103/PhysRevA.87.032324} {\bibfield  {journal}
  {\bibinfo  {journal} {Phys. Rev. A}\ }\textbf {\bibinfo {volume} {87}},\
  \bibinfo {pages} {032324} (\bibinfo {year} {2013})}\BibitemShut {NoStop}%
\bibitem [{\citenamefont {Toth}(2018)}]{toth2018lower}%
  \BibitemOpen
  \bibfield  {author} {\bibinfo {author} {\bibfnamefont {G.}~\bibnamefont
  {Toth}},\ }\href@noop {} {} (\bibinfo {year} {2018}),\ \Eprint
  {https://arxiv.org/abs/1701.07461} {arXiv:1701.07461 [quant-ph]} \BibitemShut
  {NoStop}%
\bibitem [{\citenamefont {T\'oth}\ and\ \citenamefont
  {Fr\"owis}(2022)}]{Toth2022}%
  \BibitemOpen
  \bibfield  {author} {\bibinfo {author} {\bibfnamefont {G.}~\bibnamefont
  {T\'oth}}\ and\ \bibinfo {author} {\bibfnamefont {F.}~\bibnamefont
  {Fr\"owis}},\ }\href {https://doi.org/10.1103/PhysRevResearch.4.013075}
  {\bibfield  {journal} {\bibinfo  {journal} {Phys. Rev. Res.}\ }\textbf
  {\bibinfo {volume} {4}},\ \bibinfo {pages} {013075} (\bibinfo {year}
  {2022})}\BibitemShut {NoStop}%
\bibitem [{\citenamefont {Apellaniz}\ \emph {et~al.}(2017)\citenamefont
  {Apellaniz}, \citenamefont {Kleinmann}, \citenamefont {G\"uhne},\ and\
  \citenamefont {T\'oth}}]{Apellaniz2017}%
  \BibitemOpen
  \bibfield  {author} {\bibinfo {author} {\bibfnamefont {I.}~\bibnamefont
  {Apellaniz}}, \bibinfo {author} {\bibfnamefont {M.}~\bibnamefont
  {Kleinmann}}, \bibinfo {author} {\bibfnamefont {O.}~\bibnamefont {G\"uhne}},\
  and\ \bibinfo {author} {\bibfnamefont {G.}~\bibnamefont {T\'oth}},\ }\href
  {https://doi.org/10.1103/PhysRevA.95.032330} {\bibfield  {journal} {\bibinfo
  {journal} {Phys. Rev. A}\ }\textbf {\bibinfo {volume} {95}},\ \bibinfo
  {pages} {032330} (\bibinfo {year} {2017})}\BibitemShut {NoStop}%
\bibitem [{\citenamefont {Buhrman}\ \emph {et~al.}(2001)\citenamefont
  {Buhrman}, \citenamefont {Cleve}, \citenamefont {Watrous},\ and\
  \citenamefont {de~Wolf}}]{Buhrman2001}%
  \BibitemOpen
  \bibfield  {author} {\bibinfo {author} {\bibfnamefont {H.}~\bibnamefont
  {Buhrman}}, \bibinfo {author} {\bibfnamefont {R.}~\bibnamefont {Cleve}},
  \bibinfo {author} {\bibfnamefont {J.}~\bibnamefont {Watrous}},\ and\ \bibinfo
  {author} {\bibfnamefont {R.}~\bibnamefont {de~Wolf}},\ }\href
  {https://doi.org/10.1103/PhysRevLett.87.167902} {\bibfield  {journal}
  {\bibinfo  {journal} {Phys. Rev. Lett.}\ }\textbf {\bibinfo {volume} {87}},\
  \bibinfo {pages} {167902} (\bibinfo {year} {2001})}\BibitemShut {NoStop}%
\bibitem [{\citenamefont {Abanin}\ and\ \citenamefont
  {Demler}(2012)}]{Abanin2012}%
  \BibitemOpen
  \bibfield  {author} {\bibinfo {author} {\bibfnamefont {D.~A.}\ \bibnamefont
  {Abanin}}\ and\ \bibinfo {author} {\bibfnamefont {E.}~\bibnamefont
  {Demler}},\ }\href {https://doi.org/10.1103/PhysRevLett.109.020504}
  {\bibfield  {journal} {\bibinfo  {journal} {Phys. Rev. Lett.}\ }\textbf
  {\bibinfo {volume} {109}},\ \bibinfo {pages} {020504} (\bibinfo {year}
  {2012})}\BibitemShut {NoStop}%
\bibitem [{\citenamefont {Johnson}\ \emph {et~al.}(2014)\citenamefont
  {Johnson}, \citenamefont {Clark},\ and\ \citenamefont
  {Jaksch}}]{Johnson2014}%
  \BibitemOpen
  \bibfield  {author} {\bibinfo {author} {\bibfnamefont {T.~H.}\ \bibnamefont
  {Johnson}}, \bibinfo {author} {\bibfnamefont {S.~R.}\ \bibnamefont {Clark}},\
  and\ \bibinfo {author} {\bibfnamefont {D.}~\bibnamefont {Jaksch}},\ }\href
  {https://doi.org/10.1140/epjqt10} {\bibfield  {journal} {\bibinfo  {journal}
  {EPJ Quantum Technology}\ }\textbf {\bibinfo {volume} {1}},\ \bibinfo {pages}
  {10} (\bibinfo {year} {2014})}\BibitemShut {NoStop}%
\bibitem [{\citenamefont {Daley}\ \emph {et~al.}(2022)\citenamefont {Daley},
  \citenamefont {Bloch}, \citenamefont {Kokail}, \citenamefont {Flannigan},
  \citenamefont {Pearson}, \citenamefont {Troyer},\ and\ \citenamefont
  {Zoller}}]{Daley2022}%
  \BibitemOpen
  \bibfield  {author} {\bibinfo {author} {\bibfnamefont {A.~J.}\ \bibnamefont
  {Daley}}, \bibinfo {author} {\bibfnamefont {I.}~\bibnamefont {Bloch}},
  \bibinfo {author} {\bibfnamefont {C.}~\bibnamefont {Kokail}}, \bibinfo
  {author} {\bibfnamefont {S.}~\bibnamefont {Flannigan}}, \bibinfo {author}
  {\bibfnamefont {N.}~\bibnamefont {Pearson}}, \bibinfo {author} {\bibfnamefont
  {M.}~\bibnamefont {Troyer}},\ and\ \bibinfo {author} {\bibfnamefont
  {P.}~\bibnamefont {Zoller}},\ }\href
  {https://doi.org/10.1038/s41586-022-04940-6} {\bibfield  {journal} {\bibinfo
  {journal} {Nature}\ }\textbf {\bibinfo {volume} {607}},\ \bibinfo {pages}
  {667} (\bibinfo {year} {2022})}\BibitemShut {NoStop}%
\bibitem [{\citenamefont {Wineland}\ \emph {et~al.}(1998)\citenamefont
  {Wineland}, \citenamefont {Monroe}, \citenamefont {Itano}, \citenamefont
  {Leibfried}, \citenamefont {King},\ and\ \citenamefont
  {Meekhof}}]{wineland1998experimental}%
  \BibitemOpen
  \bibfield  {author} {\bibinfo {author} {\bibfnamefont {D.~J.}\ \bibnamefont
  {Wineland}}, \bibinfo {author} {\bibfnamefont {C.}~\bibnamefont {Monroe}},
  \bibinfo {author} {\bibfnamefont {W.~M.}\ \bibnamefont {Itano}}, \bibinfo
  {author} {\bibfnamefont {D.}~\bibnamefont {Leibfried}}, \bibinfo {author}
  {\bibfnamefont {B.~E.}\ \bibnamefont {King}},\ and\ \bibinfo {author}
  {\bibfnamefont {D.~M.}\ \bibnamefont {Meekhof}},\ }\href@noop {} {\bibinfo
  {title} {Experimental issues in coherent quantum-state manipulation of
  trapped atomic ions}} (\bibinfo {year} {1998}),\ \Eprint
  {https://arxiv.org/abs/quant-ph/9710025} {arXiv:quant-ph/9710025 [quant-ph]}
  \BibitemShut {NoStop}%
\bibitem [{\citenamefont {Alton}\ \emph {et~al.}(2011)\citenamefont {Alton},
  \citenamefont {Stern}, \citenamefont {Aoki}, \citenamefont {Lee},
  \citenamefont {Ostby}, \citenamefont {Vahala},\ and\ \citenamefont
  {Kimble}}]{Alton2011}%
  \BibitemOpen
  \bibfield  {author} {\bibinfo {author} {\bibfnamefont {D.~J.}\ \bibnamefont
  {Alton}}, \bibinfo {author} {\bibfnamefont {N.~P.}\ \bibnamefont {Stern}},
  \bibinfo {author} {\bibfnamefont {T.}~\bibnamefont {Aoki}}, \bibinfo {author}
  {\bibfnamefont {H.}~\bibnamefont {Lee}}, \bibinfo {author} {\bibfnamefont
  {E.}~\bibnamefont {Ostby}}, \bibinfo {author} {\bibfnamefont {K.~J.}\
  \bibnamefont {Vahala}},\ and\ \bibinfo {author} {\bibfnamefont {H.~J.}\
  \bibnamefont {Kimble}},\ }\href {https://doi.org/10.1038/nphys1837}
  {\bibfield  {journal} {\bibinfo  {journal} {Nature Physics}\ }\textbf
  {\bibinfo {volume} {7}},\ \bibinfo {pages} {159} (\bibinfo {year}
  {2011})}\BibitemShut {NoStop}%
\bibitem [{\citenamefont {Frattini}\ \emph {et~al.}(2017)\citenamefont
  {Frattini}, \citenamefont {Vool}, \citenamefont {Shankar}, \citenamefont
  {Narla}, \citenamefont {Sliwa},\ and\ \citenamefont
  {Devoret}}]{Frattini2017}%
  \BibitemOpen
  \bibfield  {author} {\bibinfo {author} {\bibfnamefont {N.~E.}\ \bibnamefont
  {Frattini}}, \bibinfo {author} {\bibfnamefont {U.}~\bibnamefont {Vool}},
  \bibinfo {author} {\bibfnamefont {S.}~\bibnamefont {Shankar}}, \bibinfo
  {author} {\bibfnamefont {A.}~\bibnamefont {Narla}}, \bibinfo {author}
  {\bibfnamefont {K.~M.}\ \bibnamefont {Sliwa}},\ and\ \bibinfo {author}
  {\bibfnamefont {M.~H.}\ \bibnamefont {Devoret}},\ }\href
  {https://doi.org/10.1063/1.4984142} {\bibfield  {journal} {\bibinfo
  {journal} {Applied Physics Letters}\ }\textbf {\bibinfo {volume} {110}},\
  \bibinfo {pages} {222603} (\bibinfo {year} {2017})}\BibitemShut {NoStop}%
\bibitem [{\citenamefont {He}\ \emph {et~al.}(2023)\citenamefont {He},
  \citenamefont {Lu}, \citenamefont {Bao}, \citenamefont {Xue}, \citenamefont
  {Jiang}, \citenamefont {Wang}, \citenamefont {Roudsari}, \citenamefont
  {Delsing}, \citenamefont {Tsai},\ and\ \citenamefont {Lin}}]{he2023fast}%
  \BibitemOpen
  \bibfield  {author} {\bibinfo {author} {\bibfnamefont {X.~L.}\ \bibnamefont
  {He}}, \bibinfo {author} {\bibfnamefont {Y.}~\bibnamefont {Lu}}, \bibinfo
  {author} {\bibfnamefont {D.~Q.}\ \bibnamefont {Bao}}, \bibinfo {author}
  {\bibfnamefont {H.}~\bibnamefont {Xue}}, \bibinfo {author} {\bibfnamefont
  {W.~B.}\ \bibnamefont {Jiang}}, \bibinfo {author} {\bibfnamefont
  {Z.}~\bibnamefont {Wang}}, \bibinfo {author} {\bibfnamefont {A.~F.}\
  \bibnamefont {Roudsari}}, \bibinfo {author} {\bibfnamefont {P.}~\bibnamefont
  {Delsing}}, \bibinfo {author} {\bibfnamefont {J.~S.}\ \bibnamefont {Tsai}},\
  and\ \bibinfo {author} {\bibfnamefont {Z.~R.}\ \bibnamefont {Lin}},\ }\href
  {https://doi.org/10.1038/s41467-023-42057-0} {\bibfield  {journal} {\bibinfo
  {journal} {Nature Communications}\ }\textbf {\bibinfo {volume} {14}},\
  \bibinfo {pages} {6358} (\bibinfo {year} {2023})}\BibitemShut {NoStop}%
\bibitem [{\citenamefont {Arias}\ \emph {et~al.}(2023)\citenamefont {Arias},
  \citenamefont {Triana}, \citenamefont {Delgado},\ and\ \citenamefont
  {Herrera}}]{arias2023coherent}%
  \BibitemOpen
  \bibfield  {author} {\bibinfo {author} {\bibfnamefont {M.}~\bibnamefont
  {Arias}}, \bibinfo {author} {\bibfnamefont {J.~F.}\ \bibnamefont {Triana}},
  \bibinfo {author} {\bibfnamefont {A.}~\bibnamefont {Delgado}},\ and\ \bibinfo
  {author} {\bibfnamefont {F.}~\bibnamefont {Herrera}},\ }\href@noop {} {}
  (\bibinfo {year} {2023}),\ \Eprint {https://arxiv.org/abs/2309.12216}
  {arXiv:2309.12216 [quant-ph]} \BibitemShut {NoStop}%
\bibitem [{\citenamefont {Eisert}\ \emph {et~al.}(2010)\citenamefont {Eisert},
  \citenamefont {Cramer},\ and\ \citenamefont {Plenio}}]{Eisert2010}%
  \BibitemOpen
  \bibfield  {author} {\bibinfo {author} {\bibfnamefont {J.}~\bibnamefont
  {Eisert}}, \bibinfo {author} {\bibfnamefont {M.}~\bibnamefont {Cramer}},\
  and\ \bibinfo {author} {\bibfnamefont {M.~B.}\ \bibnamefont {Plenio}},\
  }\href {https://doi.org/10.1103/RevModPhys.82.277} {\bibfield  {journal}
  {\bibinfo  {journal} {Rev. Mod. Phys.}\ }\textbf {\bibinfo {volume} {82}},\
  \bibinfo {pages} {277} (\bibinfo {year} {2010})}\BibitemShut {NoStop}%
\end{thebibliography}%
%\bibliography{references,referencesAlex}

\end{document}